\documentstyle[preprint,version2,aps]{revtex}
\begin{document}
\draft
\begin{title}
Gravitational waves from merging compact binaries:\\
How accurately can one extract the binary's
parameters \\
from the inspiral waveform?
\end{title}
\author{Curt Cutler and \'Eanna E. Flanagan}
\begin{instit}
Theoretical Astrophysics, California Institute of Technology,
Pasadena, California 91125
\end{instit}

\begin{abstract}
The most promising source of gravitational waves for the planned
kilometer-size laser-interferometer detectors LIGO and VIRGO are
merging compact binaries, i.e., neutron star/neutron star (NS/NS),
neutron star/black hole (NS/BH), and black hole/black-hole (BH/BH)
binaries.  We investigate how accurately the distance to the source
and the masses and spins of the two bodies will be measured from the
inspiral gravitational wave signals by the three detector LIGO/VIRGO network
using ``advanced detectors'' (those present a few years after initial
operation).
The large number of cycles in the observable waveform increases our
sensitivity to those parameters that affect the inspiral rate, and
thereby the evolution of the waveform's phase.  These parameters are
thus measured much more accurately than parameters which affect the
waveform's polarization or amplitude.  To lowest order in a
post-Newtonian expansion, the evolution of the waveform's phase
depends only on the combination ${\cal M} \equiv (M_1 M_2)^{3/5}(M_1
+M_2)^{-1/5}$ of the masses $M_1$ and $M_2$ of the two bodies, which
is known as the ``chirp mass.''  To post-1-Newtonian order, the
waveform's phase also depends sensitively on the binary's reduced
mass $\mu \equiv M_1 M_2/ (M_1 + M_2)$, allowing, in principle, a
measurement of both $M_1$ and $M_2$ with high accuracy.  We show that
the principal obstruction to measuring $M_1$ and $M_2$ is the
post-1.5-Newtonian effect of the bodies' spins on the waveform's
phase, which can mimic the effects that allow $\mu$ to be determined.
The chirp mass is measurable with an accuracy $\Delta {\cal M}/{\cal
M} \approx 0.1\%-1\%$.  Although this is a remarkably small error bar,
it is $\sim 10$ times larger than previous estimates of $\Delta {\cal
M}/{\cal M}$ which neglected post-Newtonian effects.  The reduced mass
is measurable to $\sim 10\%-15\%$ for NS/NS and NS/BH binaries, and
$\sim 50\%$ for BH/BH binaries (assuming $10M_\odot$ BH's).
Measurements of the masses and spins are strongly correlated; there is
a combination of $\mu$ and the spin angular momenta that is
measured to within $\sim 1\%$.  Moreover, {\it if} both spins were
somehow known to be small ($\alt 0.01 M_1^2$ and $\alt 0.01 M_2^2$,
respectively), then $\mu$ could be determined to within $\sim 1\%$.
Finally, building on earlier work of Markovi\'{c}, we derive an
approximate, analytic expression for the accuracy $\Delta D$ of
measurements of the distance $D$ to the binary, for an arbitrary
network of detectors.  This expression is accurate to linear order in
$1/\rho$, where $\rho$ is the signal-to-noise ratio.  We also show
that, contrary to previous expectations, contributions to $\Delta D/D$
that are nonlinear in $1/\rho$ are significant, and we develop an
approximation scheme for including the dominant of these non-linear
effects.  Using a Monte-Carlo simulation, we estimate that distance
measurement accuracies will be $\le 15\%$ for $\sim 8\%$ of the
detected signals, and $\le 30\%$ for $\sim 60\%$ of the signals, for
the LIGO/VIRGO 3-detector network.
\end{abstract}
\newpage
\section{INTRODUCTION}
\label{intro}

Neutron star-neutron star (NS-NS) binaries with orbital periods of
less than half a day will spiral together and merge in less than a
Hubble time, due to gravitational radiation reaction.  Three such
short-period NS-NS binaries have been observed in our Galaxy; when
extrapolated to the rest of the Universe these observations result in
an estimated NS-NS merger rate in the Universe of $\sim 10^2 \,\, {\rm
yr}^{-1} {\rm Gpc}^{-3}$ \cite{Narayan,sterl}.  A strong gravitational
wave signal is emitted during the last few minutes of inspiral, before
the tidal-disruption/coalescence stage begins.  If the Laser
Interferometer Gravitational Wave Observatory (LIGO)
\cite{ligoscience}, and its French-Italian counterpart VIRGO
\cite{virgo}, achieve the so-called ``advanced detector'' sensitivity
level of Ref.~\cite{ligoscience}, then they will be able to detect
gravitational waves from the last few minutes of NS-NS inspirals out to
distances of order $\sim 1 \, {\rm Gpc}$ \cite{ligoscience}.  Hence,
event rates of order $10^2 \,\,{\rm yr}^{-1}$ may be achieved.  While
there is no direct observational evidence relevant to the merger
rates for neutron star-black hole (NS-BH) and black hole-black hole
(BH-BH) binaries, arguments based on progenitor evolution scenarios
suggest that these merger rates may also be on the order of $10^2 \,\, {\rm
yr}^{-1} {\rm Gpc}^{-3}$ \cite{sterl,tutukov}.
The merger of two $10 M_\odot$ black
holes would be detectable by LIGO/VIRGO out to cosmological distances
at redshifts of $\sim 2-3$.

The gravitational waveforms arriving at the detectors depend on the
inspiraling bodies' masses and spins, the distance to the binary,
its angular position on the sky, and the orientation of the binary's
orbital plane.  By comparing the observed
waveforms with theoretically derived templates, the observers will
extract these parameters to a level of accuracy that is determined by
the noise in the detectors, and by the detectors' relative positions
and orientations.  From the output of a single detector, there will be
sufficient information to determine the masses of the two bodies, but
not their distance or their location on the sky.  By combining the
outputs of the three LIGO/VIRGO detectors, it should be possible to
determine the location of the binary on the sky to within $\sim$one
degree \cite{schutz,tinto}, and the distance to the binary to within
$\sim 30\%$.

There are many potential applications of such measurements, as has
been emphasized by Schutz \cite{schutz}.  For example, coalescing
binaries are potentially very useful standard candles for astronomical
distance measurements --- it has been estimated that from $\sim 10^2$
detected NS-NS events it will be possible to determine the Hubble
constant $H_0$ to within $\sim 10\%$
\cite{cutleretal,draza,schutz0,finn_cosmo}.  It may also be possible
to measure NS radii, and thus constrain the NS equation of state, by
measuring the frequency at which the tidal disruption of the neutron
star causes the waves to shut off \cite{cutleretal}.  And from
gravitational wave observations of the final coalescence of two black holes,
there may follow new insights into gravitational dynamics in the
highly nonlinear regime.  The effectiveness of these and other
applications depends on the accuracy with which one can read off, from
the measured waveform, parameters such as the distance to the binary
and the masses of its two components.

The purpose of this paper is to estimate the limits on measurement
accuracies that arise from sources of noise that are intrinsic to the
detectors.  These sources of noise include, for example, thermal
vibrations in the interferometers' suspended test masses, and
randomness in the arrival times of individual photons at the
interferometers' mirrors (photon shot-noise), which simulate in the
interferometers' output the effects of gravitational waves \cite{ligoscience}.
Intrinsic detector noise is expected to be the dominant source of
error in the determination of coalescing binary parameters, in part because
gravitational waves interact very weakly with matter through which
they pass \cite{300years}.  Other possible sources of
error which we do not consider here include (i) systematic errors due
to insufficiently accurate theoretical modeling of the gravitational
waveforms, which will be important primarily for mass and spin
measurements \cite{cutleretal,cutler_finn}; and (ii)
amplification/deamplification of the wave amplitudes by gravitational
lensing effects, which will be important primarily for distance
measurements.  See Markovi\'{c} \cite{draza} for a detailed discussion of
this issue.

Many of our conclusions have already been summarized in Cutler
et al.~\cite{cutleretal}.
Initial measurement accuracy
analyses have been carried out by Finn and Chernoff \cite{finn2}, and
by Jaranowski and Krolak \cite{krolak2}, using a simplified,
``Newtonian'' model of the waveform.  While Newtonian waveforms are
adequate for predicting how accurately one can measure the distance to
the source, they do not allow one to calculate how accurately the
individual masses can be measured, as we explain
below.  For this purpose, one must include post-Newtonian corrections
to the waveform.

Much of our work was guided by the following sequence of considerations.
These considerations introduce some of the
issues addressed in this paper, motivate a number of approximations
that we make in our analysis, and give a preview of some of our
main conclusions.

First, coalescing binaries are very ``clean'' sources of gravitational
waves: the waveform is determined to high accuracy by a relatively
small number of parameters \cite{schutz}.  These parameters are the
source's location, orientation, time of coalescence, and orbital phase at
coalescence, as well as the bodies' masses and spin angular momenta.
Various other
complicating physical effects, not described by these parameters,
can be shown to be unimportant.  We can generally assume, for instance,
that the orbits are circular.  This is because radiation reaction
causes the orbit's eccentricity $\varepsilon$ to decrease during the
inspiral, according to $\varepsilon^2 \propto P^{19/9}$, where $P$ is
the orbital period \cite{peters}.  (The effect of a small eccentricity
on the phase of the waveform scales like $\varepsilon^2$.)  The
emitted gravitational waves are in the frequency band accessible to LIGO
only for the last few minutes of inspiral, when $P < 0.2 \, {\rm
sec}$.  Thus a binary born with $\varepsilon$ of order unity and $P >1
$ hour will have $\varepsilon^2 < 10^{-9}$ by the time it becomes
``visible'' to LIGO \cite{quinlan}.  Also, tidal interactions between
the bodies have been shown to be negligible
\cite{Bildsten_Cutler,Kochanek} (except for the last few orbits), so
for our purposes the bodies can be treated as structureless, spinning
point masses \cite{yes_but}.

Second, this high predictability of the gravitational waveforms means
that the technique
of matched filtering can be used to detect the waves \cite{300years}.
For the most
distant (most frequently observed) sources, this will involve
extracting the waveforms from the considerably larger
instrumental noise in which they will be imbedded.  The technique
works as follows \cite{300years}.
The measured strain amplitude in each detector
\begin{equation}
\label{hn}
s(t) = h(t) + n(t)
\end{equation}
consists of a (possibly present) signal $h(t)$, and the detector noise
$n(t)$, which we assume is Gaussian.
To detect any imbedded signal, one first suppresses
those frequency components of the signal at which the detector noise is
largest by convolving with Wiener's optimal filter $w(t)$: thus, $s(t)
\rightarrow \int w(t-\tau) s(\tau) d\tau$ \cite{WZ}.  Then, for each inspiral
waveform $\hat h(t)$ in a large set of theoretical template waveforms, one
computes the signal-to-noise ratio $S/N$, defined by
\begin{equation}
\label{sn}
{S\over N}[\hat h] =
{ \int \hat h(t)\,w(t-\tau)\,s(\tau)\, d\tau dt \over {\rm rms}
\int \hat h(t) \, w(t-\tau)\, n(\tau) \, d\tau dt  }.
\end{equation}
In Eq.~(\ref{sn}), the denominator is what {\it would} be the
root-mean-square value of the numerator,
{\it if} the detector output (\ref{hn}) consisted of noise alone.  Thus, when
no
gravitational wave is present, each $S/N [\hat h]$ is a random variable
with Gaussian distribution and root-mean-square equal to $1$.
Conversely, if $S/N [\hat h]$ is sufficiently large as to basically
preclude the possibility of
its arising from noise alone --- for {\it any} of the $\sim 10^{15}$
template waveforms that will be applied to the data each year ---
then one can assert with high confidence that a gravitational wave $h$ has been
detected, and that $h$ is close to $\hat h$.  It is easy to show
that if some template waveform  $\hat h$ yields
a signal-to-noise ratio of
$S/N [\hat h] \ge 6.0$ in each of two detectors, then
with  $> 99 \%$ confidence a gravitational wave has
been detected \cite{cutleretal}.  Defining the combined
signal-to-noise ratio $\rho$ of a
network of detectors by
\begin{equation}
\label{rho}
\rho \equiv \sqrt{\sum_a \rho_a^2},
\end{equation}
where $\rho_a$ is the $S/N$ in the $a$th detector, we see that $\rho
\approx 8.5$ represents the ``detection threshold'' for two detectors.
For a three-detector network, the detection threshold is still $\rho
\approx 8.5$, corresponding to $S/N \agt 4.9$ in each detector. Since
detections at threshold represent the most distant coalescences that
one can observe (given the binary's masses, its orientation, and its
angular position on the sky), and since coalescing binaries are
presumably distributed roughly uniformly on large scales ($\agt 100
\, {\rm Mpc}$), the mean
value of $\rho$ for detected events will be roughly $1.5$ times the
threshold value \cite{flanagan_thorne}.  Thus ``typical'' detections
will have $\rho \approx 12.7$.  Similarly, the strongest $1 \%$ of
signals should have $\rho \agt 40$; i.e., $(100)^{1/3}$ times the
threshold value.

Third, much more information is obtainable from the waveform
than one might naively expect, for the following reason.  The LIGO and
VIRGO detectors will be broad-band detectors, with good sensitivity in
the frequency range $10-500 \, {\rm Hz}$. The gravitational wave trains from
inspiraling stellar-mass binaries typically contain $\sim 10^3$ cycles
in this range.  Now, if the signal $h(t)$ and template $\hat h(t)$
lose phase with each other by just one cycle out of thousands, as they
sweep upwards in frequency from $\sim 10 \, {\rm Hz}$ to $\sim 500 \,
{\rm Hz}$, then the integral $\int {\hat h(\tau)h(t)w(t-\tau)d\tau
dt}$ will be significantly diminished.  Consequently the value of
$S/N[\hat h]$ will be small unless the phase
of the template waveform $\hat h$ is ``just right'' throughout the
inspiral.
Since the evolution of the waveform's phase is largely determined by
the masses of the two bodies (through their influence on the inspiral
rate), one might expect to measure the masses of the bodies with
fractional error $\sim 1/{\cal N}_{\rm cyc}$, where ${\cal N}_{\rm
cyc}$ is the total number of cycles in the observed waveform.  This
fractional error of $\sim 10^{-3} $ contrasts with the $\sim 20\%$
accuracy with which one can determine parameters, such as the distance
to the source, that do not affect the phase evolution (as was first pointed
out by Cutler et al.~\cite{cutleretal} and by Chernoff and Finn
\cite{finn2}).

Fourth, our extension of the measurement-error analysis to include
post-Newtonian
effects introduces the following new features.  To Newtonian order,
the gravitational wave signal depends on the two masses {\it only} through
the particular combination ${\cal M} \equiv \mu^{3/5} M^{2/5}$, where
$\mu$ is the reduced mass and $M$ is the total mass of the system.
This combination is referred to as the ``chirp mass.''  The
degeneracy in the dependence on the masses is broken, however, by
post-Newtonian effects that in principle allow one to determine the
individual masses $M_1$
and $ M_2$.  In the equation governing the evolution of
the waveform's phase [Eq.~(\ref{pndfdt}) below], the post-Newtonian
terms are $\sim M/r$ times smaller than the Newtonian terms, where $r$ is
the orbital separation. Since $ M/r \approx 1/20$ when the signal is
strongest, one might expect to determine each of the two masses $\sim
20$ times less accurately than $\cal M$.  We show in Sec.~\ref{sec3}
that this expectation is correct, {\it provided} the spins of the
bodies are known to be small.

Now, black holes and neutron stars in merging binaries may or may not
be rapidly spinning.  However if we cannot assume {\it a priori} that
their spin angular momenta are very small, then in attempting to find
the best fit to the data, we must allow for the possibility that the
spin angular momenta are of order their maximum possible values.  We
show in Sec.~\ref{spin_sec} that the extra ``confusion'' introduced by
the spin-dependence of the waveform worsens the accuracy of
individual mass measurements by more than an order of magnitude.  This
is easy to understand: the leading order spin terms in the orbital
evolution equation [Eq.~(\ref{pnsdfdt}) below] are only one-half
post-Newtonian order higher than the leading terms responsible for
splitting the mass degeneracy.  Therefore the effect on the
gravitational waveform of errors in $M_1$ and $M_2$ that keep ${\cal
M}$ fixed can be
approximately masked by somewhat larger, compensating errors in its
spins.  Hence the measured values of masses and spins will have
strongly correlated errors [cf.~Fig.~\ref{mubeta} below], thereby
increasing mass-measurement errors
\cite{correlation_effects}.  Our results for measurement accuracies are
summarized in Tables \ref{table1} and \ref{table2} and Fig.~\ref{m1m2}
below.

The rest of the paper is organized as follows.  In Sec.~\ref{sec2} we
review the anticipated detector noise levels, the basic elements of
signal processing, and the lowest-order, ``Newtonian'' waveforms.  In
Sec.~\ref{sec3} we calculate expected mass-measurement accuracies,
taking post-Newtonian effects into account.  We do this in two stages:
first neglecting spin effects in Sec.~\ref{no_spin}, then including
them in Sec.~\ref{spin_sec}.  Our emphasis is on learning {\it
roughly} what accuracies can be expected --- in part because to treat
the parameter-estimation problem in full generality would be extremely
complicated.  Therefore, we focus on a somewhat simplified ``model''
of the gravitational waveform, which nevertheless incorporates the
effects that are most important for determining the mass-extraction
accuracy.  A further approximation which we make is to use a linear
error-estimation formalism, which is valid when the errors are small
(or equivalently, when the signal-to-noise ratio is large).

Most of the information that allows one to measure the binary masses
is contained in the phase evolution of the waveform (rather than in
its amplitude or polarization).  Since all detectors in a detector
network measure very nearly the same phase evolution, for the purpose
of estimating mass measurement accuracies, to a good approximation it
is adequate to model measurements made by single detector.
The mass measurement errors for $N$ detectors are roughly those for
a single detector, divided by $\sqrt{N}$.

When measuring the distance $D$ to the binary, on the other hand, one
must also determine the position of the source on the sky and the
amplitude and polarization of the waveform.  Hence, to estimate
distance measurement accuracies, we must model measurements by an
entire detector network.  However in this case it is a needless
complication to use post-Newtonian waveforms; as we show in
Sec.~\ref{dist_errors} and Appendix \ref{decouple} below, to a good
approximation it is adequate to use Newtonian waveforms in the
analysis.  This is our approach in Sec.~\ref{dist_errors}, where we
estimate the distance measurement accuracy $\Delta D$ attainable by an
arbitrary network of detectors.  Jaranowski and Krolak have
numerically calculated in several specific cases the distance
measurement accuracy one can achieve with the LIGO/VIRGO network
\cite{krolak2}.  We provide a greatly simplified, analytic solution to
the distance-accuracy estimation problem, using an approximation due to
Markovi\'{c} \cite{draza}.  The approximation consists in neglecting
the effect on distance measurement errors of the relatively small
uncertainty in the angular position of the source on the sky.  We derive a
relatively simple formula for the rms
distance error $\Delta D$ in this approximation, which applies to any
number of detectors with arbitrary orientations.

This formula is derived using the linear error-estimation formalism
mentioned above, and consequently is accurate only to linear order in
$1/D$.  We show that, contrary to previous expectations, effects which
are nonlinear in $1/D$ have a significant effect (i.e., factors $\agt
2$) on the predicted distance-measurement accuracies, and develop an
approximate method of calculation which gives rough estimates of these
nonlinear effects.  This method is based on a Bayesian derivation of
the (non-Gaussian) probability distribution for the distance $D$,
which incorporates our {\it a priori} knowledge as well as the
information obtained from a gravitational wave measurement.  The
method also allows us to estimate values of $\Delta D$ for binaries
that are seen nearly face-on, for which, as pointed out by
Markovi\'{c} \cite{draza}, the linear error-estimation method breaks
down.  Our results for nearly face-on binaries are typically factors
of order $2$ to $3$ smaller than the upper-limit estimates given by
Markovi\'{c} \cite{draza}.

In Appendix \ref{stats} we extend the treatment of signal processing
given in Sec.~\ref{sec2} to incorporate (i) an arbitrary number of
detectors, (ii) the effects of {\it a priori} knowledge, and (iii)
estimation of measurement errors beyond the linear, Gaussian
approximation.  These extensions are required in Secs.~\ref{spin_sec}
and \ref{dist_errors}.  We also develop other tools which should be
useful in future analyses of LIGO/VIRGO measurement accuracies: we
derive an expression for the minimum signal-to-noise ratio $(S/N)_{\rm
min}$ necessary in order that the Gaussian approximation for
estimation of measurement accuracy be valid, and explain how to treat
degenerate points in parameter space at which the Gaussian approximation
breaks down.

In this paper we will focus on three fiducial types of binary ---
NS-NS, BH-NS, and BH-BH --- with fiducial masses
$M_{BH} = 10 M_\odot$ and $M_{NS} = 1.4 M_\odot$ (unless otherwise
specified).  Throughout we use units where $G=c=1$.
Thus all quantities are measured in units of seconds, except where,
for convenience, we use units of solar masses. The conversion factor
is $1 M_\odot = 4.926 \times 10^{-6} {\rm sec}$.

\section{Detection and Measurement \\
of Gravitational Waveforms}
\label{sec2}

\subsection{Detector characteristics}

In order to decide what information can be extracted from
gravitational waveforms, one must have a realistic model of the
detector noise $n(t)$.  This noise will have both Gaussian and
non-Gaussian components.  We will restrict our analysis to statistical
errors due to Gaussian noise.  It is likely that the effects of the
non-Gaussian components will be unimportant due to (i) the rejection
of events that are not simultaneously detected in two or more
detectors, and (ii) the filtering of the detector outputs with
theoretical waveform templates; however this issue needs further
study.

The remaining Gaussian noise can be described by its spectral density
$S_n(f)$, where $f$ is frequency.  The LIGO team has published an
estimate of the noise spectrum that might be attained a few years
after LIGO comes on line --- the so-called ``advanced detector'' noise
spectral density \cite{ligoscience}.  We use the
following rough analytic fit to their noise curve:
\FL
\begin{equation}
\label{snf}
S_n(f) = \left\{ \begin{array}{ll}
         \infty   & \mbox{ $f < 10 \, {\rm Hz}$,} \nonumber \\
         S_0 \left[(f_0/f)^4 +2 \left(1  + (f^2/f_0^2) \right) \right]
          & \mbox{    $f > 10 \,{\rm Hz}$} \\
		\end{array}
	\right.
\end{equation}
where $S_0 = 3 \times 10^{-48} \, {\rm Hz}^{-1}$ and $f_0 = 70 \,{\rm
Hz}$.  For frequencies $f < 10$ Hz, the noise due to seismic
vibrations is so large that we take it to be effectively infinite.
Thermal noise dominates in the frequency band $10 {\rm Hz} \alt f \alt 50
{\rm Hz}$, and
photon shot noise dominates for $f \agt 50$ Hz.
We refer the reader to
Refs.~\cite{ligoscience,300years} for more details on
the sources of noise.

The amount of detector noise determines the strength of the weakest
signals that can be detected, and thus the distance to which a given
type of source can be seen.  The noise level (\ref{snf}) will permit
the detection of NS-NS mergers out to $\sim 1 \, {\rm Gpc}$
\cite{ligoscience,finn2,flanagan_thorne}, giving an estimated
detection rate of
$\sim 10^2 \,\, {\rm yr}^{-1}$ \cite{sterl}.  In this paper we are
principally concerned {\it not} with detection issues, but rather with the
accuracy of parameter estimation. This accuracy of parameter estimation
depends only on the shape of the noise spectrum, and on the
signal-to-noise ($S/N$) of the detection; i.e., simultaneously
doubling both the noise levels and the signal strength leaves
measurement accuracy unchanged.  We normalize our results to a
fixed $S/N$, and hence our results are independent of the parameter $S_0$
appearing in the noise spectrum (\ref{snf}).

Since the LIGO team's publication \cite{ligoscience} of their estimate
of the advanced detector's noise curve, there have been new
developments in the understanding of the detector's thermal noise
which indicate that the advanced thermal noise spectrum may be flatter
than previously thought \cite{Kip}.  A modified noise-curve estimate,
reflecting this new understanding, has not yet been published.  Like
the noise spectrum in Ref.~\cite{ligoscience} on which our simplified
model (\ref{snf}) is based, the modified noise curve will depend on
the values of advanced detector parameters (such as the
quality-factors of modes of vibration of the
suspension wires and suspended masses) for which only rough estimates
are available.
Our approximate analytic
formula describing the modified advanced detector noise curve,
\FL
\begin{equation}
\label{snf2}
S_n(f) = \left\{ \begin{array}{ll}
         \infty   & \mbox{ $f < 10 \, {\rm Hz}$,} \nonumber \\
         S_m \,\alpha^{-4}\, (f/f_m)^{-5}
          & \mbox{    $10 \,{\rm Hz} \le f \le f_m/\alpha $} \nonumber \\
         S_m (f/f_m)^{-1}
          & \mbox{    $f_m/\alpha \le f \le \alpha f_m $} \nonumber \\

         S_m\, \alpha^{-3}\, (f/f_m)^2
          & \mbox{    $f \ge \alpha f_m $} \\
		\end{array}
	\right.
\end{equation}
where $S_m = 2.7 \times 10^{-47} \, {\rm Hz}^{-1}$, $f_m = 74 \ {\rm
Hz}$, and $\alpha = 3.8$ \cite{newnoise}, assumes
particular detector parameters that accentuate the
difference between Eqs.~(\ref{snf2}) and (\ref{snf}).
Since the ultimate shape of the noise curve is not yet well known, we
feel that it is useful to calculate the attainable measurement accuracies for
both of these shapes of the noise spectrum.
We shall see below [cf.~Tables II and III] that the
flatter spectrum of the modified noise curve (\ref{snf2}) leads
to a modest improvement in how accurately the binary's masses
can be measured (for fixed signal-to-noise).

\subsection{Review of parameter estimation}
\label{stats0}

In this section we give a concise summary of those elements of signal
processing that are necessary for parameter estimation.
The basic concepts of detection and measurement have also been
reviewed recently by Finn \cite{finn1} and by
Krolak {\it et al.}~\cite{krolak1}, in the specific context of laser
interferometer gravitational wave measurements.  In Appendix \ref{stats} we
give give a more
detailed treatment of parameter estimation, together with
an extensive discussion of the ways in
which the simplified linear formalism described in this section can
break down: (i) when the signal-to-noise of the detection is low, and
(ii) when our {\it a priori} knowledge of some of the binary
parameters is not negligible compared to the information obtained from
the measurement.

We assume that an inspiraling binary gravitational wave
{\it has} been
observed; i.e, that the appropriate detection criterion has been met
by the detector outputs.  We now discuss how to determine the
parameters of an inspiraling binary system that best fit the measured
signal.  The basic framework is illustrated in
Fig.~\ref{hypersurface}.  The set of all gravitational waveforms from
two inspiraling bodies can be characterized by a relatively small
number of parameters (the distance to the source; the time of merger;
five angles specifying the position of the source on the sky, the
plane of the orbit, and the orbital phase at some given time; and the
masses and spin angular momenta of the two bodies---fifteen parameters
in all, assuming that the eccentricity of the orbit is negligible).
We regard this set of waveforms as a fifteen-dimensional surface
embedded in the vector space of all possible measured signals.  In the
absence of any noise, all measured signals from inspiraling binaries
would lie on this submanifold; in practice, of course, the measured
signal consisting of waveform plus noise is displaced off the
submanifold.

The statistical properties of the noise determine a natural inner
product on the vector space of signals.  Given two signals $h_1(t)$
and $h_2(t)$, we define $\left( h_1 \, | \, h_2 \right)$ by
\cite{caveat}
\FL
\begin{equation}
\label{inner}
\left( h_1 \,|\, h_2 \right) = 2 \int_0^{\infty} \, { {\tilde h}_1^*(f)
{\tilde h}_2(f) + {\tilde h}_1(f) {\tilde h}_2^*(f) \over  S_n(f)} \,\,  df,
\end{equation}
where ${\tilde h}_1$ and ${\tilde h}_2$ are the Fourier transforms of
$h_1$ and $h_2$.  This definition is
chosen so that the probability for the noise to have some realization
$n_0(t)$ is
\begin{equation}
\label{pn0}
p(n = n_0) \, \propto \, e^{- \left( n_0\, |\, n_0 \right) /2}.
\end{equation}
Hence if the actual incident waveform is $h(t)$, then from
Eq.~(\ref{hn}) the probability of measuring a signal $s$ in the
detector output is proportional to $e^{-\left( s-h \, | \, s-h
\right)/2}$.  Correspondingly, given a measured signal $s$, the
gravitational waveform $h$ that ``best fits'' the data is the one that
minimizes the quantity $\left( s-h \, | \, s-h \right)$; see
Fig.~\ref{hypersurface}.

It also follows from Eq.~(\ref{inner}) that for any functions $g(t)$
and $k(t)$,
the expectation value of $(g|n)(k|n)$,  for an ensemble of realizations of the
detector noise $n(t)$, is just $(g|k)$.
Hence the signal-to-noise (\ref{sn}) of the detection will be
approximately given by
\begin{equation}
\label{snh}
{S\over N}[ h] = {{(h|h)}\over { {\rm rms}\  (h|n)}} = (h|h)^{1/2}.
\end{equation}
The kernel $w(t)$ of Wiener's optimal filter appearing in
Eq.~(\ref{sn}) is just the Fourier transform of $1/S_n(f)$.

For a given incident gravitational wave, different realizations
of the noise will give rise to somewhat different best-fit
parameters.  However, for large $S/N$, the best-fit parameters will have a
Gaussian distribution centered on the correct values.
Specifically, let ${\tilde \theta}^i$ be the ``correct'' values of the
parameters on which the waveforms depend,
and let ${\tilde \theta}^i + \Delta \theta^i$ be the best
fit parameters in the presence of some realization of the noise.  Then
for large $S/N$, the parameter-estimation errors $\Delta \theta^i$ have
the Gaussian probability distribution \cite{finn1}
\begin{equation}
\label{gauss}
p(\Delta \theta^i)=\,{\cal N} \, e^{-{1\over 2}\Gamma_{ij}\Delta \theta^i
\Delta \theta^j}.
\end{equation}
Here $\Gamma_{ij}$ is the so-called Fisher
information matrix  defined
by
\begin{equation}
\label{sig}
\Gamma_{ij} \equiv \bigg( {\partial h \over \partial \theta^i}\, \bigg| \,
{\partial h \over \partial \theta^j }\bigg),
\end{equation}
and ${\cal N} = \sqrt{ {\rm det}({\bf \Gamma} / 2 \pi) }$ is the
appropriate normalization factor.  It follows that the
root-mean-square error in $\theta^i$ is
\begin{equation}
\label{bardx}
\sqrt{ \left< ({\Delta \theta^i})^2 \right> } = \sqrt{\Sigma^{ii}}
\end{equation}
where ${\bf \Sigma} \equiv {\bf \Gamma}^{-1}$.

The above discussion applies to measurements made by a single
detector.  The (straightforward) generalization to a network of
detectors, which will be required in Sec.~\ref{dist_errors}, is given
in Appendix \ref{stats}.

The above discussion also neglects the effects of any {\it a priori}
constraints on the parameters that may be available.  The
incorporation of such {\it a priori} information can have a significant
effect on the predicted parameter-extraction accuracies (and also on
the best-fit parameter values themselves).  This is true not only for those
parameters to which the constraints apply, but also for the remaining
parameters because of correlations.  The effect is significant
whenever, for some parameter, the {\it a priori} information is
comparable with the information derived from the measured signal.
Hence, {\it a priori} constraints are usually important whenever we
include in an error-estimation analysis parameters which are {\it
weakly determined} by the data.  In
Appendix \ref{stats} we derive a generalization of Eq.~(\ref{sig})
[cf.~Eq.~(\ref{simple_answer}) below] which roughly incorporates the
effect of {\it a priori} information.  This generalization will be
used in Sec.~\ref{spin_sec}, where we consider the
dependence of the inspiral waveform $h$ on the spins of the two
bodies.

\subsection{The gravitational wave signal \\
in the Newtonian approximation}
\label{newtonian}

Inspiraling compact binaries can be described, to lowest order, as two
Newtonian point particles whose orbital parameters evolve secularly
due to gravitational radiation, where the gravitational waves and
corresponding energy loss rate are given by the Newtonian quadrupole
formula.  That is, the orbital frequency $\Omega$ at any instant is given by
\begin{equation}
\label{omega}
\Omega = {{M^{1/2}}\over {r^{3/2}}},
\end{equation}
where $M \equiv M_1+M_2$ is the total mass of the system and $r$
is the orbital separation.  The inspiral rate, for circular orbits,
is given by
\begin{equation}
\label{drdt}
{dr \over dt} =-{r \over E} {dE \over dt} = -{{64}\over 5} {{\mu
M^2}\over {r^3}},
\end{equation}
where $\mu \equiv M_1 M_2/M$ is the
reduced mass.
Integrating Eq.~(\ref{drdt}) we obtain
\begin{equation}
\label{rt}
r = \left( {\textstyle \frac {256}{5}} \mu M^2 \right)^{1/4}\,(t_c -t)^{1/4},
\end{equation}
where $t_c$ is the ``collision time'' at which (formally) $r
\rightarrow 0$.  Since the emitted gravitational waves are quadrupolar,
their frequency $f$  (cycles/sec) is equal to $\Omega/\pi$.
The gravitational waves induce a
measured strain $h(t)$ at the detector which is given by (see,
e.g., Ref.~\cite{300years})
\FL
\begin{equation}
\label{nh}
h(t)={{(384/5)^{1/2} \pi^{2/3} Q(\theta, \varphi, \psi, \iota) \mu M
}\over {D \,r(t)}} \cos \left( \int{2\pi f
dt}\right),
\end{equation}
where $D$ is the distance to the source. The function $Q$ and the
angles $\theta, \varphi, \psi, \iota$ (which describe
the position and orientation of the binary) are defined in
Sec.~\ref{dist_errors} below; they will not be needed
in this section.  In Eq.~(\ref{nh}) we could have included
the factor $(384/5)^{1/2} \pi^{2/3}$ in
the definition of $Q$, but choose not to for later convenience.

Because both the amplitude
and frequency of the signal increase as  $t \rightarrow t_c$, the
signal is referred to as a ``chirp.'' From Eqs.~(\ref{omega}) and (\ref{drdt}),
the frequency evolves
according to
\begin{equation}
\label{dfdt}
{df \over dt} = {{96}\over5}\pi^{8/3}\, {\cal M}^{5/3} \,f^{11/3},
\end{equation}
where ${\cal M} \equiv \mu^{3/5} M^{2/5}$ is the chirp mass parameter
discussed in Sec.~\ref{intro}.
The phase of the waveform $\phi(t)=\int^t{2\pi f(t') dt'}$ is
\begin{equation}
\label{phit}
\phi(t)=- 2\left[{\textstyle \frac {1}{5}} {\cal M}^{-1}
(t_c-t)\right]^{5/8} +\phi_c,
\end{equation}
where the constant of integration $\phi_c$ is defined by $\phi
\rightarrow \phi_c$ as $t \rightarrow t_c$.

In Eqs.~(\ref{nh})--(\ref{phit}) we have omitted the (obvious) time
delay between signal emission and detection, and we have implicitly
assumed that the detector and the binary's center-of-mass are at rest
with respect to each other. The latter requires some explanation.  If
the detector and binary are in relative motion, the detected signal is
Doppler-shifted with respect to the emitted signal. One cannot
determine this Doppler-shift from the detected signal, since $h(t)$,
as defined by Eqs.~(\ref{nh})--(\ref{dfdt}), is invariant under the
transformation
\FL
\begin{equation}
\label{invariant}
\left(f, {\cal M}, \mu, r, D , t\right) \to  \left(f/\lambda,
{\cal M}\lambda, \mu \lambda, r\lambda, D\lambda, t\lambda\right).
\end{equation}
Thus, strictly speaking, one can extract from the signal only the
``Doppler-shifted'' mass and distance parameters $\lambda \,{\cal M}$,
$\lambda \mu$, and $\lambda D$, where $\lambda$ is the Doppler-shift
factor.  This is not just a feature of our simplified, Newtonian
waveform; it also holds for the true, general-relativistic waveforms,
as can be seen on purely dimensional grounds and from the fact that
general relativity does not define any preferred mass/length scales.

Similarly, for binary sources at cosmological distances, the waves
will depend on and reveal the redshifted masses
\begin{equation}
\label{redshift}
{\cal M} = (1 + z) {\cal M}_{\rm true}, \ \ \ {\mu} = (1 + z) {\mu_{\rm
true}},
\end{equation}
where $z$ is the source's cosmological redshift, and also depend on
and reveal its so-called luminosity distance $D_L$
\cite{schutz,draza,finn1}.  Our measurement-accuracy analysis
applies to these {\it redshifted} masses and to the luminosity
distance.  The determination of the true masses for very distant
binaries will require some method of estimating redshifts; see, e.g.,
Ref.~\cite{draza}.

It is most convenient to work directly with the Fourier transform of $h(t)$,
\begin{equation}
\label{fourierT}
{\tilde h}(f) \equiv \int_{-\infty}^{\infty}\, e^{2\pi i f t}h(t)\, dt,
\end{equation}
which is easily computed using the stationary phase
approximation \cite{300years}.  Given a function $B(t)=A(t) \cos
\phi(t)$, where $d \ln A/dt \ll d\phi(t)/dt$ and $d^2\phi/dt^2 \ll
(d\phi/dt)^2$, the stationary phase approximation provides the
following estimate of the Fourier transform ${\tilde B}(f)$ for $f\ge 0$:
\FL
\begin{equation}
\label{statph}
{\tilde B}(f) \approx {1\over 2} A(t) \biggl({df \over
dt}\biggr)^{-1/2} \exp \left[i \left(2\pi f t - \phi(f)
-\pi/4 \right) \right].
\end{equation}
In this equation, $t$ is defined as the time
at which $d\phi(t)/dt = 2\pi f$, and (in a slight abuse of notation)
$\phi(f)$ is defined as $\phi\bigl(t(f)\bigr)$.   Using
Eqs.~(\ref{dfdt}) and (\ref{phit}) we obtain
\begin{eqnarray}
\label{tfpf}
t(f) & = & t_c - 5(8\pi f)^{-8/3} {\cal M}^{-5/3}
\nonumber \\
 \mbox{} \phi(f) & = & \phi_c - 2\left[8\pi {\cal M} f \right]^{-5/3}.
\end{eqnarray}
Hence from Eq.~(\ref{statph}), the Fourier transform of the Newtonian
waveform is
\begin{equation}
\label{fourh}
{\tilde h}(f) = {Q\over
{D}} {\cal M}^{5/6}
f^{-7/6}\exp \left[i \Psi(f) \right]
\end{equation}
for $f \ge 0$, where the phase $\Psi(f)$ is
\begin{equation}
\label{fphase}
\Psi(f) = 2\pi f t_c -\phi_c -{{\pi}\over 4 }
+{3\over4}(8\pi {\cal M} f)^{-5/3}.
\end{equation}

Equation (\ref{fourh}) for ${\tilde h}(f)$ is clearly invalid at very
high frequencies, because the real inspiral will terminate at some
finite orbital frequency.  For BH-BH and BH-NS mergers, there will be
a transition from inspiral to a final plunge \cite{Kip_Amos} near the
location of the last stable circular orbit, which is roughly at $r = 6
M$ for non-spinning bodies \cite{lastorbit}.  The final plunge will
last roughly one orbital period.  [Neutron stars merging with rapidly
spinning black holes may instead tidally disrupt, thereby shutting off
the waves, outside the horizon \cite{Bildsten_Cutler}.]
For NS-NS mergers, the two bodies will collide and coalesce at
roughly $r = 6M$.  Generally therefore the inspiral
gravitational wave $h(t)$ will ``shut off'' at roughly
$r = 6M$, and correspondingly  ${\tilde h}(f)$ will shut off
at roughly $f = (6^{3/2}\pi M)^{-1}$.
We therefore ``correct'' the waveform (\ref{fourh}) by setting
${\tilde h}(f) = 0$ for $f> (6^{3/2}\pi M)^{-1}$.  We
note that when $r>6M$,
\begin{equation}
\label{statok}
{{|r^{-1} dr/dt|} \over {d\phi/dt} } = {2 \over 3} {d^2 \phi / dt^2
\over (d \phi / dt)^2} < {1\over {55}} ({{4\mu}\over {M}}),
\end{equation}
so the stationary phase approximation should reproduce the
Fourier transform of $h(t)$ with good accuracy throughout the
inspiral.  Note that, as advertised in Sec.~\ref{intro}, in the
Newtonian approximation the signal (\ref{fourh}) depends on $M_1$
and $M_2$ only through the chirp mass ${\cal M}$.

Using Eqs.~(\ref{snf}) and (\ref{fourh}), we can see how the
signal-to-noise squared accumulates as the frequency sweeps upwards:
\begin{eqnarray}
\label{sf}
(S/N)^2 (f) & \equiv & 4 \int_0^f  { |{ \tilde h}(f^\prime) |^2 \over
S_n(f^\prime) } \, df^\prime \nonumber \\
\mbox{}     &  = & 4 {{Q^2}\over {D^2}} {\cal M}^{5/3} \int_0^f
{(f^\prime)^{-7/3}
 \over S_n(f') } \, d f^\prime .
\end{eqnarray}
In Fig.~\ref{snrf} we plot the integrand $ d (S/N)^2 / df = 4 |{\tilde
h}(f)|^2 /S_n(f)$, using the advanced detector noise spectrum
(\ref{snf}).  The shape of this curve is universal once the noise
spectrum is given: the masses, the relative angles, the distance to
the source, etc.~affect only the overall amplitude.  (This is strictly
true only for the ``Newtonian'' signal, but will remain true to a good
approximation when post-Newtonian effects are taken into
account.)  While $90 \%$ of the cycles come between $10$ and $40\ {\rm Hz}$,
and while most of the energy is released in the last few orbits at $f
> 200 \, {\rm Hz}$, we find that $\sim 60 \%$ of the total signal-to-noise
squared accumulates between $40$ and $100 \,{\rm Hz}$, the frequency band in
which LIGO is most sensitive.

We now evaluate the Fisher information matrix (\ref{sig}).
For measurements using a single detector, there are only four parameters
on which the Newtonian signal depends: an overall amplitude ${\cal A}
\equiv (Q/D) {\cal M}^{5/6}$, and ${\cal M},
\,t_c$, and $\phi_c$.  The derivatives of $\tilde h(f)$ with respect to
these parameters (for $f > 0$) are given by
\begin{mathletters}
\label{dh}
\begin{eqnarray}
&&{\partial {\tilde h} \over \partial\, {\rm ln} {\cal A} }  =  {\tilde
h}, \ \ \ \ \  { \partial {\tilde h} \over \partial t_c}  =  2 \pi i
f \,{\tilde h}, \\
&&{ \partial {\tilde h} \over \partial \phi_c}  =   -i \,\tilde h,\ \
\ \  {\partial {\tilde h} \over \partial\, {\rm ln} {\cal M}}  =
-{{5i}\over 4} (8 \pi {\cal M} f )^{-5/3} \, \tilde h.
\end{eqnarray}
\end{mathletters}

 From Eqs.~(\ref{dh}) and the noise spectrum (\ref{snf}), it is
straightforward to evaluate the Fisher information matrix (\ref{sig})
and its inverse $\Sigma^{ij}$ \cite{innerprod}.  General expressions
for the elements of $\Gamma_{ij}$ using Newtonian waveforms, valid for
any detector noise spectrum, are given in Ref.~\cite{finn2}. We will
not reproduce them here.  However, for purposes of comparison to our
post-Newtonian results in Sec.~\ref{sec3}, we list the rms errors
$\Delta {\cal A},\,\Delta {\cal M}, \,\Delta t_c$ and $\Delta \phi_c$
for the case of low-mass (e.g., NS-NS) binaries, assuming the
approximate waveform (\ref{fourh}) and the detector noise spectrum
(\ref{snf}):
\begin{mathletters}
\label{newtresults}
\begin{eqnarray}
\Delta({\rm ln}\,{\cal A}) & =& 0.10 \, \biggl({{10}\over
{S/N}}\biggr)\\
\Delta t_c & = & 0.40\,\biggl({{10}\over{S/N}}\biggr) \,{\rm msec}, \\
\Delta\phi_c &=& 0.25 \biggl({{10}\over {S/N}}\biggr) \, {\rm rad} \\
\label{newtcalM}
\Delta ({\rm ln}\, {\cal M}) & = & 1.2 \times
10^{-5}\,\biggl({{10}\over {S/N}}\biggr)\,
\left({{\cal M} \over {M_\odot} }\right)^{5/3}.
\end{eqnarray}
\end{mathletters}
For low-mass binaries, the fact that we ``cut off'' the waveform
$\tilde h(f)$ for $f>(6^{3/2}\pi M)^{-1}$ has little effect on the rms
errors (\ref{newtresults}), due to the sharp rise in $S_n(f)$ at high
frequency.  The exact scaling of $\Delta ({\rm ln}\, {\cal M})$ as
${\cal M}^{5/3}$, and the fact that $\Delta t_c$ and $\Delta \phi_c$
are independent of $M$, strictly hold only when the cut-off is
unimportant. For BH-BH binaries with $S/N = 10$, one has $\Delta t_c =
0.60\, {\rm msec}$, $\Delta \phi_c = 0.32 \,{\rm rad}$, and $\Delta
({\rm ln}\, {\cal M}) = 1.3 \times 10^{-5}\,\left({{\cal M}/{M_\odot}
}\right)^{5/3}$.

The rather phenomenal accuracy attainable for the chirp mass
${\cal M}$ is due to the large number ${\cal N}_{\rm
cyc}$ of cycles in the detectable portion of the gravitational
waveform.   We see from Eq.~(\ref{dfdt}) that ${\cal N}_{\rm cyc}$
scales like ${\cal M}^{-5/3}$, so $ \Delta ({\rm ln}\, {\cal M})$ is
proportional $1/{\cal N}_{\rm cyc}$, as one would expect.

The rms errors (\ref{newtresults}) apply to single-detector
measurements. In practice, one will have a network of detectors, with
different locations and orientations. For a network, $\Delta ({\rm
ln}\, {\cal M})$, will be roughly given by Eq.~(\ref{newtcalM}),
but with $S/N$ replaced by the combined signal-to-noise $\rho$ of the
detector network, defined by Eq.~(\ref{rho}) above.  This is
because independent estimates of ${\cal M}$ are obtained from each detector.
The same argument does {\it not} apply to the rms errors in
$t_c$, $\cal A$, and $\phi_c$, because the gravitational waves will
arrive at the different detectors at different times, and because
detectors with different orientations measure different values of
$\cal A$ and $\phi_c$ [cf.~Sec.~\ref{dist_errors} below].

We conclude this section by noting that from the measured value of the
chirp mass $\cal M$ alone, one already obtains a lower limit on the
larger of the individual masses, and upper limits on the smaller mass
and on the reduced mass.  We adopt the convention that $M_1 \ge M_2$;
i.e., $M_1$ always refers to the {\it larger} of the two masses.  Then
it follows by definition that
\FL
\begin{equation}
\label{ineq}
M_1 \ge  2^{1/5} {\cal M},  \ \ \  M_2 \le  2^{1/5} {\cal M}, \
\ \ \mu \le  2^{-4/5} {\cal M}.
\end{equation}
However, if $\mu$ is unknown, then the mass ratio $M_1/M_2$ is
unconstrained.  The bounds (\ref{ineq}) that follow from measuring
${\cal M}$ may themselves be of astrophysical interest.  For instance,
if one determines using (\ref{ineq}) that $M_1 \ge 3 M_\odot$, then
one may conclude that the heavier body is a black hole (assuming the
redshift is small, cf.~Eq.~(\ref{redshift}) above and associated
discussion).  Also, it has been suggested \cite{cutleretal} that from
LIGO/VIRGO measurements of NS-BH coalescences where the BH is rapidly
spinning, it may be possible to constrain the neutron-star equation of
state by measuring the frequency at which the NS's tidal disruption
causes the waves to shut off.  Knowledge of this tidal-disruption
frequency, coupled with an upper limit on the neutron star mass $M_2$
determined from the inspiral waveform, would allow one to place an
upper limit on the stiffness of the equation of state.

\section{Post-Newtonian Effects\\
 and Parameter Estimation}
\label{sec3}

We now extend the analysis of the previous section to include
post-Newtonian effects.  We continue to treat the bodies as point
masses, since tidal interactions have a negligible effect.  Also, for
the moment we will neglect the effects of the bodies' spin angular
momenta.

The post-Newtonian approximation provides the most accurate
description currently available of the gravitational radiation from
inspiraling, stellar-mass binaries.  Corrections of order $M/r$
($P^1N$ corrections) to the lowest-order, Newtonian waveform
(\ref{nh}) were calculated almost twenty years ago by Wagoner and Will
\cite{wagoner1}.  Calculations of the inspiral rate have recently been
extended to $P^{1.5}N$ order, for the case of non-spinning bodies, by
Wiseman \cite{wiseman} (after Cutler {\it et al}. \cite{cutler_finn}
and Poisson \cite{eric} had determined the form of the $P^{1.5}N$
correction for the case $\mu/M \ll 1$).  By ``$P^xN$ order'' we mean
that corrections to the quadrupole-formula radiation field and
corresponding inspiral rate that are of order $(M/r)^x$ have been taken into
account, along with order $(M/r)^x$ corrections to the non-radiative
orbital equations which determine, e.g., the orbital frequency at a
given separation.  [There is no standard convention for ``counting''
post-Newtonian orders in calculations involving radiation; e.g., some
authors refer to the lowest-order radiation field as $P^{2.5}N$. Our
own terminology is motivated by the application considered here: since
radiation reaction effects {\it cause} the inspiral, $O(M/r)$
corrections to the quadrupole formula accumulate secularly and have
just as large an effect on the phase of the orbit $\phi(f)$ as do
$O(M/r)$ corrections to the orbital frequency at a given radius.]

The post-Newtonian waveforms improve upon their Newtonian counterparts
in three respects \cite{krolak,wagoner2}.  First, they include
contributions from higher-order multipoles of the stress-energy tensor
(e.g., mass-octupole and current-quadrupole radiation in addition to
the mass-quadrupole term), whose frequencies are different harmonics
of the orbital frequency.  Second, they include post-Newtonian
corrections to the lowest-order expressions for the amplitude of each
multipole component.  And, most importantly for our purposes,
post-Newtonian corrections to the energy $E(r)$ and gravitational wave
luminosity $dE/dt(r)$ modify the inspiral rate and thereby the
accumulated orbital phase $\Phi(t)$.  We can write $h(t)$ schematically
as
\begin{equation}
\label{sum}
h(t) = \Re \, \left\{ \sum_{x,m} h_m^x(t) e^{i m \Phi(t)} \right\}
\end{equation}
where ``$\Re$'' means ``the real part of'', $x$ indicates the term's
post-Newtonian order, the integer $m$ labels the different harmonics,
and $\Phi(t)$ is the orbital phase. Each amplitude $h_m^x$ has the form
\begin{equation}
\label{hxm}
h_m^x(t) \equiv {{\mu M}\over {D\, r(t)}}\,g_m^x(M_1/M_2)\,
Q_m^x(\theta,\varphi, \psi, \iota)
\end{equation}
where $r(t)$ is the orbital separation, $g_m^x$ is some function of
the mass ratio, and $Q_m^x$ is a function of the source's position
on the sky and the orientation of the orbital plane.  To connect with
the notation of Sec.~\ref{sec2} and below, we note that the phase
$\phi(t)$ of the quadrupole part of the waveform is essentially
twice the orbital phase: $\phi(t) = 2\Phi(t) + k$, for some
constant $k$ that depends on the relative positions and
orientations of the detector and the binary \cite{k_var}.
Thus, the expansion for $h(t)$
through $P^{1.5}N$ order is given by
\FL
\begin{eqnarray}
\label{schem}
h(t) & = & \Re [ \,  (h_2^0+ h_2^1+ h_2^{1.5})e^{2i\Phi} + ( h_1^{0.5}+
h_1^{1.5})e^{i\Phi} \nonumber \\
\mbox{} & &  + (h_3^{0.5}+ h_3^{1.5})e^{3i\Phi} + h_4^1
e^{4i\Phi} + h_5^{1.5} e^{5i\Phi} ]
\end{eqnarray}
where $\Phi(t)$ has the post-Newtonian expansion
\begin{equation}
\label{psiexp}
\Phi(t) = \Phi^0 + \Phi^1 + \Phi^{1.5} + O(M/r)^2.
\end{equation}
In Eq.~(\ref{psiexp}), $\Phi^x$ refers to the $P^xN$ order
contribution to the orbital phase.
As indicated by Eq.~(\ref{psiexp}), the term $\Phi^{0.5}$ vanishes
identically, as do several omitted terms in Eq.~(\ref{schem}). The term
$h_2^0 e^{2 i \Phi^0}$ is just the Newtonian, mass-quadrupole waveform
given by Eq.~(\ref{nh}), while the terms $h_1^{0.5}e^{i \Phi^0(t)}$
and $h_3^{0.5}e^{3 i \Phi^0(t)}$ are the lowest order
current-quadrupole pieces of the waveform.  The term
$h_2^{1.5}\,e^{2i\Phi}$ is the so-called ``hereditary'' or ``tail''
term produced by the interaction of the outgoing wave with the binary's
gravitational potential \cite{eric,Damour}.  The interested
reader can find explicit expressions for the amplitudes $h_m^x$
through $P^1N$ order in Krolak \cite{krolak}.

In Sec.~\ref{intro} we argued that the waveform's accumulated phase
$\Phi$ contains most of the ``information'' that allows
sensitive measurement of the masses of the bodies.  Since this paper
aims at only an approximate calculation of parameter-estimation
accuracies, rather
than use the full $P^{1.5}N$ waveform (\ref{schem}), we calculate the
Fisher information matrix (\ref{sig}) using the following ``model''
waveform:
\begin{equation}
\label{pnht}
h(t)=\Re\biggl\{h_2^0\, e^{2 i\left[\Phi^0 + \Phi^1 +
\Phi^{1.5}\right]} \biggr\}.
\end{equation}
That is, we include $P^1N$ and $P^{1.5}N$ corrections to the phase of
the waveform, since these are decisive for extracting the mass and
spin parameters of the binary, but we neglect the other post-Newtonian
effects that are nominally of the same order.  We expect
that the values of $\Delta M_1$ and  $\Delta M_2$  calculated using
Eq.~(\ref{pnht}) will be a reasonable approximation to the
error bars one would calculate using the true, general
relativistic waveforms (assuming one had access to them).

There is another, practical, reason for the use of the truncated
waveform (\ref{pnht}).  As explained in Sec.~\ref{sec2}, we can
simplify the error-estimation analysis by considering only
single-detector measurements, and still obtain a reasonable estimate
of the accuracies attainable for mass and spin measurements.  However,
as stated above, each of the amplitudes $h_m^x$ has a different
dependence on the angles $(\theta, \varphi,\psi, \iota)$.  These
angles cannot be measured using one detector alone. The position of
the source $\theta, \varphi$ is determined from differences in signal
arrival times at (at least) three widely separated detectors
\cite{schutz}.  Moreover at least two of the detectors must have
different orientations to obtain even a crude estimate of the angles
$\psi,\iota$ (which describe the principal polarization axis of the
wave and the angle between the line of sight and the normal to the
orbital plane --- see Sec.~\ref{dist_errors} below).  Thus, to make use of the
extra information contained
in the post-Newtonian terms that we are omitting in Eq.~(\ref{pnht}),
a full detector network would have to be modeled.  Hence, for
simplicity, in our model waveform (\ref{pnht}) we omit all of the terms in
Eq.~(\ref{schem}) except for the largest one.  [Although we do analyze a
general network of detectors in Sec.~\ref{dist_errors} below, that
analysis takes advantage of the fact that the phase-evolution
information and the amplitude/polarization information in the measured
waveforms are largely independent, and --- complementary to this
section's analysis --- focuses on the amplitude/polarization
information alone.]

\subsection{Parameter estimation neglecting spin effects}
\label{no_spin}

In this section we estimate how well the masses $M_1$ and $M_2$ could
be determined from the waveform, if we knew {\it a priori} (or {\it a
posteriori} by some independent means), that both bodies had negligible
spin.  Note that this is different from the situation where
the spins {\it happen} to be zero, but where we have no knowledge of
this fact apart from the information contained in the gravitational
waveform.

In fact, it would not be justified to assume {\it a priori} that
compact objects found in binaries have negligible spins.  For one
thing, the formation of close binaries generally involves a period of
mass transfer, which would tend to spin up the accreting body.
Observationally, there are three known NS-NS binaries that
will merge within a Hubble time; at the time of merger, the pulsars
in these binaries will all be spinning at roughly $1-2\%$
of their maximum possible angular velocities \cite{sterl}.
(The spin rates at merger
will be roughly  a factor of two smaller than current values, due
to magnetic dipole radiation.)
We show in Sec.~\ref{spin_sec} below that allowing
for spins of this magnitude increases the resulting error
bars for mass measurements by roughly a factor of two, compared
to the error bars obtained if spins are assumed to vanish. Nevertheless, we
feel it is instructive to calculate the Fisher information matrix
neglecting spin effects (i.e., assuming the spins are negligible {\it
a priori}), both to illustrate the inclusion of post-Newtonian terms
and to provide a basis for comparison with the results obtained when we
include spins.

We now briefly derive the $P^{1.5}N$ corrections to the phase of the
waveform.  Through $P^{1.5}N$ order, the
orbital frequency, energy, and energy-loss rates (for non-spinning
bodies) are
\cite{wagoner1,wiseman}:
\FL
\begin{eqnarray}
\label{basic1}
\Omega(r) & = & {M^{1/2}\over r^{3/2}}
\left[1 + \left({{-3}\over 2} +{{\mu}\over{2 M}} \right) {M \over r} +
 O\left({M\over r}\right)^2
\right] \\
\label{basic2}
E(r) & = & {{-\mu M}\over {2r}} \left[ 1 + \left({{-7}\over 4}
+{{\mu}\over{4M}} \right){M\over r} + O\left({M\over r}\right)^2  \right] \\
\label{basic3}
{dE \over dt}(r) & = &  -{{32}\over 5} ({\cal M} \Omega)^{10/3}
\biggl[1 +\left( {{-1247}\over{336}} + {{35\mu}\over{12M}} \right)\left({M\over
r}\right)
\nonumber \\
\mbox{} & & + 4\pi\,\left({M\over r}\right)^{3/2} +  O\left({M\over r}\right)^2
\biggr],
\end{eqnarray}
where $r$ is the orbital separation in DeDonder gauge (the standard
gauge choice for post-Newtonian calculations), and $t$ refers to time
measured at infinity.

Defining $f \equiv \Omega/\pi$, the frequency (in cycles/sec) of the
quadrupolar part of the gravitational waves, we combine
Eqs.~(\ref{basic1})--(\ref{basic3}) to obtain
\FL
\begin{eqnarray}
\label{pndfdt}
df/dt & = & {{96}\over 5}\pi^{8/3}\,{\cal M}^{5/3} f^{11/3}
\biggl[1 - \left({{743}\over{336}} + {{11\mu}\over{4M}} \right)(\pi M
f)^{2/3} \nonumber \\
\mbox{} & & + 4\pi \, (\pi M f) +  O(\pi M f)^{4/3} \biggr].
\end{eqnarray}
In Eq.~(\ref{pndfdt}) and below, we use $(\pi M f)^{1/3}$
as our post-Newtonian expansion
parameter, instead of $(M/r)^{1/2}$.  We note that $(\pi M f)^{1/3}$
equals $(M/r)^{1/2}$ up to but not including terms of order $(M/r)^{3/2}$.
This change of variables is
advantageous because the frequency of the wave is a directly
measurable, gauge-independent quantity (unlike the radius of the
orbit).  Equation (\ref{pndfdt}) can be easily integrated to obtain
$t(f)$ and $\phi(f)$, where $\phi \equiv \pi \int{f dt}$ is the
phase of the waveform.  Defining $x \equiv (\pi M f)^{2/3} $,
we find that
\FL
\begin{eqnarray}
\label{pntf}
t(f) & = & t_c - 5(8\pi f)^{-8/3} {\cal M}^{-5/3}
\biggl[1 + {4\over 3}\left({{743}\over{336}} +
{{11\mu}\over{4M}}\right)x \nonumber \\
\mbox{} & & - {{32\pi}\over 5} x^{3/2} +  O(x^2) \biggr], \\
\label{pnpf}
\phi(f) & = & \phi_c - 2\left[8\pi {\cal M} f \right]^{-5/3}
\biggl[1 + {5\over 3}\left({{743}\over{336}} + {{11\mu}\over{4M}}
\right)\,x \nonumber \\
\mbox{} & & - 10\pi\, x^{3/2}   + O(x^2) \biggr],
\end{eqnarray}
where, as in Sec.~\ref{sec2}, we define $t_c$ and $\phi_c$ by $t
\rightarrow t_c$ and $\phi \rightarrow \phi_c$ as $f
\rightarrow \infty$.

Using Eqs.~(\ref{pntf}) and (\ref{pnpf}) and the stationary phase
approximation, we can repeat the analysis of Sec.~\ref{sec2} to obtain
$\tilde h(f)$.  As before, we (crudely) model the end of the inspiral
at $r \approx 6M$ by setting $\tilde h(f) = 0$ for $f > (6^{3/2}\pi
M)^{-1}$.  The stationary phase result then becomes
\FL
\begin{equation}
\label{pnmodel} {\tilde h}(f) = \left\{ \begin{array}{ll}
{\cal A}\, f^{-7/6}\, e^{i \Psi}\ \ \ \ \ \ & \mbox{$0 < f <
(6^{3/2}\pi M)^{-1}$} \\
0				 & \mbox{$(6^{3/2} \pi M)^{-1} < f$},
\end{array} \right.
\end{equation}
where ${\cal A} =(Q/D)\,{\cal M}^{5/6}$ and
\FL
\begin{eqnarray}
\label{pnpsi}
\Psi(f) & = & 2\pi f t_c -\phi_c -\pi/4 +{3\over 4}(8 \pi {\cal M} f
)^{-5/3} \nonumber \\
\mbox{} & & \times \, \left[1 + {20\over 9}\left({743\over
336}+{{11\mu}\over{4M}}\right) x -16 \pi x^{3/2} \right].
\end{eqnarray}
Note that the post-Newtonian correction terms in square brackets in
Eq.~(\ref{pnpsi}) have their greatest effect on the phase of $\tilde
h(f)$ at {\it low} frequencies, because they are multiplied by the
overall factor $f^{-5/3}$.  This may seem counterintuitive, since the
post-Newtonian corrections to the inspiral rate are largest at small
$r$, or high $f$; however the high-frequency portion of the waveform
contains far fewer cycles, so the cumulative effect of PN corrections
on the waveform's phase is smaller there.

Our model waveform (\ref{pnpsi}) for non-spinning bodies depends on
five parameters: ${\cal A}$, $\phi_c$, ${\cal M}$, $\mu$, and $t_c$.
It is actually somewhat simpler to compute and interpret the Fisher
information matrix $\Gamma_{ij}$ in terms of the following modified
parameters for which the rms errors are rescaled: $\ln
{\cal A}$, $\phi_c$, $\ln {\cal M}$, $\ln \mu$, and $ f_0 t_c$, where
$f_0$ is some fiducial frequency. With respect to these parameters,
the derivatives of $\tilde h(f)$ are \cite{cutoff}
\begin{mathletters}
\label{pnderivs}
\FL
\begin{eqnarray}
\label{pnderiv1}
{\partial {\tilde h}(f) \over \partial  \ln {\cal A} } & = & {\tilde
h}(f) \\
{\partial {\tilde h}(f) \over \partial  f_0 t_c } & = & 2 \pi i \,(f/f_0)
{\tilde h}(f) \\
{\partial \tilde h(f) \over \partial \phi_c} & = & - i {\tilde h}(f)
\end{eqnarray}
\FL
\begin{eqnarray}
{\partial {\tilde h}(f) \over \partial  \ln {\cal M} } & = &
-{{5 i }\over{4 }}(8\pi {\cal M} f)^{-5/3}
{\tilde h}(f) \nonumber \\
\mbox{} & & \times \,
\left[1 + {{55 \mu}\over{6M}}\,x + 8\pi\,x^{3/2} \right] \\
{\partial {\tilde h}(f) \over \partial \ln \mu} & = & {{3i}\over
{4}} (8\pi {\cal M} f)^{-5/3} {\tilde h}(f) \nonumber \\
\label{pnderiv2}
\mbox{} & & \times \, \left[({{-3715}\over {756}} +{{55 \mu}\over{6M}})
\,x  + 24 \pi \, x^{3/2} \right].
\end{eqnarray}
\end{mathletters}
Using Eqs.~(\ref{pnderivs}) and the noise spectrum (\ref{snf}), we
have numerically computed $\Gamma_{ij}$, its inverse $\Sigma^{ij}$,
and the corresponding errors $\Delta \phi_c = \sqrt{\Sigma^{\phi_c\,
\phi_c}}$, etc.  Since our model waveform includes
post-Newtonian corrections to the phase but not to the amplitude,
$\Sigma^{ij}$ is block diagonal: $\Sigma^{\ln{\cal A}\, j} = 0$ for
$j$ = $\phi_c$, $\ln {\cal M}$, $\ln \mu$, or $ f_0 t_c$.  Hence,
$\Delta {\cal A}/{\cal A} = (S/N)^{-1}$, while errors in ${\cal A}$
are uncorrelated with errors in the other parameters.  Table
\ref{table1} lists $\Delta \phi_c$, $\Delta t_c $, $\Delta {\cal
M}/{\cal M} $, and $\Delta \mu/{\mu}$ for a range of values of $M_1$
and $M_2$.  The results in Table \ref{table1} are for a single
detector and are normalized to $S/N=10$.  For measurements by a
detector network, the rms errors $\Delta {\cal M}/{\cal M} $, and
$\Delta \mu/{\mu}$ will be approximately those given Table
\ref{table1}, but with $S/N$ replaced by $\rho$, the combined
signal-to-noise (\ref{rho}) of the network.  As explained in
Sec.~\ref{sec2}, this is because each detector provides
almost-independent estimates of $\cal M$ and $\mu$.  The result we
particularly wish to draw attention to is: {\it if spins can be
treated as negligible then $\mu$ can typically be measured to $\sim
1\%$}, while ${\cal M}$ can be determined to $\sim 0.01-0.1\%$

Table \ref{table1} also lists the correlation coefficient $c_{{\cal M}\, \mu}
\equiv \Sigma^{{\cal M}\, \mu}/(\Sigma^{{\cal M}\,{\cal M} } \
\Sigma^{\mu\, \mu})^{1/2}$, a
dimensionless ratio indicating the degree to which errors in ${\cal
M}$ and $\mu$ are correlated.  The quantity $c_{{\cal M}\, \mu}$ is
independent of $S/N$, and by definition satisfies $c_{{\cal M}\, \mu}
\in \bigl[-1,1\bigr]$.  We find that typically $|c_{{\cal M}\, \mu}| >
0.90 $, indicating that the errors in $\cal M$ and $\mu$ are strongly
correlated.  This strong correlation implies that there exists a
linear combination of $\cal M$ and $\mu$ which can be determined much more
accurately than either  $\cal M$ or  $\mu$ individually
\cite{correlation_effects}.  In
particular, $\Delta \left({\cal M} - (\Sigma^{{\cal
M}\,\mu}/\Sigma^{\mu\, \mu}) \, \mu\right)$ is smaller than $\Delta
{\cal M}$ by a factor of $\sim (1-c_{{\cal M}\,
\mu}^2)^{-1/2}$.    Indeed, the value of $\Delta \left({\cal M} -
(\Sigma^{{\cal
M}\,\mu}/\Sigma^{\mu\, \mu}) \, \mu\right)$ computed using our
$P^{1.5N}$ waveform (\ref{pnmodel}) is
approximately the same as $\Delta \cal M$
[cf.~Eqs.~(\ref{newtresults}) above] computed using the Newtonian
waveform (\ref{fourh}) \cite{correlation_effects}.

How accurately can $M_1$ and $M_2$ be
determined?  While it is
straightforward to answer this question when the mass ratio is large,
we shall see that some care is required when
$M_1$ and $M_2$ are comparable, since in this case the distribution of
errors
 in $M_1, M_2$ is non-Gaussian.  Recall that we have adopted the convention
that $M_1 \ge M_2$. Then we have
\FL
\begin{eqnarray}
\label{m1}
M_{1,2} & = & {1\over 2}\biggl[{\cal M}^{5/2}\,\mu^{-3/2} \pm
\nonumber \\
\mbox{} & & \biggl({\cal M}^5\,\mu^{-3} \ -4{\cal
M}^{5/2}\,\mu^{-1/2}\biggr)^{1/2}\ \biggr].
\end{eqnarray}
Using Eq.~(\ref{m1}),
$\Sigma^{M_1 M_1}$ and $\Sigma^{M_2 M_2}$ can be expressed
as linear combinations of $\Sigma^{{\cal M} {\cal M} }$,
$\Sigma^{{\cal M} \mu} $, and
$\Sigma^{\mu \mu} $.  However it is clear from Table \ref{table1} that in
practice the $\Sigma^{\mu \mu}$ term will give the dominant
contribution.  Neglecting the terms
proportional to $\Sigma^{{\cal M} {\cal M} }$ and
$\Sigma^{{\cal M} \mu}$, we find that
\begin{mathletters}
\label{dm1dm2}
\begin{eqnarray}
\label{dm1}
\Sigma^{M_1 M_1} &=&  \Sigma^{\mu \mu}\left[{{M (\mu-3M_1)}\over
{2\mu (M_1-M_2)}}\right]^2 \\
\label{dm2}
\Sigma^{M_2 M_2} &=& \Sigma^{\mu \mu} \left[{{M (\mu-3M_2)}\over {2\mu
(M_1-M_2)}} \right]^2.
\end{eqnarray}
\end{mathletters}
For example, if $M_1 = 10 M_\odot$ and $M_2 = 1.4 M_\odot$,
Eqs.~(\ref{dm1dm2}) imply that
$\Delta M_1 /M_1 \approx 1.9 \Delta \mu /\mu$ and
$\Delta M_2 /M_2 \approx 1.4 \Delta \mu /\mu$.

While the expressions (\ref{dm1dm2}) for $\Sigma^{M_1 M_1}$ and
$\Sigma^{M_2 M_2}$ should be adequate
for estimating the distribution of errors when $M_1 \gg M_2$, these
expressions unfortunately diverge when $M_1=M_2$.  This divergence is due to
the fact that the Jacobian of the transformation $(M_1, M_2) \rightarrow
({\cal M}, \mu)$ vanishes when $M_1=M_2$.
Of course, the rms mass measurement errors do not actually become infinite.
Rather, the {\it linear} approximation that one
typically uses to estimate rms errors loses its validity.  That is,
the approximation that
\begin{eqnarray}
\label{linear}
\Delta \tilde h & \approx & {{\partial \tilde h}\over{\partial {\cal A}}}
\Delta {\cal A}
+ {{\partial \tilde h}\over{\partial\phi_c}} \Delta \phi_c
+ {{\partial \tilde h}\over{\partial t_c}} \Delta t_c \nonumber \\
\mbox{} && + {{\partial \tilde h}\over{\partial M_1}} \Delta M_1
+ {{\partial \tilde h}\over{\partial M_2}} \Delta M_2,
\end{eqnarray}
for variations $\Delta {\tilde h}$ of a size determined by typical
realizations of the noise, becomes inaccurate when $M_1 - M_2 \to 0$,
as $\partial \tilde h/\partial M_1 + \partial \tilde h/\partial M_2
\rightarrow 0$ in this limit.


To overcome this problem we proceed as explained in
Sec.~\ref{degenerate} below, and use the PDF for the best-fit values
${\hat {\cal M}}$, ${\hat \mu}$ of the parameters ${\cal M}$, $\mu$,
which is a simply a Gaussian centered on the true parameters
${\tilde {\cal M}}$, ${\tilde \mu}$.  [Thus, we are considering
so-called frequentist
errors, cf.~Sec.~\ref{twoerrors} below].  Let ${\hat M}_1$ and ${\hat
M}_2$ be the corresponding best-fit values for the individual masses.
Substituting into this PDF the transformation ${\hat {\cal M}} = {\cal
M}({\hat M}_1,{\hat M}_2)$ and ${\hat \mu} = \mu({\hat M}_1,{\hat
M}_2)$ yields a non-Gaussian PDF for ${\hat M}_1, {\hat M}_2$, from
which we can calculate the $95 \%$ confidence limits for ${\hat M}_1$
and ${\hat M}_2$.  The use of confidence limits is somewhat crude, in
the sense that it leaves out much of the information contained in the
PDF, but it is suitable for our purpose of determining {\it roughly}
how accurately these quantities can be measured.  Since $\Delta {\cal
M}$ is very small, for the purposes of this discussion we can assume
${\cal M}$ has been measured exactly.  Let $\tilde\mu$ is the true
value of the binary's reduced mass.  Then with $95 \%$
confidence ${\hat \mu}$ lies in the interval
\begin{equation}
\label{muconf}
\tilde\mu - 2\Delta \mu < {\hat \mu} < \tilde\mu + 2\Delta \mu
\end{equation}
where $\Delta \mu \equiv (\Sigma^{\mu\,\mu})^{1/2}$ is determined from
the variance-covariance matrix.  Roughly speaking, a necessary
condition for the distribution of ${\hat M}_1$ and ${\hat M}_2$ to be
Gaussian is that ${\tilde \mu} + 2 \Delta\mu < 2^{-4/5}\,{\cal M}$ (so
that the $\mu$'s $95\%$ confidence interval does not include the
equal-mass case).

 From Eqs.~(\ref{m1}) and  (\ref{muconf}) we obtain  the following
$95 \%$ confidence limits on ${\hat M}_1$ and ${\hat M}_2$:
\begin{mathletters}
\label{conf_limits}
\FL
\begin{eqnarray}
M_1({\cal M}, \tilde\mu +  2\Delta\mu ) < {\hat M_1} < M_1({\cal M}, \tilde\mu
-
2\Delta\mu) \label{conf1} \\
M_2({\cal M}, \tilde\mu -  2\Delta\mu) < {\hat M_2}
< M_2({\cal M}, \tilde\mu +  2\Delta\mu )
\label{conf2}
\end{eqnarray}
\end{mathletters}
where the functions $M_1$ and $M_2$ are given by Eq.~(\ref{m1}) above.
If ${\tilde \mu} + 2\Delta\mu$ is greater than the maximum allowed
value of $\mu$, then one should replace ${\tilde \mu} + 2\Delta\mu $
by $2^{-4/5} {\cal M}$ in Eqs.~(\ref{conf_limits}).  For example, if
${\cal M} = 1.219 M_\odot$, ${\tilde \mu} = 0.7 M_\odot$, and $\Delta
\mu / {\tilde \mu} = 0.004$ (the NS-NS case), then one can state with
$95\%$ confidence that ${\hat M}_1$ and ${\hat M}_2$ lie in the ranges
$1.4 M_\odot < {\hat M_1} < 1.65 M_\odot$ and $1.2 M_\odot < {\hat
M_2} < 1.4 M_\odot$.  Thus $M_1$ and $M_2$ are determined with much
less accuracy than $\mu$ when the two masses are roughly equal.

Figure \ref{m1m2} below illustrates the meaning of the confidence
limits in $M_1$ and $M_2$ (in the context of more accurate
calculations incorporating spin effects).

\subsection{Parameter estimation including spin effects}
\label{spin_sec}

We now present a rough calculation of the degree to which mass
measurement accuracy is degraded when the spins cannot be assumed to be
negligible.  For the
same reasons as in Sec.~\ref{no_spin}, we incorporate the effects of
spins on the phase of the waveform, but neglect their effects on the
waveform amplitude.

Let $\vec S_1$ and $\vec S_2$ be the spin angular momenta of the two
bodies, and let $\vec L$ be the total orbital angular momentum. We
define the unit vector $\hat L$ by $\hat L \equiv \vec L/|\vec L|$.
Then Kidder, Will, and Wiseman \cite{kidder} have shown that, due to an
``$\vec L \cdot \vec S$'' term in the two-body force law as well as
spin corrections to the expressions for the system's mass-quadrupole
and current-quadrupole moments, Eq.~(\ref{pndfdt}) becomes modified at
$P^{1.5}N$ order as follows:
\FL
\begin{eqnarray}
\label{pnsdfdt}
df/dt & = & {{96}\over 5}\pi^{8/3}\,{\cal M}^{5/3}\,f^{11/3}
\biggl[1 - \left({{743}\over{336}} + {{11\mu}\over {4M}}\right)\,x
\nonumber \\
\mbox{} & & + (4\pi-\beta)\, x^{3/2} + O(x^2) \biggr]
\end{eqnarray}
where again $x \equiv (\pi M f)^{2/3}$, and where
\begin{eqnarray}
\label{defbeta}
\beta & \equiv & M^{-2} \hat L \cdot \bigg[ \left({{113}\over{12}} +{{25}\over
4}{{M_2}\over{M_1}}\right)\vec S_1 \nonumber \\
\mbox{} && + \left({{113}\over{12}} +{{25}\over
4}{{M_1}\over{M_2}} \right)\vec S_2 \bigg].
\end{eqnarray}
Through $P^{1.5}N$ order, the six components of $\vec S_1$ and $\vec
S_2$ affect the waveform's phase only via the particular combination
(\ref{defbeta}).  (Of course, other combinations appear at higher order.)

We now discuss the magnitude of the correction due to $\beta$.  For
black holes, one has a strict upper limit on the magnitude of the
spins: $|\vec S_i| \le M_i^2$.  This is also roughly the upper limit
for neutron stars, though the actual upper limit depends on the
(uncertain) nuclear equation of state.  We can therefore estimate the
maximum size of $\beta$ by considering the case where the spins are
aligned with $\vec L$, and where $|\vec S_1|/M_1^2 =|\vec S_2|/M_2^2=1 $.
In this case $\beta= {{113}\over{12}} - {{19}\over{12}}\bigl(4\mu/
M\bigr)$.  This maximum value $\beta_{max}$ is always within $10\%$ of
$8.5$, regardless of the mass ratio.

The $P^{1.5}N$ order equations of motion also contain
``$\vec L \times \vec S$'' terms, which do not directly affect
$df/dt$, but do so indirectly by causing the directions of
$\hat L$, $\vec S_1$ and $\vec S_2$ to precess during the inspiral ---
essentially the Lense-Thirring effect.
The equations describing the secular evolution of
$\hat L$, $\vec S_1$ and $\vec S_2$
through $P^2N$ order are \cite{spins_paper}:
\begin{mathletters}
\label{spineqs}
\FL
\begin{eqnarray}
{d\hat L \over dt} &=&  r^{-3} \bigg[a_1 \vec S_1 +
 a_2 \vec S_2 \nonumber \\
\mbox{} & & - {3\over 2}{{(\vec S_2 \cdot \hat L)
\vec S_1 + (\vec S_1 \cdot \hat L)\vec S_2}\over L} \bigg]\times \hat L \\
{d\vec S_1 \over dt} &= &r^{-3}\left[ a_1 L \hat L
+{1\over 2}\vec S_2- {3\over 2}(\vec S_2 \cdot \hat L)\hat L
\right]\times \vec S_1 \\
{d\vec S_2 \over dt} &=&  r^{-3}\left[a_2 L \hat L +{1\over 2}\vec S_1-
{3\over 2}(\vec S_1 \cdot \hat L)\hat L  \right]\times \vec S_2,
\end{eqnarray}
\end{mathletters}
where $a_1 = 2 + (3 M_2)/(2 M_1)$, $a_2 = 2 + (3 M_1)/(2 M_2)$, $L =
|{\vec L}| = \mu \sqrt{M r}$, and where, to this order, one can use the
expression (\ref{rt}) for $r(t)$.

The precession of $\hat L$, $\vec S_1$ and $\vec S_2$ causes $\beta$
to evolve; $d\beta/dt$ as calculated from Eqs.~(\ref{spineqs}) does
{\it not} vanish identically.  Fortuitously, however, $\beta$ is {\it
almost} conserved by Eqs.~(\ref{spineqs}), in the following sense. We
integrated these equations numerically from $f=10 {\rm Hz}$ to $f
=(6^{3/2}\pi M)^{-1}$, for a wide variety of spin magnitudes, initial
spin directions, and mass ratios; we found that $\beta$ never deviates
from its average value by more than $\sim 0.25$ (or $\sim 0.03
\,\beta_{max}$).  Moreover, the non-constant part of $\beta$ is
oscillatory, which further diminishes its integrated effect on the
waveform's phase.  These properties of the evolution of $\beta$ are
explored analytically and numerically in Appendix \ref{beta_const}.

The near-constancy of $\beta$ allows a considerable simplification
of our model waveform: in Eq.~(\ref{pnsdfdt}), we simply take $\beta$
to be a constant. That is, we treat $\beta$ as just another parameter,
like ${\cal M}$ and $\mu$, on which the signal depends. The Fourier
transform of our model
waveform, including spin effects, is therefore given by
\FL
\begin{equation}
\label{pnshf}
\tilde h(f) = \left\{ \begin{array}{ll}
		{\cal A}\, f^{-7/6}\, e^{i \Psi}\ \ \ \ \  & \mbox{$0 < f < (6^{3/2}\pi
M)^{-1}$} \\
		0				 & \mbox{$(6^{3/2}\pi M)^{-1}
< f$}
		\end{array} \\
                 \right.
\end{equation}
where now
\FL
\begin{eqnarray}
\label{pnspsi}
\Psi(f) & = &2\pi f t_c -\phi_c -\pi/4 +{3\over 4}(8 \pi {\cal M} f
)^{-5/3} \\
\mbox{} & & \times \, \biggl[1+ {20\over 9}\left({743\over 336}+{{11
\mu}\over {4M}}\right)x +\left(4\beta -16 \pi \right)x^{3/2} \biggr]. \nonumber
\end{eqnarray}

Now, for spinning bodies it is not really correct to treat the
amplitude ${\cal A} \equiv Q(\theta, \phi, \psi, \iota) D^{-1} {\cal
M}^{5/6}$ as constant.  The precession of the orbital plane described
by Eqs.~(\ref{spineqs}) causes the angles $\psi$ and $\iota$ and to
vary---and hence $Q(\theta,\varphi,\psi,\iota)$ to vary---throughout the
inspiral.  Typically, the orbital plane precesses around the total
angular momentum vector $\vec J \equiv \vec L + \vec S_1 + \vec S_2$
roughly $20$ times during the observable portion of the inspiral.  The
result is a sinusoidal modulation of the waveform envelope
\cite{cutleretal,spins_paper1}, and the amplitude of the modulation can
be large when $|\vec S_1|$ or $|\vec S_2|$ is comparable to $|\vec
L|$.  Nevertheless, in the interest of simplifying the calculation,
{\it in our model waveform} (\ref{pnshf}) {\it we take ${\cal A}$ to
be a constant}.  We discuss further below the implications of this
simplification.

The derivatives $\partial {\tilde h} / \partial \, {\rm ln} {\cal A}$,
$\partial \tilde h/\partial(f_0 t_c)$, and $\partial \tilde h/
\partial\phi_c$ of the signal (\ref{pnshf}) are given by the same
expressions as in Eqs.~(\ref{pnderivs}).  The derivatives of $\tilde
h(f)$ with respect to ${\rm ln} \,{\cal M}$, ${\rm ln} \,\mu$, and
$\beta$ are:
\begin{mathletters}
\label{pnsderivs}
\FL
\begin{eqnarray}
{\partial {\tilde h}(f) \over \partial \ln \, {\cal M}} & = & -{{5
i}\over{4}}(8\pi {\cal M} f)^{-5/3} {\tilde h}(f) \biggl[1 + {{55
\mu}\over{6M}} x \nonumber \\
\mbox{} & & + (8\pi-2\beta) \,x^{3/2} \biggr] \\
{\partial {\tilde h}(f) \over \partial \ln \mu } & = & {{3i}\over
{4}} (8\pi {\cal M} f)^{-5/3} \,{\tilde h}(f) \,\biggl[({{-3715}\over
{756}} +{{55 \mu}\over{6M}}) \,x \nonumber \\
\mbox{} & & + (24 \pi-6\beta) \,x^{3/2} \biggr] \\
{\partial {\tilde h}(f) \over \partial \beta} & = &
3i\,(8\pi {\cal M} f)^{-5/3} (\pi M f) \,{\tilde h}(f).
\end{eqnarray}
\end{mathletters}
Using Eqs.~(\ref{pnsderivs}) we again compute the variance-covariance
matrix $\Sigma^{ij}$ for a range of values of $M_1$ and $M_2$.
One can show that $\Delta \phi_c$,
$\Delta t_c $, $\Delta {\cal M}/{\cal M} $, and $\Delta \mu/{\mu}$ do
not depend the value of $\beta$.  The simple way to prove this is to
make a change of variables from $({\cal A},\phi_c,t_c,{\cal
M},\mu,\beta)$ to $({\cal A},\phi_c,t_c,{\cal M},\mu,\beta^\prime)$,
where
\begin{equation}
\beta^\prime \equiv \left(4\beta -16 \pi \right) {\cal M}^{5/6}
{\mu}^{-3/2}.
\end{equation}
Since the waveform phase $\Psi(f)$ [Eq.~(\ref{pnspsi})] depends
linearly on $\beta^\prime f^{-2/3}$, the Fisher information matrix
calculated with respect to the new variables is independent of
$\beta^\prime$.  This implies that the rms errors in the {\it other}
parameters, and their correlation coefficients, are independent of the
value of $\beta$.  The values of $\Delta \beta$, $c_{{\cal M} \beta}$
and $c_{\mu \beta}$ do depend on $\beta$, however.

In Table \ref{table2} we list the rms errors $\Delta \phi_c$, $\Delta
t_c $, $\Delta {\cal M}/{\cal M} $, $\Delta \mu/{\mu}$, and $\Delta
\beta$ for the same fiducial binaries that appear in Table \ref{table1}.
In computing the results in Table \ref{table2} we use the model of the
advanced detector noise curve given by Eq.$~(2.1)$.  Since we are
principally concerned with how our lack of knowledge of the bodies'
spins affects how well we can determine the {\it other} parameters, we
take the ``true'' value of $\beta$ to be zero in all cases.  As in
Table \ref{table1}, the results in Table \ref{table2} are for a single
detector and are normalized to $S/N = 10$; for a detector network, the
rms errors $\Delta {\cal M}/{\cal M} $, and $\Delta \mu/{\mu}$ and
$\Delta \beta$ will be approximately those given in Table
\ref{table2}, but with $S/N$ replaced by the combined signal-to-noise
ratio $\rho$.

Summarizing the results of Table \ref{table2}, we find that
${\Delta {\cal M}}/{\cal M} $ is roughly an order
of magnitude larger than predicted by the Newtonian analysis of
Sec.~\ref{sec2}, but still typically less than $0.1\%$.
Thus, despite the ``confusion'' introduced by the extra parameters that
enter at post-Newtonian order, we conclude that ${\cal M}$
can still be measured with remarkable accuracy.
However, compared to the case where the bodies
are assumed to have negligible spin {\it a priori}, we see that $\Delta \mu$
has increased by a factor which ranges from $20$ to $60$!

Table \ref{table2} also reveals the ``reason'' for this loss of
accuracy: the correlation coefficient $c_{\mu \beta}$ is extremely
close to $-1$ \cite{correlation_effects}. Clearly the strong
correlation is due to the fact that the frequency dependence of the
``$\vec L \cdot \vec S$'' term in the expression (\ref{pnspsi}) for
the waveform phase $\Psi(f)$ is very similar to the
frequency dependence of the other post-Newtonian terms in
Eq.~(\ref{pnspsi}).  This strong correlation implies that there is a
combination of $\mu$ and $\beta$ which can be determined to much
higher accuracy than $\mu$ itself \cite{correlation_effects}.
Specifically, $\Delta\left(\mu - (\Sigma^{\mu\,
\beta}/\Sigma^{\beta\,\beta}) \, \beta\right)$ is smaller than
$\Delta \mu$ by a factor of $\sim (1-c_{\mu\,\beta}^2)^{-1/2}$, which
is approximately $20-60$ for the cases in Table \ref{table2}.  Thus
the combination $\mu - (\Sigma^{\mu\,\beta}/\Sigma^{\beta\,\beta}) \,
\beta$ can be determined with approximately the same accuracy that
one {\it could} achieve for $\mu$, {\it if} spin effects could be
neglected [cf. Table \ref{table1}].  Since both ${\cal M}$ and this
particular combination of $\mu$ and $\beta$ can be determined to high
accuracy, the inspiral gravitational wave measurement essentially
constrains the parameters to lie near a thin two-dimensional strip in
$({\cal M}, \mu, \beta)$ space.  This is illustrated in
Fig.~\ref{mubeta}, for the case of a BH-NS binary.

Up to this point, the formalism we have been using to calculate
measurement accuracies neglects {\it a priori} constraints on the
parameters, and thus implicitly assumes that $\beta$ can take on
arbitrary values.  This assumption should be adequate as long as the
$95 \%$ confidence intervals determined from $\Sigma^{\beta \beta}$
are well within the ``allowed'' range: $|\beta| \le \beta_{max}
\approx 8.5$.  However we see from Table \ref{table2} that this
criterion is not satisfied when both bodies are heavier than a few
solar masses.  For example, when $M_1 =M_2 = 10 M_\odot$ we calculate
$\Delta \beta = 19.5$.  We can (somewhat crudely) incorporate the
restricted range of $\beta$ into our formalism, as follows.  We
replace the {\it a priori} information $|\beta| < \beta_{\rm max}$
at hand by an assumed Gaussian distribution $p^{(0)}(\beta) \propto
e^{-{1\over 2}(\beta/5)^2}$ for $\beta$.  In Appendix \ref{stats}\ we
derive an expression for the variance-covariance matrix which
incorporates the effect of an (assumed Gaussian) {\it a priori}
probability distribution for the signal parameters.  We have used this
result [Eq.~(\ref{simple_answer}) below] to re-evaluate the variance-covariance
matrix for the two high-mass binaries shown in Table \ref{table2}.
(Taking the restricted range of $\beta$ into account makes little
difference to the other cases in Table \ref{table2}.)  These
re-evaluated results are marked in Table \ref{table2} with a dagger
$(\dag)$.  Again, the rms errors listed are for $S/N = 10$; note
however that since $p^{(0)}(\beta)$ is fixed, the rms errors no longer scale
simply as $(S/N)^{-1}$.  We see that taking the restricted range of
$\beta$ into account leads to the improved estimate $\Delta
\mu/\mu \approx 50\%$ in both the high-mass cases.

We mentioned above that in the three known short-period
NS-NS binaries,  the radio pulsars will, at the time of merger,
all have spin angular momenta that are $\alt 2\%$ of their
maximum possible values.  We feel
it is an interesting exercise to calculate what measurement accuracies
could be attained if we knew that NS's in nature were slowly spinning
in general, e.g., if we knew {\it a priori} that $\beta < .02
\beta_{max} $ for NS-NS mergers. Repeating the procedure used above,
we take $p^{(0)}(\beta) \propto e^{-{1\over 2}(\beta/0.1)^2}$, and we
use Eq.~(\ref{simple_answer}) to calculate the variance-covariance
matrix for the NS-NS case, for $S/N = 10$.  We find $\Delta \mu/\mu
\approx 0.9 \%$, which is roughly twice the value obtained in
Sec.~\ref{no_spin}, where spin effects were taken to be completely
negligible.

We turn again to the question of how accurately the individual masses
can be measured. The procedure for calculating $\Delta M_1$ and
$\Delta M_2$ in terms of $\Delta {\cal M}$ and $\Delta \mu$ is of
course the same as described in Sec.~\ref{no_spin}.
Thus for the BH-NS case, using the fact that
$\Delta\mu/\mu \approx 15\%$, we find from Eqs.~(\ref{dm1dm2}) that
$\Delta M_1/M_1 \approx 30\%$ and $\Delta M_2/M_2 \approx 20\%$.
In Sec.~\ref{no_spin} we explained that the distribution of errors in
$M_1$ and $M_2$ will be non-Gaussian if $\mu + 2 \Delta\mu
> 2^{-4/5}\,{\cal M}$.  By this criterion, if $\Delta
\mu/\mu \approx 15\%$, then we can reliably estimate $\Delta M_1$ and
$\Delta M_2$ by using Eqs.~(\ref{dm1dm2}) only if $M_1/M_2 \ge
5.5$

Again, even when the Gaussian approximation is invalid, one can still
use Eqs.~(\ref{conf_limits}) to place $95\%$ confidence limits on
$M_1$ and $M_2$.  Consider again the NS-NS case, which we looked at in
this context in Sec.~\ref{no_spin}, with the {\it true} values of the
masses being $M_1 = M_2 = 1.4 M_\odot$.  Then ${\cal M} = 1.22
M_\odot$, and, using the $2\sigma$ error bar indicated by Table
\ref{table2} (for $S/N = 10$) we see that, $95 \%$ of the time, the
observers would measure $\mu$ to be between $0.56 M_\odot$ and $0.70
M_\odot$.  Correspondingly, the measured values of $M_1$ and $M_2$
would lie in the ranges $1.4 M_\odot < {\hat M_1} < 3.2 M_\odot$ and
$0.7 M_\odot < {\hat M_2} < 1.4 M_\odot$.  Thus, in the NS-NS case,
measuring $\mu$ to within $20\%$ means determining the individual
masses only to within a factor of $\sim 2$.  The constraints obtained
on $M_1$ and $M_2$, for this case and the BH-NS case, are illustrated
in Fig.~\ref{m1m2}.

Finally, we repeat these calculations using the flatter spectrum
(\ref{snf2}) instead of (\ref{snf}) as our model of the advanced
detector noise.  The results are shown in table \ref{table3}.  We see
that the main conclusions which we drew from table \ref{table2} are
unchanged, but that (for fixed signal-to-noise) the relative errors
$\Delta {\cal M}/{\cal M}$ and $\Delta\mu/\mu$ are a factor of $\sim
1.5$ times smaller with the flatter noise spectrum (\ref{snf2}).  This
is presumably due to the fact that noise spectrum (\ref{snf2})
exhibits better sensitivity at low frequencies, where most of
the gravitational wave cycles (and hence most of the sensitivity)
come from.

\subsection{Caveats and future work}
\label{caveats}

Since the results in Tables \ref{table1}--\ref{table3} were obtained
using several
approximations and simplifying assumptions,
we feel that it is useful to collect the most important
of these in one place.  They are as follows:

First, we restricted attention to statistical errors arising from
detector noise.  In practice, theoretical template waveforms will be
quite difficult to compute accurately \cite{cutleretal,cutler_finn}.
Hence some systematic error may also arise from fitting the data to
imperfect template waveforms.  Currently a large effort is underway in
the relativity community to calculate templates sufficiently
accurately that these systematic errors will be at most comparable to
the statistical errors that we have obtained --- at least $P^2N$ and
possibly higher order templates will be required.  We note that
template inaccuracies, while giving rise to important systematic
errors in parameter extraction, will not significantly diminish our
ability to {\it detect} the waves \cite{cutleretal}.

Second, we assumed the ``advanced LIGO'' noise-curve shape, for which
we have used two estimates: Eq.~(\ref{snf}) and Eq.~(\ref{snf2}).  As
emphasized above, these are only rough estimates of the spectral shape
that the LIGO/VIRGO detectors will actually achieve.  We have seen,
however, that our results do not depend very sensitively on the exact
shape of the noise spectrum.

Third, we have used the approximate, linearized error-estimation
formalism described in Sec.~\ref{stats0} and Appendix
\ref{stats}; the rms errors so calculated are guaranteed to be
accurate only in the limit that the errors are small. When the errors
are so large that the linearized approach is invalid, then our
approach will probably generally underestimate the true variances.  To
avoid the limitations of the linearized error analysis, we are
currently performing a Monte-Carlo simulation of the parameter
extraction process.

Fourth, we calculated the variance-covariance matrix $\Sigma^{ij}$
using the simplified model waveform (\ref{pnshf}), which is
qualitatively inaccurate in a number of respects.  In particular, our
model waveform depends on the spins of the two bodies only through a
single parameter, $\beta$.  We have neglected the spin-induced
precession of the orbital plane, which also arises at $P^{1.5}N$
order, and we have neglected the effect of the spin-spin coupling on
inspiral rate, which arises at $P^2N$ order.  We have also neglected
higher-order multipole
radiation (except insofar as the energy carried away by the higher
multipoles affects the inspiral rate), and have only crudely
modeled the cut-off of the waveform during the bodies' final
tidal-disruption, plunge, or coalescence. [In particular, we have made
no attempt to model the spin dependence of the cut-off.]

It is unclear to us whether the inadequacies of our model waveform
have led us to underestimate or overestimate parameter-extraction
accuracies.  On the one hand, the $P^2N$ spin-spin interaction term
that we have neglected would, if included, inevitably lead to some
degradation of parameter-extraction accuracy (as always happens when
there are more parameters to fit for).  On the other hand, it seems
clear that the effect on $\Delta \mu$ of adding the spin-spin term to
the waveform will be far less dramatic than the inclusion of the
spin-orbit term $\beta$, for two reasons: (i) as shown by Kidder {\it
et al.} \cite{kidder}, for the BH-NS and NS-NS cases, the effect of
the spin-spin term on the accumulated phase of the waveform is a
factor of at least $20$ smaller than the effect of the spin-orbit
term, and (ii) the correlation coefficient $|c_{\mu \beta}|$ is so
close to $1$ because the frequency dependences of the $\mu$ and
$\beta$ terms in the waveform phase $\Psi(f)$ are so similar; the
$\mu$ and spin-spin terms are less similar in their
frequency dependence.

Finally, by neglecting higher-order multipoles, spin-precession effects,
and the details of the final plunge, we have effectively thrown away
information that would be contained in the true waveform.  In a more
complete analysis, this ``additional'' information could perhaps
decrease measurement uncertainties.  In particular, if in some cases the
spin-related modulation of the waveform carries substantial
information about $\beta$, then it is clear from Fig.~\ref{mubeta}
that $\Delta \mu$ and consequently $\Delta M_1$ and $\Delta M_2$ could
be reduced by large factors.  This is an important possibility which
we are currently investigating.

\section{Accuracy of distance measurements}
\label{dist_errors}

\subsection{Overview}
\label{dist_overview}

In the previous sections we have investigated how accurately the
masses of the inspiraling compact objects can be measured from the
phase evolution of the detected gravitational waveforms.  The other
interesting parameters that are measurable from the outputs of a
network of detectors are the distance $D$ to the source, and its position
on the sky.  These parameters will be encoded in the amplitudes, phases and
arrival times of the signals $h_a(t)$ read out from the detectors.  At
least three geographically separated detectors will be needed in order
to determine the distance \cite{schutz}.  We start by describing, in
detail, the dependence of the signals $h_a(t)$ on the binary's
distance and sky location.

Let ${\bf x}_a$ be the position and ${\bf d}_a$ be the polarization
tensor of the $a$th detector in a detector network.  By polarization
tensor we mean that tensor ${\bf d}_a$ for which the detector's
output is given in terms of the waves' transverse traceless strain
tensor ${\bf h}({\bf x},t)$ by
\begin{equation}
\label{polardef}
h_a(t) = {\bf d}_a : {\bf h}({\bf x}_a,t).
\end{equation}
Here the colon denotes a double contraction.  If the arms of the
detector are in the directions of the unit vectors ${\bf {\it l}}$ and
${\bf m}$, then ${\bf d}_a = ({\bf {\it l}}
\otimes {\bf {\it l}} - {\bf m} \otimes {\bf m})/2$ \cite{Forward}.

We introduce a spherical polar coordinate system $(\theta,\varphi)$
centered at the Earth so that the axis $\theta = 0$ is the Earth's
axis of rotation.  The angle $\varphi$ is longitude and $\pi/2 -
\theta$ is North latitude for $\theta < \pi/2$.  Let ${\bf
n} = (\sin \theta \, \cos \varphi, \sin \theta \, \sin \varphi, \cos
\theta)$ be the unit vector in the direction $(\theta$, $\varphi)$,
and let ${\bf e}^+_{\bf n}$ and ${\bf e}^\times_{\bf n}$ be a basis
for the transverse traceless tensors perpendicular to ${\bf n}$.  If
we demand in the usual way that ${\bf e}^A_{\bf n} : {\bf e}^B_{\bf n}
= 2 \delta^{AB}$, for $A,B = +,\times$, then this basis is unique up
to rotations of the form
\begin{equation}
\label{rot}
{\bf e}^+ + i {\bf e}^\times \to e^{2 i \Delta \psi} ({\bf e}^+ + i
{\bf e}^\times).
\end{equation}
The quantities
\begin{equation}
\label{formfactor}
F^A_a({\bf n})  \equiv {\bf e}^A_{\bf n} : {\bf d}_a,
\end{equation}
for $A = +,\times$, are the so-called detector beam-pattern functions
for the $a$th detector \cite{300years}.

Consider a coalescing binary source in the direction ${\bf n}$.  As in
Sec.~\ref{spin_sec}, let ${\hat {\bf L}}$ denote the unit vector in the
direction of the binary's orbital angular momentum, and let
\begin{equation}
v = \cos \iota = {\hat {\bf L}} \cdot {\bf n},
\end{equation}
so that $\iota$ is the inclination angle of the orbit to the line of sight.  As
seen from
the Earth, the orbit looks elliptical, and the principal axes of the
ellipse give a preferred polarization basis ${\bf e}^{\prime+}$, ${\bf
e}^{\prime \times}$ for the waves.  Specifically, we define
\begin{equation}
{\bf e}_x^\prime = {{\bf n} \times {\hat {\bf L}} \over || {\bf n} \times
{\hat {\bf L}}||},
\end{equation}
\begin{equation}
{\bf e}_y^\prime = {-{\bf n} \times {\bf e}_x^\prime \over || {\bf n} \times
{\bf e}_x^\prime||},
\end{equation}
where the minus sign is inserted to accord with standard conventions; the waves
propagate in the direction $-{\bf n}$.  The preferred basis is
${\bf e}^{\prime+} = {\bf e}_x^\prime \otimes {\bf e}_x^\prime -
{\bf e}_y^\prime \otimes {\bf e}_y^\prime$,
${\bf e}^{\prime\times} = {\bf e}_x^\prime \otimes {\bf e}_y^\prime +
{\bf e}_y^\prime \otimes {\bf e}_x^\prime$.  In terms of this basis,
the waves' strain tensor is ${\bf h}(t) = h_+(t) {\bf e}^{\prime+} +
h_\times(t) {\bf e}^{\prime\times}$, where in the quadrupole-moment
approximation the waveforms $h_+(t)$ and $h_\times(t)$
are as given in, e.g., Ref.~\cite{300years}.  Taking the Fourier
transform we find
\begin{equation}
\label{waves_def}
{\tilde h}_A(f) = \chi_A(v) \, {\tilde h}_0(f),
\end{equation}
where $\chi_+(v) = (1+v^2)/2$, $\chi_\times(v) = -i v$, and
\FL
\begin{equation}
\label{tildeh0}
{\tilde h}_0(f) = \sqrt{5\over{24}} \pi^{-2/3} D^{-1} {\cal M}^{5/6}
f^{-7/6}\, \exp \left[i \Psi(f) \right]
\end{equation}
for $f \ge 0$.  The phase $\Psi(f)$ is the same as previously given in
Eq.~(\ref{pnspsi}), and depends only on the parameters ${\cal M}$,
$\mu$, $\beta$, $t_c$, and $\phi_c$.

If we fix a polarization basis ${\bf e}^+$, ${\bf e}^\times$, then we
have
\begin{equation}
\label{basis_relation}
{\bf e}^{\prime A} = R^A_{\,\,\,B}(2 \psi) \, {\bf e}^B
\end{equation}
for some polarization angle $\psi$, where $R^A_{\,\,\,B}$ is the
rotation matrix
\begin{equation}
R^A_{\,\,\,B}(2 \psi) =
\left(
\begin{array}{cc}
 \cos(2 \psi) &  \sin(2 \psi) \\
 - \sin(2 \psi) \,\,\,&  \cos(2 \psi)
\end{array}\right).
\end{equation}
The conventional definitions of ${\bf e}^+$, ${\bf e}^\times$ and the
corresponding definition of $\psi$ for a single detector are given in
Refs.~\cite{300years,spins_paper1}.  A network of several detectors,
however, determines a different preferred basis ${\bf e}^+$, ${\bf
e}^\times$ (see below), so for the moment we allow the basis to be
arbitrary and define $\psi$ via Eq.~(\ref{basis_relation}).  By combining
Eqs.~(\ref{polardef}), (\ref{formfactor}), (\ref{waves_def}), and
(\ref{basis_relation}) we obtain the signal read out from the $a$th
detector:
\begin{equation}
\label{fullexpr}
{\tilde h}_a(f) = R^A_{\,\,\,B}(2 \psi) \, \chi_A(v) F^B_a({\bf n}) \,
e^{2 \pi i \tau_a f} \, {\tilde h}_0(f),
\end{equation}
where $\tau_a = - {\bf n} \cdot {\bf x}_a$  \cite{caveat4}.
The first three factors in Eq.~(\ref{fullexpr}) taken together are
proportional to the quantity $Q(\theta, \varphi, \psi, \iota)$ that
appears in Eqs.~(\ref{nh}) and (\ref{fourh}).

Now the overall amplitude ${\cal A}$ of the signal at one detector can be
measured to an accuracy (cf.~Sec.~\ref{sec2})
\begin{equation}
\label{naive}
{\Delta {\cal A} \over {\cal A}} = {1 \over \rho_a},
\end{equation}
where $\rho_a = \left( h_a \, | \, h_a \right)^{1/2}$ is the signal to
noise ratio (SNR) measured at that detector.  Since
\begin{equation}
\label{amplitude1}
{\cal A} \propto \, {Q(\theta, \varphi, \psi, \iota) \over D}
\end{equation}
we expect the accuracy of distance measurements to be very roughly
$\Delta D / D \approx 1 / \rho$, where $\rho^2 = \sum \rho_a^2$ is the
 SNR (\ref{network_snr}), giving an accuracy of $ \sim 10\%$ for typical
detected signals.  However from Eq.~(\ref{amplitude1}) the signal
amplitudes are also strongly affected by the angles $\theta$,
$\varphi$, $\psi$, and most importantly the inclination angle $\iota$.
Hence, there will be correlations between the measured values of $D$
and of these angles, and the accuracy of distance measurement will be
reduced relative to the above naive estimate based on
Eq.~(\ref{naive}).

It is straightforward in principle to calculate the effect of all the
correlations by calculating the Fisher information matrix (\ref{sig})
from the waveform (\ref{fullexpr}) for all of the variables $D$, ${\bf
n}$, $v$, and $\psi$ together with the variables ${\cal M}$, $\phi_c$, $t_c$,
$\mu$ and $\beta$ discussed in Secs.~\ref{sec2} and \ref{sec3}.  An
analysis of this sort, but without including the post-Newtonian parameters
$\mu$ and $\beta$, has been carried out by Jaranowski and Krolak
\cite{krolak2},
who numerically calculate the rms error $\Delta D$ for various
different values of the angular variables.  They use the three
detector network consisting of the two LIGO detectors and the VIRGO
detector in Pisa, Italy, with their planned orientations.  Although
these authors do not take into account post-Newtonian effects, it seems likely,
for reasons which we
discuss below in Sec.~\ref{draza_approx} and Appendix \ref{decouple},
that their results for $\Delta D$ will not be sensitive to this
restriction.
Similar numerical calculations have been carried out by Markovi\'{c}
\cite{draza}, who assumed the same network of detectors.  He
identified a useful approximation for calculating $\Delta D/ D$, based
on identifying those variables with which the distance measurement is
most strongly correlated, and neglecting the effect of the much
smaller correlations with the other variables.

In Sec.~\ref{draza_approx} below we present an analytic calculation of
$\Delta D / D$
which simplifies the treatments given in Refs.~\cite{draza,krolak2}.
Because the rms error $\Delta D$ depends on
several angular variables, it is difficult to explore its behavior
over the whole parameter space using numerical calculations of the
type in Refs.~\cite{draza,krolak2}.  Here, by using Markovi\'{c}'s
approximation, we derive an approximate analytic expression for
$\Delta D$, which is valid for any network of detectors.

We also
extend the analysis of Refs.~\cite{draza,krolak2} in the following two
respects.
First, we parametrize the dependence of the result on the positions
and orientations of all of the detectors in the following useful way.
We show that, for a given position $\theta, \varphi$ on the sky, the
detector network parameters influence $\Delta D$ {\it only} through
(i) the selection of a preferred polarization basis $({\bf e}^+, {\bf
e}^\times)$ [or equivalently a preferred polarization angle $\psi({\bf
n})$, cf.~Eq.~(\ref{rot}) above], and (ii) two quantities
$\sigma_D({\bf n})$ and $\varepsilon_D({\bf n})$, where we call
$\sigma_D({\bf n})$ the amplitude sensitivity and $1 -
\varepsilon_D({\bf n})$ the polarization sensitivity \cite{timing}.
We discuss these ``network sensitivity functions'' in detail in
Sec.~\ref{fns} below.  They are defined in such a way that
the total signal-to-noise ratio squared (\ref{network_snr}) of a
detected signal coming from the direction ${\bf n}$ with polarization
$\psi$ is of the form [cf.~Eq.~(\ref{rho3}) below]
\FL
\begin{equation}
\rho^2 \propto \sigma_D({\bf n}) \left[ 1 + \varepsilon_D({\bf n})
\cos( 4 \psi + \mbox{const} ) \,f(v) \, \right],
\end{equation}
where the function
\begin{equation}
f(v) \equiv { (1 - v^2)^2 \over 1 + 6 v^2 + v^4}
\end{equation}
is independent of ${\bf n}$ and $\psi$.  The values of $\sigma_D$ and
$\varepsilon_D$ are, as an example, $\sigma_D = 1$ and
$\varepsilon_D=0$ for the case of two detectors at the same location, rotated
with respect to each other by $45^\circ$, and for vertically incident
waves.  In Figs.~\ref{ligovirgo1} and \ref{ligovirgo2} below we show
plots of these quantities as functions of the angles $\theta$ and
$\varphi$, for the 3-detector, LIGO/VIRGO network.

Second, we extend in Sec.~\ref{complex} the analysis beyond the
linear, Gaussian
approximation outlined in Appendix \ref{stats}, which is normally used
to estimate the rms errors.  We do this by calculating the exact
(within the Markovi\'{c} approximation), non-Gaussian probability
distribution for the distance $D$ which incorporates both our {\it a
priori} knowledge and the information obtained from a gravitational
wave measurement.  This extension becomes important in two different
regimes.  The first regime is when $v = \cos \iota \to 1$,
corresponding to binaries that we perceive to be almost face-on.  In
the limit $v \to 1$, the value of $\Delta D$ predicted by the linear
approximation becomes infinite.  As shown by Markovi\'{c}
\cite{draza}, this is because two of the signal parameters become
degenerate (i.e., the derivatives $\partial {\bf h}/ \partial
\theta^i$ become linearly dependent) as $v \to 1$.  Markovi\'{c} gave
rough estimates of the effect of this breakdown of the linear
formalism on the predicted value of $\Delta D$; the effect is not
treated in the exact numeric calculations of Ref.~\cite{krolak2}.
Here, using the non-Gaussian distribution for $D$, we obtain an
improved approximation to $\Delta D$ near the points of degeneracy.
The second regime where our non-Gaussian extension of the
error-estimation method is important is the limit of low
signal-to-noise, and correspondingly of large relative errors in the
measured binary parameters.  Since the Fisher matrix method of
calculating the rms errors in the measured parameters gives
essentially the leading order term in an expansion in powers of
$(S/N)^{-1}$, this method will be inaccurate at low values of $S/N$.
By using an approximation which takes into account the dominant
effects that are non-linear in $(S/N)^{-1}$, we numerically estimate
$\Delta D$ for different values of the parameters
${\bf n}$, $v$, and $\psi$.
We show that the linear estimates for $\Delta D$ are typically off
by factors $\agt 2$, even for signal-to-noise ratios of more than
twice the threshold value for detection, due in some cases to large
non-Gaussian tails in the PDF for $D$.  Thus, effects that are
non-linear in $(S/N)^{-1}$ are often {\it not} a small correction for
typical detected signals.

Finally, in Sec.~\ref{simul} we apply our non-linear error estimation
method to calculate the distribution of measurement accuracies for the
LIGO/VIRGO network, using a Monte-Carlo simulation.  We estimate that
$\sim 8 \, \%$ of the distance measurements will be accurate to $\le
15 \, \%$, and $\sim 60 \, \%$ to  $\le 30 \, \%$ (see
Fig.~\ref{histogram} below).

Our analyses are applicable to binaries at cosmological distances,
provided we interpret $D$ as the luminosity distance to the source,
and ${\cal M}$ as $(1 + z)$ times the true chirp mass, where $z$ is
the source's redshift \cite{schutz,draza,finn2}.  However, a
potentially important effect that we neglect is the spin-induced
modulation of the signal amplitudes discussed in Sec.~\ref{spin_sec}
and Ref.~\cite{spins_paper1}.  Hence, our results for $\Delta D$
should be regarded as rough estimates (and probably also as lower
limits, since it seems most likely that including spin effects in the
computation will increase $\Delta D$).  However, the tools we develop
below will be useful in future, more complete analyses of distance
measurement accuracies.

We use throughout this section the notations of Appendix \ref{stats}.

\subsection{The network functions $\sigma_D({\bf n})$ and
$\varepsilon_D({\bf n})$}
\label{fns}

The overall SNR (\ref{network_snr}) and the Fisher information matrix
(\ref{gamma_def}) are determined by inner products involving the
signal ${\bf h}(t)$ and its derivatives $\partial {\bf h} / \partial
\theta^i$ with respect to the signal parameters $\theta^i$.  We now
show that a large class of these inner products depends on the network
properties [i.e., the detector positions ${\bf x}_a$ and polarization
tensors ${\bf d}_a$] only through the two functions of sky location,
$\sigma_D({\bf n})$ and $\varepsilon_D({\bf n})$.  We start by
defining the complex amplitudes
\begin{equation}
\label{ampdef1}
{\cal A}_B \equiv  R^A_{\,\,\,B}(2 \psi) \, \chi_A(v) \, e^{-i \phi_c}
/ D
\end{equation}
which are intrinsic to the incident waves, and the detector amplitudes
\begin{equation}
\label{ampdef2}
{\cal A}_a = \sum_{B = +,\times} {\cal A}_B F^B_a({\bf n})
\end{equation}
which characterize the signals seen at the various detectors.  In terms
of these quantities, the signal (\ref{fullexpr}) can be written as
\begin{equation}
\label{signalsimp}
{\tilde h}_a(f) = {\cal A}_a \, e^{2 \pi i \tau_a f} \, {\tilde k}(f),
\end{equation}
where ${\tilde k}(f) \equiv D e^{i \phi_c} {\tilde h}_0(f)$ is
independent of $D$ and $\phi_c$.  The inner product of two signals
${\bf h}$ and ${\bf h}^\prime$ written in this way, with amplitude
parameters ${\cal A}_A$ and ${\cal A}_A^\prime$, is given by
Eqs.~(\ref{tildeh0}), (\ref{signalsimp}) and (\ref{product_def}):
\begin{equation}
\label{product_amp}
\left( {\bf h} \, | \, {\bf h}^\prime \right)
= \Re \left[ {\cal A}_a^* {\cal A}_b^\prime \, \kappa^{ab} \right] \big(
{\tilde k} \, \big| \, {\tilde k} \big).
\end{equation}
Here the positive definite Hermitian matrix $\kappa^{ab}$ is
\FL
\begin{equation}
\label{kappa_def}
\kappa^{ab} =  {\int_0^\infty df \, f^{-7/3} \, \left[ {\bf
S}_n(f)^{-1} \right]^{ab} \,\, e^{2 \pi i f (\tau_b - \tau_a)}
\over \int_0^\infty df \, f^{-7/3} / S_n(f) },
\end{equation}
and $S_n(f)$ in the denominator is the average of the spectral noise
densities in all the detectors.  If the detectors are all identical,
and correlated sources of noise [represented by the off-diagonal
elements of ${\bf S}_n(f)$] are unimportant, then $\kappa^{ab}$ is
just $\delta^{ab}$.

In terms of the wave amplitudes ${\cal A}_A$, the inner product
(\ref{product_amp}) is, from Eq.~(\ref{ampdef2}),
\begin{equation}
\label{product_amp1}
\left( {\bf h} \, | \, {\bf h}^\prime \right)
= \Re \left[ {\cal A}_A^* {\cal A}_B^\prime \, \Theta^{AB} \right] \big(
{\tilde k} \, \big| \, {\tilde k} \big),
\end{equation}
where the matrix ${\bf \Theta}$ is given by
\begin{equation}
\label{Theta_def}
\Theta^{AB}({\bf n}) = \sum_{a,b} F^A_a({\bf n}) F^B_b({\bf n}) \, \kappa^{ab}.
\end{equation}
We see that all inner products of the type (\ref{product_amp}) depend
on the network parameters only through the $2 \times 2$ Hermitian
matrix ${\bf \Theta}$.  Two key simplifications now arise.  First,
correlated sources of noise will presumably be limited to pairs of
detectors at the same detector site, so that the detector-network
noise matrix (\ref{s_h_def}) will have a block-diagonal form with each
block corresponding to a detector site.  If the detectors at each site
are all oriented the same way, as is likely, then the product of beam
pattern functions appearing in Eq.~(\ref{Theta_def}) will be constant
over each block in the indices $a,b$ that corresponds to a non-zero
subblock of the matrix ${\bf S}_n(f)$.  Hence, from
Eqs.~(\ref{kappa_def}) and (\ref{Theta_def}), we see that the
imaginary part of ${\bf \Theta}$ will vanish.  Second, if we change
the basis ${\bf e}^+$, ${\bf e}^\times$ by a transformation of the
form (\ref{rot}), which amounts to redefining the polarization angle
$\psi$ by
\begin{equation}
\label{barpsi}
\psi \to {\bar \psi} = \psi + \Delta \psi,
\end{equation}
then ${\bf \Theta}$ will transform according to ${\bf \Theta}
\to {\bf R}( 2 \Delta \psi) \cdot {\bf \Theta} \cdot {\bf R}(- 2
\Delta \psi)$.  For fixed ${\bf n}$, we can by choosing $\Delta \psi$
suitably make ${\bf \Theta}$ diagonal, and so be of the form
\begin{equation}
\label{diagonal_form}
{\bf \Theta} = \sigma_D \left(
\begin{array}{cc}
 1 + \varepsilon_D & 0 \\
 0 & 1 - \varepsilon_D
\end{array}\right),
\end{equation}
where $0 \le \varepsilon_D \le 1$.  This defines the network functions
$\sigma_D({\bf n})$ and $\varepsilon_D({\bf n})$.  The required value
of $\Delta \psi = \Delta \psi({\bf n})$ is given by
\begin{equation}
\label{deltapsi}
\tan (4 \Delta \psi) = {2 \Theta_{+\times} \over \Theta_{++} +
\Theta_{\times \times}}.
\end{equation}

The combined SNR (\ref{network_snr}) can be determined in terms of
these network functions by combining Eqs.~(\ref{ampdef1}),
(\ref{product_amp1}) and (\ref{diagonal_form}) to give
\FL
\begin{equation}
\label{rho3}
\rho^2 = \rho_0^2 \, \sigma_D({\bf n}) \left[ c_0(v) +
\varepsilon_D({\bf n}) \, c_1(v) \cos(4 {\bar \psi}) \,\right],
\end{equation}
where $c_0(v) =  (1 + v^2)^2/4 + v^2$, $c_1(v) = (1 + v^2)^2/4 - v^2 =
(1 - v^2)^2/4$, and ${\bar \psi}$ is given by Eqs.~(\ref{barpsi})
and (\ref{deltapsi}).  The quantity
\FL
\begin{equation}
\label{rho02}
\rho_0^2 \equiv D^{-2} \big( {\tilde k} \, \big| \, {\tilde k} \big)
\,= \, \big( {\tilde h}_0 \, \big| \, {\tilde h}_0 \big) = 4
\int_0^\infty \, {|{\tilde h}_0(f) |^2 \over S_n(f)} \, df
\end{equation}
appearing in Eq.~(\ref{rho3}) is the SNR that would apply to one
detector if a face-on ($v =1$) binary were directly overhead.  From
Eqs.~(\ref{tildeh0}) and (\ref{rho02}), we find $\rho_0 = r_0 / D$,
where for the noise spectrum (\ref{snf}) the distance $r_0$ is
\FL
\begin{equation}
\label{r0}
r_0 = 6.5 \, {\rm Gpc} \left( {{\cal M} \over M_\odot} \right)^{5/6}
\left( {f_0 \over 70 \, {\rm Hz} } \right)^{-2/3} \left( {S_0 \over 3
\times 10^{-48} \, {\rm sec} } \right)^{-1/2}.
\end{equation}
The fiducial values of the detector parameters $S_0$ and $f_0$ used
here are those appropriate for the advanced LIGO detectors
\cite{ligoscience}, cf.~Sec.~\ref{sec2} above.  Note that the
dependence of the SNR (\ref{rho3}) on the polarization angle $\psi$
vanishes when the binary is perceived to be face-on ($v = 1$), as we
would expect physically due to rotational invariance about the line of
sight.  In the opposite limit of edge-on binaries ($v \to 0$), the
incident waves are highly linearly polarized, and the SNR typically
depends strongly on $\psi$, varying by factors of $\sim 10$ or more as
$\psi$ is varied [see Fig.~\ref{ligovirgo2} below].

We now derive simple formulae for the functions $\sigma_D$ and
$\varepsilon_D$.  From Eq.~(\ref{diagonal_form}), it is clear that
$\sigma_D$ is just half of the trace of the matrix ${\bf \Theta}$,
which is invariant under rotations.  Using Eqs.~(\ref{formfactor}) and
(\ref{Theta_def}) gives
\FL
\begin{equation}
2 \sigma_D = \left\{ \sum_A \, ( {\bf e}^A_{\bf n} )_{ij} \,
({\bf e}^A_{\bf n} )_{kl} \right\} \, \, \,
\left\{ \sum_{a,b} \, \kappa^{ab} \,\, ( {\bf d}_a )_{ij} \, ({\bf
d}_a)_{kl} \right\}.
\end{equation}
If we denote the first term in curly brackets by $S_{ijkl}$, then it is
straightforward to show that
\FL
\begin{eqnarray}
\label{bigS}
S_{ijkl} & = & - \delta_{ij} \delta_{kl} + ( \delta_{ik} \delta_{jl} +
\delta_{il} \delta_{jk}) + (\delta_{ij} n_k n_l + \delta_{kl} n_i n_j)
\nonumber \\
& & - ( \delta_{ik} n_j n_l + \delta_{il} n_j n_k + \delta_{jk} n_i
n_l + \delta_{jl} n_i n_k) \nonumber \\
& & + n_i n_j n_k n_l.
\end{eqnarray}
This yields for the amplitude sensitivity function the formula
\begin{eqnarray}
\label{sigmaDdef}
\sigma_D({\bf n}) & = & {1 \over 2} \kappa^{ab} \,[ \,2 \, {\bf
d}_a : {\bf d}_b - 4 \, {\bf n}  \cdot
({\bf d}_a \cdot {\bf d}_b ) \cdot {\bf n} \nonumber \\
\mbox{} & & + \, ( {\bf n} \cdot {\bf d}_a
\cdot {\bf n} ) ( {\bf n} \cdot {\bf d}_b
\cdot {\bf n} ) \,]
\end{eqnarray}
where we have used the property ${\rm Tr}\,{\bf d}_a = 0$.

It is similarly straightforward to evaluate the polarization
sensitivity $1 - \varepsilon_D({\bf n})$.
We introduce the notation
\begin{eqnarray}
\label{tensorprod}
\left< {\bf d}_a \, | \, {\bf d}_b \right>_n & \equiv & ({\bf
d}_a)_{ij} \, S_{ijkl} \, ( {\bf d}_b)_{kl} \\
\label{perpform}
& = & 2 \, {\bf d}_a^\perp : {\bf d}_b^\perp - ({\rm Tr} \, {\bf d}_a^\perp)
\, ({\rm Tr} \, {\bf d}_b^\perp),
\end{eqnarray}
where ${\bf d}_a^\perp$ denotes the projection $(\delta_{ik} - n_i
n_k) (\delta_{jl} - n_j n_l) \, ({\bf d}_a)_{kl}$ of ${\bf d}_a$
perpendicular to ${\bf n}$.  Then, using the relation from
Eq.~(\ref{diagonal_form}) that ${\rm Tr} \, {\bf \Theta}^2 = 2
\sigma_D^2 ( 1 + \varepsilon_D^2)$, and Eqs.~(\ref{formfactor}) and
(\ref{Theta_def}), gives
\FL
\begin{equation}
\label{varepsilonDdef}
\varepsilon_D({\bf n})^2 = {1 \over 2 \sigma_D({\bf n})^2} \sum_{abcd}
\left< {\bf d}_a \, | \, {\bf d}_b \right>_n
\, \left< {\bf d}_c \, | \, {\bf d}_d \right>_n \kappa^{ac}
\kappa^{bd} \,\,\,\,\, -1.
\end{equation}

We now evaluate $\sigma_D$ and $\varepsilon_D$ for the LIGO/VIRGO detector
network.  Let ${\bf e}_{\hat r} = {\bf n}$, ${\bf e}_{\hat \theta}$
and ${\bf e}_{\hat \phi}$ be the usual basis of orthonormal vectors.
Then for a detector at position $\theta,\varphi$ on the Earth's
surface, such that the angle measured anticlockwise from the local
eastwards directed meridian to the bisector of the detector arms is
$\alpha$, the polarization tensor is
\begin{eqnarray}
\label{ptensordef}
{\bf d} & = & - \sin (2 \alpha) ( {\bf e}_{\hat \theta}
\otimes {\bf e}_{\hat \theta} - {\bf e}_{\hat \phi} \otimes {\bf
e}_{\hat \phi})/2 \nonumber \\
\mbox{} & & + \cos (2 \alpha) ( {\bf e}_{\hat \theta}
\otimes {\bf e}_{\hat \phi} + {\bf e}_{\hat \phi} \otimes {\bf
e}_{\hat \theta})/2.
\end{eqnarray}
The values of $(\theta,\varphi,\alpha)$ for the various detectors are
$(59.4^\circ, -90.8^\circ, 243^\circ)$ for the LIGO detector in
Hanford, Washington, $(43.5^\circ, -119.4^\circ, 171^\circ)$ for the
LIGO detector in Livingston, Louisiana, and $(46.4^\circ, 10.25^\circ,
117^\circ)$ for the VIRGO detector in Pisa, Italy \cite{krolak2}.  We
assume that the detectors at all three sites are identical and that
noise sources are uncorrelated, so that from Eq.~(\ref{kappa_def}),
$\kappa^{ab} = \delta^{ab}$.  The resulting plots of $\sigma_D$ and
$1-\varepsilon_D$ are shown in Figs.~\ref{ligovirgo1} and \ref{ligovirgo2}.

\subsection{The Markovi\'{c} approximation}
\label{draza_approx}

We now explain the approximation method used by Markovi\'{c}
\cite{draza}, which we modify slightly below.  We start by considering
the accuracy $\Delta {\bf n}$ with which a given source can be located
on the sky.  The location ${\bf n}$ will be largely determined by
``time of flight'' measurements between the various detectors,
i.e., measurements of the quantities $\tau_a - \tau_b =- {\bf n} \cdot
({\bf x}_a - {\bf x}_b)$ in Eq.~(\ref{fullexpr})
\cite{schutz,krolak2}.  Hence, the variables ${\bf n}$ and $\tau_a$
will be strongly correlated, and $\Delta {\bf n}$ will be largely
determined by the ratio of the timing accuracies $\Delta (\tau_a -
\tau_b)$ to the light travel times between the various detectors.
Schutz \cite{schutz} has estimated the resulting angular resolution to
be $\sim 1$ square degree for typical detected signals, which is roughly in
agreement with the
recent detailed coalescing binary calculations of Jaranowski and
Krolak \cite{krolak2}.  It is also in rough agreement with numerical
simulations of G\"ursel and Tinto \cite{tinto}, which were carried out
in the context of arbitrary bursts of gravitational waves.  Hence, we
see from Eq.~(\ref{fullexpr}) that typical variations in ${\bf n}$
will give rise to variations in the measured value of $D$ that are
small compared to $\Delta D$.  Thus, the correlations between $D$ and
${\bf n}$ should be small, and to a good approximation we can treat
${\bf n} $ as known when calculating $\Delta D$ \cite{draza}.

In the approximation that ${\bf n}$ is constant, we can divide the
remaining parameters into two groups.  The first consists of the four
``amplitude'' parameters $D$, $v$, $\psi$, and $\phi_c$, which
determine the two complex amplitudes ${\cal A}_+$ and ${\cal
A}_\times$ via Eq.~(\ref{ampdef1}).  The second group of parameters
consists of ${\cal M}$, $t_c$, $\mu$, together with some spin
parameters, which enter only in the phase $\Psi(f)$ of the Fourier
transform of the signal, and control the evolution in time of the
phase of the waveform \cite{caveat5}.  In Appendix \ref{decouple} we
show that the
second group of parameters decouples from the first to linear order in
$1/\rho$, and in the constant ${\bf n}$ approximation.  More
precisely: if one calculates the Fisher matrix (\ref{gamma_def}) for
all of the variables except ${\bf n}$, inverts it to obtain the
covariance matrix $\Sigma^{ij}$, and takes the $4 \times 4$ subblock
of $\Sigma^{ij}$ corresponding to the amplitude group of parameters,
then the result is the same as if one computes the Fisher matrix for
just the four amplitude parameters alone, and then inverts that.
Heuristically what this means is that the effect of the correlations
between $(D, v, \psi)$ and all of the parameters $\phi_c, t_c, {\cal
M}$, etc., can be computed by considering the correlations with just
one phase variable, namely $\phi_c$, the orbital phase at coalescence
\cite{caveat5}.

We now calculate the Fisher information matrix (\ref{gamma_def}) for
the four amplitude parameters $D$, $v$, $\psi$ and $\phi_c$, and for
an arbitrary detector network, as this should yield a good
approximation to $\Delta D$.  The approximation that was used in
Ref.~\cite{draza} was in fact to consider only $D$, $v$, and $\psi$;
below we find [cf.~Eqs.~(\ref{gamma_ans}) and (\ref{bigG})] that the
fractional corrections due to also including $\phi_c$ are of order
$\varepsilon_D({\bf n}) \, \sin( 4 {\bar \psi})$, where ${\bar \psi}$
is given by Eqs.~(\ref{barpsi}) and (\ref{deltapsi}).  Since
$0 \le \varepsilon_D \le 1$ always, the fractional
corrections are always $\alt 1$ \cite{drazacorr}.

 From Eq.~(\ref{gamma_def}), it is clear that the Fisher matrices
calculated using two different sets of variables are simply related by
transforming with the Jacobian matrix of the variable transformation.
Hence, we can use any convenient set of variables to evaluate
$\Gamma_{ij}$ and $\Sigma^{ij}$, and afterwards transform to the
physical variables of interest.  We define the variables $\alpha$ and
$\beta$ by
\begin{equation}
\label{transform}
\left( \begin{array}{c}
\alpha \\
\beta  \end{array} \right)= {1 \over D}
\left( \begin{array}{c}
v  \\
(1 + v^2)/2 \end{array} \right).
\end{equation}
The waveform (\ref{fullexpr}) depends linearly on these variables,
which simplifies the computation.

Using Eqs.~(\ref{ampdef1}) -- (\ref{signalsimp}), (\ref{product_amp1}),
(\ref{diagonal_form}), (\ref{rho02}), (\ref{gamma_def}) and the
relation $\rho_0 = r_0 / D$, we obtain
\begin{equation}
\label{gamma_ans}
\Gamma_{ij} = r_0^2 \, \sigma_D({\bf n}) \left[ F_{ij} +
\varepsilon_D({\bf n}) \, G_{ij} \,\right],
\end{equation}
where the variables are $\theta^i = (\psi, \alpha, \beta, \phi_c)$.
Defining $c_0 = \alpha^2 + \beta^2$, $c_1 = \beta^2 - \alpha^2$, $c_4
= \cos(4 {\bar \psi})$, and $s_4 = \sin(4 {\bar \psi})$, the matrices
${\bf F}$ and ${\bf G}$ are given by
\begin{equation}
{\bf F} = \left[ \begin{array}{cccc}
4 c_0  & 0 & 0 &  -4 \alpha \beta \\
0 & 1 & 0 & 0 \\
0 & 0 & 1 & 0 \\
 -4 \alpha \beta & \,\,\,\,0 \,\,\,\,& \,\,\,\,0 \,\,\,\,& c_0
\end{array} \right]
\end{equation}
and
\begin{equation}
\label{bigG}
{\bf G} = \left[ \begin{array}{cccc}
-4 c_1 c_4  & \,\, 2 \alpha s_4 & - 2 \beta s_4 & 0 \\
 2 \alpha s_4 & - c_4 & 0  & - \beta s_4 \\
- 2 \beta s_4 & 0 & c_4 &  \alpha s_4 \\
0 & - \beta s_4 & \alpha s_4 & c_1 c_4 \end{array} \right].
\end{equation}
Inverting the matrix (\ref{gamma_ans}) and taking the $2 \times 2$
subblock corresponding to the variables $\alpha, \beta$, we find with
the help of {\it Mathematica} that
\FL
\begin{equation}
\label{alphabetaGamma}
\Sigma^{ij} = {1 \over r_0^2 \, \sigma_D \, (1 -
\varepsilon_D^2)} \left[ \begin{array}{cc}
1 + \varepsilon_D c_4 & 0 \\
0 & 1 - \varepsilon_D c_4
\end{array} \right].
\end{equation}
Finally, transforming this with the Jacobian of the transformation
(\ref{transform}) and taking the $(D,D)$ element of the resulting
matrix yields
\begin{equation}
\label{final_ans}
\Delta D^2 = {8 D^4 \over n_d \, r_0^2 } \,\, \Upsilon({\bf n},v,\psi)^2,
\end{equation}
where
\FL
\begin{equation}
\label{upsilon}
\Upsilon({\bf n},v,\psi)^2 = { n_d \left[ (1 + v^2) -
\varepsilon_D \cos(4 {\bar \psi}) \,(1 - v^2) \right] \over 2 \sigma_D
\, (1 - \varepsilon_D^2) \,(1 - v^2)^2},
\end{equation}
and $n_d$ is the number of detectors.  The dimensionless function
$\Upsilon$ satisfies
\begin{equation}
\Upsilon({\bf n},v,\psi) \ge 1,
\end{equation}
since from Eqs.~(\ref{sigmaDdef}), (\ref{perpform}), and
(\ref{diagonal_form}) it follows that $\sigma_D \le n_d /2$ and
$0 \le \varepsilon_D \le 1$ always, for any detector network.

Equation (\ref{final_ans}) is the main result of this subsection.  We
now discuss its properties and range of applicability.
%
%
It clearly breaks down and overestimates $\Delta D$ when $v
\to 1$.  As shown by Markovi\'{c}, this is because $\partial {\bf h} /
\partial D \, \propto \, \partial {\bf h} / \partial v$ at $v=1$, so
that the linear error-estimation method breaks down.  However, it will
{\it underestimate} the true measurement error for sufficiently small
values of the SNR $\rho$, because of the inadequacy of the linear
error-estimation formalism in this regime (cf.~Appendix \ref{stats}).
In Sec.~\ref{complex} below we numerically calculate more accurate
values of $\Delta D /D$, and show that even for small values of $v$,
and even for relatively large values of $\rho$ (e.g., $\rho \agt 20$,
more than twice the threshold), the results predicted by the formula
(\ref{final_ans}) can be off by factors $\agt 2$.

Hence, the formula (\ref{final_ans}) is of only limited applicability.
Its main virtue is that it allows one to understand qualitatively how
the distance measurement accuracy is influenced by the parameters
$\sigma_D$, $\varepsilon_D$, $\psi$, and (to a more limited extent)
$v$; and thereby by using Figs.~\ref{ligovirgo1} and
\ref{ligovirgo2} how it varies with sky location ${\bf n}$.  We now
discuss the dependence of $\Delta D / D$ on these parameters.

As the polarization angle $\psi$ is varied, it can be seen that
\begin{equation}
\Upsilon_{\rm min}({\bf n},v) \le \Upsilon({\bf n},v,\psi) \le
\Upsilon_{\rm max}({\bf n},v),
\end{equation}
where $\Upsilon_{\rm min}$ and $\Upsilon_{\rm max}$ are given by
substituting $\cos(4 {\bar \psi}) = \pm 1$ in Eq.~(\ref{upsilon}).  As
an illustration, Figs.~\ref{upmax} and \ref{upmin} show $\Upsilon_{\rm
min}$ and $\Upsilon_{\rm max}$ as functions of ${\bf n}$ at $v = 1 /
\sqrt{2}$.  It can be seen that the distance measurement accuracy can vary
over the sky by factors of order $\sim 20$, for binaries at a fixed
distance and with fixed inclination angle.  The reason for this strong
variation of more than an order of magnitude is easy to understand.  A
key feature of the result (\ref{upsilon}) is the factor of $1 / (1 -
\varepsilon_D^2)$, which diverges in the limit $\varepsilon_D \to 1$.
This divergence is {\it not} an artifact of our approximate, linear
error-estimation method (unlike the divergence in $\Upsilon$ at $v \to
1$).  The physical reason for the divergence as $\varepsilon_D \to 1$
is that for directions ${\bf n}$ such that $1 - \varepsilon_D({\bf n})
\ll 1$, the detector network has very poor ability to disentangle the
two polarization components of the incident waves, both of which are
needed in order to determine $D$.  As shown in Fig.~\ref{ligovirgo2},
there are large regions on the sky in which the polarization
sensitivity $1 - \varepsilon_D$ of the LIGO/VIRGO network is poor,
which correspond to the regions of high $\Delta D / D$ in
Figs.~\ref{upmax} and \ref{upmin} \cite{alternative}.

Part of the reason for the low values of $1 - \varepsilon_D$ for the
LIGO/VIRGO network is that the two LIGO detectors are nearly parallel,
so that they access essentially a single polarization component of the
gravitational wave field.  [They were chosen in this way in order to
enhance the reliability of detection of burst sources].  The addition of a
fourth detector would
greatly improve the polarization sensitivity of the network.  In
Fig.~\ref{aus} we plot that fraction $\Omega(\varepsilon_D)/4 \pi$
of the sky in which the polarization sensitivity is $\le 1 -
\varepsilon_D$, for the LIGO/VIRGO network.  We also as an
illustration plot the same quantity for a hypothetical 4-detector network
consisting of the LIGO and VIRGO detectors together with a  detector in
Perth, Australia, whose parameters [cf.~Eq.~(\ref{ptensordef}) above]
are assumed to be $(\theta,\varphi,\alpha) =
(121^\circ,116^\circ,90^\circ)$.

The quantity $\Upsilon({\bf n},v,\psi)$ shown in Figs.~\ref{upmax} and
\ref{upmin} gives distance-measurement accuracy as a function of {\it
Earth-fixed} coordinates $(\theta, \varphi)$.  The
distance-measurement accuracy for coalescing binaries at a given right
ascension and declination, averaged over many sources with different
arrival times, will clearly be given by the average over $\varphi$ of
$\Upsilon$ (due to the Earth's rotation).  Values of this averaged
accuracy in the band $0.15 \alt | \cos \theta | \alt 0.65$ are
typically a factor of $\sim 2$ better than those outside this band,
over the poles and near the celestial equator.  Similarly, the average
over $\varphi$ of $1 / \sigma_D(\theta,\varphi)$ is roughly
proportional to the average maximum distance to which sources can be
seen at a given declination; it does not vary by more than $\sim 20 \,
\%$.  Note that the distribution of sources on the sky is expected to
be approximately isotropic because the large distance ($\agt 200 \,
{\rm Mpc}$) to typical coalescences.

Lower bounds for $\Delta D / D$ can be obtained by combining
Eqs.~(\ref{rho02}), (\ref{final_ans}), and (\ref{upsilon})
and minimizing over ${\bf n},\,v$, and $\psi$.  If we define
\begin{eqnarray}
\sigma_{\rm max} & = & \max_{\bf n} \,\, \sigma_D({\bf n}) \\
& = & 1.04 \mbox{\ \ \     (for LIGO/VIRGO)},
\end{eqnarray}
we obtain the following lower bounds on $\Delta D/D$:
\begin{eqnarray}
\label{deltadmin}
{\Delta D \over D} & \ge &{2 \over \sqrt{\sigma_{\rm max}}} \, {D \over
r_0}, \\
\label{deltadmin1}
{\Delta D \over D} & \ge & {1 \over \rho},
\end{eqnarray}
together with the upper bound for the overall SNR $\rho$
\begin{equation}
\label{rhomax}
\rho \le \sqrt{2 \sigma_{\rm max}}\, \, {r_0 \over D}.
\end{equation}
These bounds remain roughly valid when effects that are
nonlinear in $r_0 / D$ are approximately taken into account
[cf.~Figs.~\ref{scatter1}-\ref{scatter4} below].

\subsection{Extension of analysis to beyond\\
the Gaussian approximation}
\label{complex}

As explained in Appendix \ref{stats}, the Fisher matrix approach to calculating
the probability distribution function (PDF) for the measured parameters is an
approximation whose validity depends in part on the particular set of
variables one uses to evaluate the Fisher matrix.  In particular, the
approximation works best for parameters on which the signal ${\bf
h}(t)$ depends {\it linearly}.  The key idea for dealing with the
degeneracy limit $v \to 1$ is to calculate the Gaussian probability
distribution for the amplitudes ${\cal A}_A$, which is exact because
the signal depends linearly on these amplitudes (see
Sec.~\ref{apriori} below).  Substituting Eq.~(\ref{ampdef1}) into this
PDF then yields the exact, non-Gaussian distribution for the
parameters $D$, $v$, $\psi$ and $\phi_c$, where we mean ``exact'' in
the context of the Markovi\'{c} approximation $\Delta {\bf n} =0 $.
{}From this non-Gaussian distribution, values of $\Delta D$ can be
determined which are more accurate than those given by
Eq.~(\ref{final_ans}) in the regime $v \to 1$ and for low
signal-to-noise ratios.

In this subsection we calculate so-called Bayesian errors instead of
frequentist errors.  The distinction is carefully explained in
Sec.~\ref{twoerrors} of Appendix \ref{stats}.  The distinction is important
only beyond leading order in $1/\rho$, and hence unimportant
elsewhere in this paper.  In practical terms, the use of Bayesian
errors means that the rms errors will be expressed as functions of the
measured, best-fit values for the source parameters, instead of their
true values.

By using Eqs.~(\ref{product_amp}) and (\ref{rho02}) one finds that the
exponential factor in Eq.~(\ref{pdf1}), given
a gravitational wave measurement, is proportional to
\FL
\begin{equation}
\exp \left[ - {r_0^2 \over 2}\,\,
({\cal A}_a - {\hat {\cal A}}_a )^* ({\cal A}_b - {\hat {\cal A}}_b )
\, \kappa^{ab} \right].
\end{equation}
Here the quantities ${\hat {\cal A}}_a$ are the amplitudes that we
measure at each detector (by using matched filtering).  The
corresponding PDF for the intrinsic amplitudes ${\cal A}_A$ is, from
Eqs.~(\ref{ampdef2}) and (\ref{pdf1}),
\FL
\begin{eqnarray}
\label{pdfa}
p({\cal A}_A) & = & {\cal N} p^{(0)}({\cal A}_A) \\
\mbox{} & & \times \, \exp \left[ - {r_0^2 \over 2}\,\,
({\cal A}_A - {\hat {\cal A}}_A )^* ({\cal A}_B - {\hat {\cal A}}_B )
\, \Theta^{AB} \right], \nonumber
\end{eqnarray}
where
\begin{equation}
\label{estdef}
{\hat {\cal A}}_A \equiv ({\bf \Theta}^{-1})_{AB} F^B_a {\hat {\cal
A}}_b \, \kappa^{ab}.
\end{equation}
Here $p^{(0)}({\cal A}_A)$ is our {\it a priori} PDF for the
amplitude parameters, and ${\cal N}$ is a normalization constant.

As an aside, Eq.~(\ref{estdef}) provides us with the
maximum-likelihood estimator ${\hat D}$ (in the constant ${\bf n}$
approximation) of the distance to the binary in terms of the measured
amplitudes ${\hat {\cal A}}_a$.  This is because Eq.~(\ref{ampdef1}),
re-expressed in terms of hatted quantities, may be inverted to
determine ${\hat D}$ in terms of the ${\hat {\cal A}}_A$'s:
\begin{equation}
\label{Destimator}
{\hat D} = { {\hat \beta} - \sqrt{{\hat \beta}^2 - {\hat \alpha}^2}
\over {\hat \alpha}^2},
\end{equation}
where [cf.~Eq.~(\ref{transform}) above]
\begin{equation}
\label{hatalpha}
{\hat \alpha}^2 = {1 \over 2} \left[ | {\hat {\cal A}}_+ |^2 + | {\hat {\cal
A}}_\times| ^2 - |{\hat {\cal A}}_+^2  + {\hat {\cal A}}_\times^2 |
\right]
\end{equation}
and
\begin{equation}
\label{hatbeta}
{\hat \beta}^2 = {1 \over 2} \left[ | {\hat {\cal A}}_+ |^2 + | {\hat {\cal
A}}_\times| ^2 + |{\hat {\cal A}}_+^2  + {\hat {\cal A}}_\times^2 |
\right].
\end{equation}

The PDF $p^{(0)}({\cal A}_A)$ in Eq.~(\ref{pdfa}) represents our {\it
a priori} information about the distribution of the parameters ${\cal
A}_A$ [or equivalently from Eq.~(\ref{ampdef1}) of the parameters
$(D,\,v,\,\psi,\,\phi_c)$ ], {\it given} that a signal has been
detected.  Since we expect sources to be uniformly distributed in
orientation and in space (on the relevant scales of $\agt 100 \, {\rm
Mpc}$), we take
\begin{eqnarray}
\label{APprob}
d\,p^{(0)} & \ \propto \ & d\psi \, d\phi_c \, \Theta(1-v^2) \, dv
\nonumber \\
\mbox{} & & \times \, \Theta(D) \, \Theta(D_{\rm max} - D) \, D^2  d D.
\end{eqnarray}

Here $\Theta$ is the step function, and the cutoff for distances
greater than $D_{\rm max}$ is a (somewhat crude) representation of our
knowledge that very distant sources would not have been detected.  A
suitable choice for $D_{\rm max}$ is the distance $r_0$,
cf.~Eqs.~(\ref{rho3}) and (\ref{r0}) above.  Our results below are
insensitive to the exact location of this cutoff, but it must be
included to make the PDF (\ref{pdfa}) formally normalizable.  Now let
$D_0$, $v_0$, $\psi_0$ and $\phi_{c0}$ be the parameters obtained from
the amplitudes ${\hat {\cal A}}_A$ by inverting Eq.~(\ref{ampdef1}),
so that, in particular, $D_0 = {\hat D}$.  Substituting
Eqs.~(\ref{ampdef1}) and (\ref{APprob}) into (\ref{pdfa}) yields a
non-Gaussian PDF for the variables $(D,\,v,\,\psi,\,\phi_c)$ which
depends on the parameters $(D_0,\,v_0,\,\psi_0,\,\phi_{c0})$.  From
this PDF it is straightforward in principle to calculate $\Delta D$,
by first integrating over $v$, $\psi$, and $\phi_c$ to determine the
reduced PDF $p(D)$ for $D$ alone.  If one first expands the argument
of the exponential to second order in the quantities $D - D_0$,
$v-v_0$, $\psi - \psi_0$, and $\phi_c - \phi_{c0}$, the result
obtained is just Eq.~(\ref{final_ans}) above, which is accurate to
linear order in $1/\rho$.

Thus, in order to go beyond this linear approximation, one has to
integrate the PDF (\ref{pdfa}) over $v$, $\psi$ and $\phi_c$.  Because
this is difficult to do exactly, we now make an approximation which
treats the correlations between $D$ and $(\psi,\phi_c)$ to linear
order in $1/\rho$, but treats more precisely the strong
correlations between $D$ and $v$.  This approximation should give
rough estimates of effects that are nonlinear in $1/\rho$, and
moreover removes the singularity in our previous result
(\ref{final_ans}) at $v = 1$.  The approximation consists of expanding
the argument of the exponential in Eq.~(\ref{pdfa}) to second order in
$\psi - \psi_0$ and $\phi_c - \phi_{c0}$, and integrating over $\psi$
and $\phi_c$.  One then obtains a function of $v$, $D$, $v_0$, $D_0$,
and $\psi_0$; the dependence on $\phi_{c0}$ drops out.  This function
is of the form (prefactor) $\times$ (exponential factor).
The prefactor depends only weakly on $D$ and $v$ in comparison to the
exponential factor, so we can approximate it to be constant
\cite{prefactor}.  We then obtain the following PDF, which may also be
obtained by substituting the transformation (\ref{transform}) into the
Gaussian PDF for the variables $\alpha, \beta$ that corresponds to the
variance-covariance matrix (\ref{alphabetaGamma}).

The result is, in terms of the
rescaled distance ${\cal D} = D / D_0$,
\FL
\begin{eqnarray}
\label{PDFvcalD}
dp(v,{\cal D}) & = & {\cal N} \, {\cal D}^2 \, \exp \bigg\{ -{1 \over 2
\Delta_1^2}
\left({v \over {\cal D}} - v_0\right)^2 \nonumber \\
\mbox{} & & - {1 \over 2 \Delta_2^2} \left[{1 + v^2 \over 2 {\cal D}} -
{1 + v_0^2 \over 2} \right]^2 \bigg\} \nonumber \\
& & \times \Theta({\cal D}) \Theta(D_{\rm max}/ D_0 - {\cal D}) \Theta(1 -
v^2) dv d{\cal D}.
\end{eqnarray}
Here $\Theta$ is the step function, ${\cal N}$ is a normalization
constant,
\begin{mathletters}
\label{deltad12}
\begin{eqnarray}
\Delta_1 & = & {D_0 \over r_0} \sqrt{1 + \varepsilon_D \cos(4 {\bar
\psi}_0) \over \sigma_D (1 - \varepsilon_D^2)}, \\
\mbox{} \Delta_2 & = & {D_0 \over r_0} \sqrt{1 - \varepsilon_D \cos(4 {\bar
\psi}_0) \over \sigma_D (1 - \varepsilon_D^2)},
\end{eqnarray}
\end{mathletters}
and ${\bar \psi}_0 =  \psi_0 + \Delta \psi({\bf n})$
[cf.~Eq.~(\ref{deltapsi}) above].  In terms of these quantities, the
previous, approximate result (\ref{final_ans}) is
\begin{equation}
{\Delta D \over D_0} = \, {2 \sqrt{ v_0^2 \Delta_1^2
+ \Delta_2^2} \over 1 - v_0^2}.
\end{equation}

 From the PDF (\ref{PDFvcalD}) one can numerically calculate the reduced
PDF for ${\cal D}$ alone,
\begin{equation}
\label{PDFcalD}
p({\cal D}) = \int_{-1}^1 dv \, p(v,{\cal D}),
\end{equation}
and thereby determine $\Delta D$.  As an example we show in
Fig.~\ref{prob} a plot of $p(D)$ for a particular choice of the
parameters $D_0$, $v_0$, $\psi_0$, and for a particular direction on
the sky.  The non-Gaussian fall off in this figure at large values of
$D$ is a general feature, although its magnitude in this example is
larger than is typical.  It can be seen that the distance measurement
accuracy is a factor of $\sim 2$ worse than that predicted by
Eq.~(\ref{final_ans}).

Now from Fig.~\ref{prob} it can be seen that the value of $D$ which
maximizes $p(D)$ is {\it not} the same as $D_0$, i.e., the $D$-component of the
point $(v_0,D_0)$ which maximizes $p(v,D)$.  Hence, the
``maximum-likelihood'' method for estimating signal parameters is
ambiguous --- the results obtained for one variable depend on whether
or not other variables are integrated out before the maximum is taken.
As explained in Appendix \ref{stats}, we advocate as the ``best-fit''
value of $D$ the expected value
\begin{equation}
\langle D \rangle \, = \, \int { D}\, p({ D})\,d{
D} = \int D p(v,D) dv dD,
\end{equation}
instead of the maximum-likelihood estimate $D_0$.  [Maximum-likelihood
estimation will need to be used, however, to obtain initial estimates
of the signal parameters].  Correspondingly, to estimate distance
measurement errors we use the quantity
\begin{equation}
\label{final_ans1}
{\Delta D \over D}\, \equiv \, {\sqrt{\langle {\cal D}^2 \rangle -
\langle {\cal D} \rangle^2} \over \langle {\cal D} \rangle }.
\end{equation}
This can be calculated numerically from Eqs.~(\ref{PDFvcalD}) and
(\ref{PDFcalD}), and in general will depend in a complicated way on
the parameters $\Delta_1$, $\Delta_2$, and $v_0$, and very weakly on
the rescaled cutoff $D_{\rm max}/ D_0$.  For the binary merger example of
Fig.~\ref{prob}, we show in Fig.~\ref{vdep} how the accuracy
(\ref{final_ans1}) varies with $v_0$, and in Fig.~\ref{ddep} how it varies
(through the parameters $\Delta_1$ and $\Delta_2$) with the distance
$D_0$.

The merger of a BH-NS binary of masses $10 M_\odot$ and $1.4 M_\odot$
would produce a signal whose amplitude is $2.11$ times stronger than
the NS-NS merger of Fig.~\ref{ddep} [from Eq.~(\ref{r0}) above].  Hence,
taking also into account a cosmological enhancement factor of $(1 +
z)^{5/6}$ \cite{draza}, a plot of $\Delta D / D$ versus $D_0$ for a
BH-NS binary otherwise the same as the binary in Fig.~\ref{prob}
would look roughly the same as Fig.~\ref{ddep}, but rescaled to extend
to luminosity distances $\sim 2 \, {\rm Gpc}$ (the exact value
depending on the cosmological model) \cite{draza}.

\subsection{Simulation of what LIGO/VIRGO will measure}
\label{simul}

In order to explore more completely the distance measurement accuracy
(\ref{final_ans1}) over the whole parameter space, we carried out the
following Monte-Carlo calculation.  Random values of $D_0$, $v_0$,
${\bar \psi}_0$, $\theta$, and $\varphi$ were chosen, distributed
according to the measure $d D_0^3 \, d v_0 \, d {\bar \psi}_0 \, d \cos
\theta \, d \varphi$.
Those parameter choices for which the combined SNR (\ref{rho02}) (for
NS-NS binaries) was less than the threshold of $8.5$ were discarded,
and samples were generated until 1000 NS-NS signals had been
``detected.''  Because of this thresholding procedure (which roughly
corresponds to the actual thresolding procedure that will be used),
the distribution of values of $D_0$, $v_0$ etc. for {\it detected}
events will not be given by $d D_0^3 \, d v_0 \, d {\bar \psi}_0 \, d
\cos \theta \, d \varphi$.  For example, there is a significant bias
in detected events towards high values of $v_0$, i.e., towards face-on
binaries.

Scatter plots of the distances $D_0$, signal-to-noise ratios $\rho$,
and distance measurement accuracies (\ref{final_ans1}) for these
randomly generated data points are shown in Figs.~\ref{scatter1} --
\ref{scatter3}.  We used the LIGO/VIRGO network functions shown in
Figs.~\ref{ligovirgo1} and \ref{ligovirgo2}.  Figures \ref{scatter1}
-- \ref{scatter3} give some idea of the potential capability of the
LIGO/VIRGO network.  The distance scale in these graphs
is determined by the detector sensitivity level (\ref{snf}) that we
have assumed, which is uncertain to within a factor of $\sim 2$.  The
distance scale would also be $2$ to $3$ times larger for NS-BH
binaries, as mentioned above.  By contrast, the distribution of
measurement accuracies, which we show in Fig.~\ref{histogram}, is
independent of the scale of the detector noise.  This figure shows
that the measurement accuracy will be better than $30\%$ for over half
of the detected sources.

A relatively large fraction, about $1/5$, of detected
events have poor ($\ge 50\%$) measurement accuracies.  This is
primarily due to the effect discussed in Sec.~\ref{draza_approx}: low
values of the detector network polarization sensitivity $1 -
\varepsilon_D({\bf n})$ over much of the sky.  The effect of the
polarization sensitivity can be clearly seen in Fig.~\ref{scatter4},
which is a scatter plot of polarization sensitivity versus distance
measurement accuracy.

Finally, we emphasize that our results should be regarded as fairly
rough estimates, because we have neglected the following effects: (i)
The spin-related modulation of the amplitudes ${\cal A}_A$ mentioned
in in Sec.~\ref{dist_overview} and discussed in
Ref.~\cite{spins_paper1}; (ii) the correlations between the variables
$D,v$ and $\psi,\phi_c$, except to linear order in $1/\rho$
\cite{caveat7}; and (iii) the correlations between the parameters
$D,v,\psi,\phi_c$ and the ``phase parameters'' ${\cal M}, \, \mu,\,
\beta$.  As discussed above, we show in Appendix \ref{decouple} that
these correlations vanish to linear order in $1/\rho$, but there will
be some correlation effects at higher order.  Despite these neglected
effects, we feel that the approximation method that we have used based
on Eqs.~(\ref{PDFvcalD}) and (\ref{PDFcalD}) gives results that are
more considerably more accurate than previous linear treatments [as
summarized by Eq.~(\ref{final_ans})], because the dominant
correlations at linear order in $1/\rho$ are those between $D$ and
$v$, and we have treated these correlations exactly.

\section{Conclusions}
\label{sec5}

Modulo the caveats in Sec.~\ref{caveats}, we have confirmed the
general conclusion that one can measure the binary's chirp mass ${\cal
M}$ with rather astonishing accuracy.  While our estimates of $\Delta
{\cal M}/{\cal M}$ are a factor of $\sim 20$ greater than those
obtained from the less accurate Newtonian analysis
\cite{finn2,krolak2}, we have found that $\Delta {\cal M}/{\cal M}$
should still be $0.01\% -1\%$ for typical measurements.

We have investigated the idea that detailed phase information might
also allow accurate determination of the binary's reduced mass $\mu$.
A calculation that neglected the effects of the bodies' spins on the
waveform suggested that $\mu$ might typically be measured to within
$\sim 1 \%$.  However a more complete analysis showed that errors in
$\mu$ can be substantially masked by compensating errors in the spin
parameter $\beta$.  Including the correlations with $\beta$, we
estimated that $\Delta \mu/\mu \approx 10\%$ for low-mass (NS-NS)
binaries and that $\Delta \mu/\mu \approx 50\%$ for high-mass (BH-BH)
binaries.  Moreover, $\Delta M_1/M_1$ and $\Delta M_2/M_2$ are
generally much greater than $\Delta \mu/\mu$ unless $M_1/M_2 \gg 1$
(BH-NS case).

These results are somewhat disappointing; it would have been more
exciting to find that post-Newtonian effects allow both masses to be
determined to within a few percent. In this regard, however, it is
useful to keep two points in mind.  First, since typical measurements
will have $S/N \approx 10$, one should detect events with $S/N \ge 50$
roughly $1 \%$ of the time.  For the advanced-detector noise curve
(\ref{snf}), and assuming the coalescence rates estimated in
Ref.~\cite{sterl}, such strong events should be seen $\sim$ once per
year for NS-NS binaries, and several times per year for NS-BH and
BH-BH binaries
\cite{flanagan_thorne}.  For these strongest sources, measurement
errors will be a factor of $\sim 5$ lower than their typical values.
Second, the measurement-derived PDF on the parameter space constrains
the values of $M_1$, $M_2$, and $\beta$ much more strongly than is
indicated by their individual variances, as illustrated in
Fig.~\ref{mubeta} above.  The large rms errors are due to correlations
between the measured parameters; certain linear combinations of the
parameters (eigenvectors of the covariance matrix) are determined with
high accuracy \cite{correlation_effects}.  This may be useful when
combined with information obtained by other means.

With regard to potential accuracy of distance measurements, our key
conclusions are the following:

(i) We have confirmed the general conclusion reached previously
\cite{krolak2,draza} that correlations between the distance $D$ and
other angular variables (primarily the angle of inclination of the
binaries orbit) will reduce $\Delta D$ by a factor of typically $2$ or
$3$ from the naive estimate $\Delta D / D = 1 /
(\mbox{signal-to-noise})$; see Fig.~\ref{scatter3} above.

(ii) Distance measurement accuracy will depend strongly on the
direction towards the source relative to the detectors, as shown in
Figs.~\ref{upmax} and \ref{upmin}.  This is because of the different
{\it polarization sensitivities} of the detector network in different
directions, and the fact that the complex amplitudes of both
polarization components of the incident waves are needed in order to
determine the distance.  The polarization sensitivity of the
LIGO/VIRGO network is somewhat poor in this regard (because the two
LIGO detectors are almost parallel); it would be substantially
improved by the addition of a fourth detector.  This provides
additional motivation for the construction of additional
interferometers around the world, which would also improve the angular
accuracy of sky-location measurements \cite{tinto}.

(iii) Previous estimates of distance measurement errors have been
accurate only to linear order in $1/D$.  Our results indicate that
this linear approximation will be inadequate for typical detected
signals, so that the incorporation of non-linear effects will be
necessary in order to accurately ascertain measurement errors [and
also to accurately estimate the distances themselves; see
Sec.~\ref{how_filter} of Appendix \ref{stats}].

(iv) We have carried out a Monte-Carlo simulation of distance
measurement accuracies for a large number of randomly chosen sources,
using a method of calculation which roughly estimates the non-linear
effects, and incorporating the amplitude sensitivity and polarization
sensitivity of the LIGO/VIRGO detector network.  Our results suggest
that $\sim 8\%$ of measured distances will be accurate to better than
$\sim 15\%$, and that $\sim 60\%$ of them will be accurate to better
than $30 \%$.

\nonum
\section{ACKNOWLEDGMENTS}

It is a pleasure to thank the following people for helpful
discussions: Theocharis Apostolatos, Lars Bildsten, Sam Finn, Daniel
Kennefick, Andrzej Krolak, Dragoljub Markovi\'{c}, Eric Poisson,
Bernard Schutz, Gerald Sussman, Cliff Will, Alan Wiseman, and
especially Kip Thorne, who provided much of the inspiration for this
paper.  We are grateful to Finn for sharing with us the results of his
research, on which much of our work is based, at a preliminary stage.
We thank Thorne and Bildsten for carefully reading the manuscript, and
for detailed comments.  Our understanding of some subtle issues in
parameter estimation owes much to a recent review article by Tom
Loredo \cite{loredo}.  This work was supported in part by NSF grant
PHY-9213508.

\appendix{Estimation \\
   of Signal Parameters}
\label{stats}

In this appendix we review some aspects of the statistical theory of
estimation of signal parameters as applied to gravitational wave
astronomy.  This subject has been concisely summarized in Appendix A
of Ref.~\cite{krolak1}, and has recently been treated in detail by
Finn \cite{finn1}.  Hence in many places we merely write down the key
results, without proof, in order to establish notation and equations
for use in the text.  However, we also present some extensions to the
formalism developed by Finn \cite{finn1}: We carefully distinguish
between Bayesian and frequentist estimates of errors, and discuss the
validity of these two methods of error calculation.  We show that
maximum-likelihood parameter estimation, while useful, is not the
optimal data-processing strategy, and, following Davis \cite{Davis},
suggest the use of the so-called Bayes estimator.  We derive an
expression for the minimum signal-to-noise ratio $(S/N)_{\rm min}$
necessary in order that the usual Gaussian approximation for
estimation of measurement accuracy be valid, and explain how to treat
degenerate points in parameter space at which the Gaussian
approximation breaks down.  Finally we give a discussion of the
effects of including {\it a priori} information, which corrects the
corresponding material in Ref.~\cite{finn1}.

\subsection{Basic Formulae}

The output of a network of detectors can be represented as a vector
${\bf s}(t) = (s_1(t), \ldots , s_{n_d}(t))$, where $n_d$ is the
number of detectors, and $s_a(t)$ is the strain amplitude read out
from the $a$th detector.  There will be two contributions to the
detector output ${\bf s}(t)$ --- the intrinsic detector noise ${\bf
n}(t)$ (a vector random process), and the true gravitational wave
signal ${\bf h}(t)$ (if present):
\begin{equation}
{\bf s}(t) = {\bf h}(t) + {\bf n}(t).
\end{equation}
We assume that the signal is a burst of known form, but depending on
several unknown parameters ${\bf \theta} = (\theta^1, \ldots
,\theta^k)$, so that ${\bf h}(t) = {\bf h}(t;{\bf \theta})$. Thus, we
do not consider the cases of periodic or stochastic waves
\cite{300years}.  We also assume for simplicity that the detector
noise is stationary and Gaussian.  For the LIGO and VIRGO detectors,
the stationarity assumption is justified for the analysis of short,
burst waves
\cite{krolak1}.  However, the actual noise may have important
non-Gaussian components, the implications of which for the purposes of
signal detection thresholds and data analysis are not yet fully
understood.  We do not deal with this issue here.

With these assumptions, the statistical properties of the detector
noises can be described by the auto-correlation matrix
\FL
\begin{eqnarray}
\label{c_n_def}
C_n(\tau)_{ab} & = & \langle s_a(t + \tau) s_b(t) \rangle -
\langle s_a(t + \tau) \rangle \, \langle s_b(t) \rangle \nonumber \\
\mbox{} & = & \langle n_a(t + \tau) n_b(t) \rangle -
\langle n_a(t + \tau) \rangle \, \langle n_b(t) \rangle,
\end{eqnarray}
where the angular brackets mean an ensemble average or a time average.
The Fourier transform of the correlation matrix, multiplied by two, is
the power spectral density matrix:
\begin{equation}
\label{s_h_def}
S_n(f)_{ab} = 2 \int_{-\infty}^{\infty} d \tau \, e^{2 \pi i f \tau}
C_n(\tau)_{ab}.
\end{equation}
This satisfies the formal equation
\begin{equation}
\label{p1}
\langle {\tilde n}_a(f) \, {\tilde n}_b(f^\prime)^* \rangle = {1 \over 2}
\delta(f - f^\prime) S_n(f)_{ab},
\end{equation}
or more generally and precisely
\FL
\begin{equation}
\left< \exp \left\{ i \int dt \,{\bf w}(t) \,\cdot {\bf n}(t) \right\}
\right> =  \exp \left\{ - {1 \over 2} \int_0^\infty d f \, {\tilde
{\bf w}}^\dagger \cdot {\bf S}_n \cdot {\tilde {\bf w}} \right\},
\end{equation}
for any sufficiently well-behaved test functions $w_a(t)$.  Here
tildes denote Fourier transforms, according to the convention that $$
{\tilde h}(f) = \int_{-\infty}^{\infty} e^{2 \pi i f t} h(t) dt.  $$
We note that there are two different commonly used definitions of
power spectral density in the literature.  The above convention is
used in Refs.~\cite{ligoscience,300years,finn1,krolak,Meers_all}.  The
alternative convention is to use a spectral noise density defined by
$S^{(2)}_n(f) \equiv S_n(f)/2$, as used in
Refs.~\cite{WZ,krolak1,DS,Helstrom,Schutz2}.

The Gaussian random process ${\bf n}(t)$ determines a natural inner
product $\left( \ldots | \ldots \right)$ and associated distance or
norm on the space of functions ${\bf h}(t)$.  As discussed in
Sec.~\ref{stats0}, this is defined so that the probability that the
noise takes a specific value ${\bf n}_0(t)$ is
\begin{equation}
\label{noise_distribution}
p[{\bf n} = {\bf n}_0] \, \propto \, e^{- \left( {\bf n}_0 | {\bf n}_0
\right) / 2 },
\end{equation}
and it is given by \cite{caveat}
\FL
\begin{equation}
\label{product_def}
\left( {\bf g} \, | \, {\bf h} \right) \equiv 4 \, \Re \int_0^\infty df
\,\, {\tilde g}_a(f)^* \left[ {\bf S}_n(f)^{-1} \right]^{ab}
{\tilde h}_b(f),
\end{equation}
where $\Re$ means ``the real part of''.  It also satisfies the
equation \cite{finn1}
\begin{equation}
\label{identity}
\langle \, \left( {\bf n} | {\bf g} \right) \, \left( {\bf n} | {\bf
h} \right) \, \rangle \, = \, \left( {\bf g} | {\bf h} \right),
\end{equation}
for any functions ${\bf g}$ and ${\bf h}$.

In this paper we are interested only in the estimation of signal
parameters once a gravitational wave burst has been detected.  Thus,
we suppose that we have measured some detector output ${\bf s}(t)$,
and that it satisfies the appropriate criterion for us to conclude
that it contains a signal of the form ${\bf h}(t;{\tilde {\bf
\theta}})$ for some unknown set of parameters ${\tilde {\bf \theta}}$:
\begin{equation}
\label{signal_present}
{\bf s}(t) = {\bf h}(t;{\tilde {\bf \theta}}) + {\bf n}(t).
\end{equation}
The central quantity of interest is then the probability distribution
function (PDF) for ${\tilde {\bf \theta}}$ given the output ${\bf
s}(t)$.  As Finn has shown \cite{finn1}, this is given by
\begin{equation}
\label{pdf1}
p[{\tilde {\bf \theta}} \, | \, {\bf s},\, \mbox{detection} ] = {\cal
N} \, p^{(0)} ({\tilde {\bf \theta}})
\,\, e^{ - {1 \over 2} \big( {\bf h}({\tilde {\bf \theta}}) - {\bf s}
\, \big| \, {\bf h}({\tilde {\bf \theta}}) -  {\bf s} \big) }.
\end{equation}
Here ${\cal N} = {\cal N}({\bf s})$ is a normalization constant, and
$p^{(0)}({\tilde {\bf \theta}})$ is the PDF that represents our {\it a
priori} knowledge.

\subsection{Two types of measurement accuracy}
\label{twoerrors}

We now discuss how to characterize the accuracy of measurement of the
parameters ${\bf \theta}$.  Normally statistical ``one sigma''
experimental errors are defined operationally in terms of the average
of the actual errors over an ensemble of repeated identical
measurements (which corresponds mathematically to the width of an
appropriate PDF).  Now in practice one cannot repeat or duplicate a
given gravitational wave measurement, but in principle one can do so
by waiting a sufficiently long time and throwing away all detected
signals that do not match the original one.  In this manner one can
operationally define an ensemble of ``identical'' measurements.  The
notion of error which results then depends in a crucial way on what is
meant by ``identical.''  One can either demand that the signals ${\bf
h}(t)$ incident on the detectors be identical and consider the
resulting spread in the values of the detector outputs ${\bf s}(t)$
given by Eq.~(\ref{signal_present}), or demand that the detector
outputs ${\bf s}(t)$ be identical and consider the resulting spread in
the values of the incident signals ${\bf h}(t)$.  The two notions of
error which result can be called Bayesian errors and frequentist
errors, adopting the terminology from common usage in a more general
context \cite{loredo}.  We now discuss in more detail the definition
and meaning of these two types of error, in order to clarify the
relationship between our method of calculating measurement error and
previous work in this area
\cite{finn1,krolak1,krolak2,krolak3}.  The following discussion
is based on that of Loredo \cite{loredo}.

In the frequentist approach, one first specifies the algorithm the
experimenters should use to determine the ``best-fit'' values ${\hat
\theta}$ of the parameters $\theta$ from the gravitational wave
measurement ${\bf s}$:
\begin{equation}
\label{estimator}
{\hat {\bf \theta}} = {\hat {\bf \theta}}( {\bf s}).
\end{equation}
This is also called a statistic or estimator.  Next, one assumes that
Eq.~(\ref{signal_present}) holds for some value of $\tilde \theta$,
and by substituting this equation into Eq.~(\ref{estimator}), and
using Eq.~(\ref{noise_distribution}), one derives the PDF $p({\hat
{\bf \theta}} \, | {\tilde {\bf \theta}})$ for ${\hat \theta}$ given
${\tilde \theta}$.  Then the expected value with respect to this PDF
of ${\hat \theta}^i - {\tilde \theta}^i$,
\begin{eqnarray}
\label{bias}
b^i & = &\langle {\hat \theta}^i \rangle - {\tilde \theta}^i,
\nonumber \\
\mbox{} &= & \int  {\hat \theta}^i p({\hat \theta} |
{\tilde \theta}) \, d {\hat \theta} \, - {\tilde \theta}^i,
\end{eqnarray}
gives the ``bias'' $b^i$ of the estimator ${\hat \theta}({\bf s})$.
The diagonal elements of the expected value of $({\hat \theta}^i -
{\tilde \theta}^i) \, ({\hat
\theta}^j - {\tilde \theta}^j)$ characterize the measurement error.
More specifically, we define
\FL
\begin{eqnarray}
\label{sigma_frequentist}
\Sigma_{\rm FREQ}^{ij} & = & \Sigma_{\rm FREQ}^{ij}[ \tilde \theta; {\hat
\theta}(\cdot)] \\
\nonumber
\mbox{} &= & \biggl< \left\{ {\hat \theta}^i[{\bf h}(\tilde \theta) +
{\bf n}] - \tilde \theta^i \right\} \,
\left\{ {\hat \theta}^j[{\bf h}(\tilde \theta) +
{\bf n}] - \tilde \theta^j \right\} \biggr>_{\bf n}. \nonumber
\end{eqnarray}
Here the notation on the first line indicates that ${\bf \Sigma}_{\rm
FREQ}$ depends on the functional form of the estimator ${\hat \theta}$
as well as the assumed signal parameters ${\tilde \theta}$, and the
angular brackets on the second line denote expectation value with
respect to the noise ${\bf n}$.  The matrix (\ref{sigma_frequentist})
is a measure of parameter-extraction accuracy that includes the effect
of the bias, since
\begin{equation}
\label{sigma_frequentist2}
\Sigma_{\rm FREQ}^{ij}  = \langle \delta {\hat
\theta}^i \, \delta {\hat \theta}^j \rangle + b^i b^j,
\end{equation}
where $\delta {\hat \theta}^i \equiv {\hat \theta}^i - \langle {\hat
\theta}^i \rangle$.

The physical meaning of the quantity (\ref{sigma_frequentist}) is the
following.  Suppose that a large number of identical gravitational
wave trains, described by the parameter values $\tilde \theta$,
impinge on the detector network.  For each measured signal, the
experimenters calculate using the algorithm ${\hat \theta}$ the
best-fit values of the source parameters.  Then the rms average
deviation of these best-fit values from the true value $\tilde \theta$
is given by Eq. (\ref{sigma_frequentist}).  Moreover, the usual method
of implementing a Monte-Carlo simulation of the measurement process
would also predict errors given by (\ref{sigma_frequentist})
\cite{krolak3}.

In the frequentist method, one focuses attention on a particular
incident signal ${\bf h}(t;\tilde \theta)$, and considers different
possible measured detector outputs ${\bf s}(t)$.  By contrast, in the
Bayesian approach, one focuses attention on a particular measured
detector output ${\bf s}$.  The error in measurement is simply taken
to be the width (variance-covariance matrix) of the PDF (\ref{pdf1})
for the true value ${\tilde \theta}$ of $\theta$ given the measurement
${\bf s}$.  Thus,
\FL
\begin{eqnarray}
\label{sigma_bayesian}
\Sigma_{\rm BAYES}^{ij} & = & \Sigma_{\rm BAYES}^{ij}[ {\bf s};
p^{(0)}(\cdot)] \\
\nonumber
\mbox{} &= & \int ({\tilde \theta}^i - \langle {\tilde \theta}^i
\rangle) \, ({\tilde \theta}^j - \langle {\tilde \theta}^j
\rangle) \, p({\tilde \theta} \, | \, {\bf s}) \, d {\tilde \theta},
\end{eqnarray}
where $\langle {\tilde \theta}^i \rangle = \int {\tilde \theta}^i\,
p({\tilde \theta} \, | \, {\bf s}) d {\tilde \theta}$.  Note that this
measure of error depends on different quantities than its frequentist
counterpart (\ref{sigma_frequentist}) --- the measured signal ${\bf
s}$, and the {\it a priori} PDF $p^{(0)}$.

The physical meaning of the quantity (\ref{sigma_bayesian}) is the
following.  Suppose that a large number of different gravitational
wave trains are incident upon the detector network, where the
distribution of the wave parameters ${\tilde \theta}$ is given by the
PDF $p^{(0)}$.  Only a small fraction of these will produce, at the
output of the detectors, the signal ${\bf s}(t)$.  In this small
fraction, however, there will be some spread of values of the
parameters ${\tilde \theta}$, because of different realizations of the
detector noise ${\bf n}(t)$ that combine with the incident waves to
produce the measured signal according to Eq.~(\ref{signal_present}).
This spread is characterized by the matrix (\ref{sigma_bayesian}).

The measure of error (\ref{sigma_bayesian}) characterizes the total
amount of information that is contained in the measured signal ${\bf
s}$, which is independent of how the experimenters choose to process
this signal.  In practical situations, however, one typically would
like to know what accuracy can be achieved by a given, imperfect,
data-processing algorithm (e.g., one which takes a manageable amount
of computer time).  It is possible to define a more general Bayesian
error that is appropriate for the situation where a particular
algorithm or statistic ${\hat \theta}(\cdot)$ is chosen to estimate
the signal parameters ${\theta}$ from the measured signal ${\bf s}$.
This measure of error is
\FL
\begin{eqnarray}
\label{sigma_bayesian1}
\Sigma_{\rm BAYES}^{ij} & = & \Sigma_{\rm BAYES}^{ij}[ {\bf s}, {\hat
\theta}({\bf s}) ; p^{(0)}(\cdot)] \\
\nonumber
\mbox{} &= & \int ({\tilde \theta}^i - {\hat \theta}({\bf s})) \,
({\tilde \theta}^j - {\hat \theta}({\bf s})) \, p({\tilde \theta} \, |
\, {\bf s}) d {\tilde \theta}.
\end{eqnarray}
Physically this quantity is just the (square of the) rms average, over
the small fraction of incident waves discussed above, of the
difference between the true value ${\tilde \theta}$ of the parameters
and the ``measured value'' ${\hat \theta}({\bf s})$.  It is clear that
the rms errors $\Sigma_{\rm BAYES}^{ii}[ {\bf s}, {\hat \theta}({\bf
s}) ; p^{(0)}(\cdot)]$ will be minimized and take on their minimum
values $\Sigma_{\rm BAYES}^{ii}[ {\bf s}; p^{(0)}(\cdot)]$ when one
chooses for ${\hat \theta}$ the so-called Bayes estimator \cite{Davis}
\begin{equation}
\label{Bayesestimator}
{\hat \theta}_{\rm BE}^i({\bf s}) \equiv \int {\tilde \theta}^i\,
p({\tilde \theta} \, | \, {\bf s}) d {\tilde \theta}.
\end{equation}

One final point about Bayesian errors is the following.  Suppose that
the experimenters calculate from the measured signal ${\bf s}$ the
best-fit value ${\hat \theta} = {\hat \theta}({\bf s})$, and then
discard all the remaining information contained in the signal ${\bf
s}$.  Then there are very many signals ${\bf s}^\prime$ that could
have been measured and that are compatible with the experimenters'
measurements, in the sense that ${\hat \theta}({\bf s}^\prime) = {\hat
\theta}$ \cite{finn1}.  Correspondingly, there is a larger spread of
possible values of ${\tilde \theta}$, and hence the predicted rms
measurement errors based on the measurement ${\hat \theta}$ alone are
given by the following modification of Eq.~(\ref{sigma_bayesian1}):
\FL
\begin{eqnarray}
\label{sigma_bayesian2}
\Sigma_{\rm BAYES}^{ij} & = & \Sigma_{\rm BAYES}^{ij}[ {\hat \theta};
p^{(0)}(\cdot)] \\
\nonumber
\mbox{} &= & \int ({\tilde \theta}^i - {\hat \theta}) \,
({\tilde \theta}^j - {\hat \theta}) \, p({\tilde \theta} \, |
\, {\hat \theta}) d {\tilde \theta}.
\end{eqnarray}
Here $p({\tilde \theta} \, | \, {\hat \theta})$ is the probability
distribution introduced by Finn \cite{finn1} for the true parameter
values ${\tilde \theta}$ given the estimated values ${\hat \theta}$.
It is given by the standard Bayesian formula
\begin{equation}
\label{tildegivenhat}
p({\tilde \theta} \, | \, {\hat \theta}) = {\bar {\cal N}}\,
p^{(0)}({\tilde \theta}) \, p({\hat \theta} \, | \, {\tilde \theta}),
\end{equation}
where ${\bar {\cal N}} = {\bar {\cal N}}({\hat \theta})$ is a
normalization constant that depends on ${\hat \theta}$.  Note that the
matrix (\ref{sigma_bayesian2}) depends only on the measured value
${\hat \theta}$ of the estimator and not on its functional form ${\hat
\theta}(\cdot)$.

The predicted measurement error (\ref{sigma_bayesian2}) differs from
the previously defined measurement error (\ref{sigma_bayesian1})
because the measured signal ${\bf s}$ contains information about the
likely size of the error, so that discarding ${\bf s}$ makes a
difference.  For example, suppose that a detector-output data train
contains a signal from a coalescing binary, and that by some standard
algorithm the experimenters determine best-fit values of the binaries
parameters.  Then given these best-fit values, one can estimate the
likely size of the measurement error --- this is given by
Eq.~(\ref{sigma_bayesian2}).  However, if they also determined that
the data train contains an uncommonly large non-Gaussian burst of
noise that accounts for $20 \%$ of the estimated signal amplitude, the
estimates of the likely parameter-extraction errors would clearly have
to be modified.

Which of the above-defined measurement errors is appropriate to assess
the capability of the LIGO/VIRGO detector network?  It is generally
accepted that, if one has a given measurement ${\bf s}$, the Bayesian
approach is the fundamental and correct one, and that the frequentist
approach is justified only to the extent that it reproduces the
results of Bayesian analyses.  This is essentially because, given a
particular measurement ${\bf s}$, it is irrelevant to consider an
ensemble of other, different measurements ${\bf s}^\prime$
\cite{loredo}.  However, for our purpose of trying to {\it anticipate} the
capability of gravitational wave detectors before any measurements are
available, it seems that this message looses its bite.  It certainly
seems reasonable to imagine a fixed gravitational wave-train incident
upon the detector network, and to inquire about the spread
(\ref{sigma_frequentist}) in measured values of the source parameters
due to differing realizations of the detector noise.

In fact, there is a certain sense in which frequentist errors and
Bayesian errors are equivalent, which is well known: the average of
the predicted frequentist error over the whole parameter space is the
same as a suitable average of the predicted Bayesian error.  Thus, in
a sense the same errors are being calculated in each case; it is just
their dependence on parameters that is being changed.  In particular,
if the predicted errors do not vary strongly with the parameters
${\tilde \theta}$, then the two types of error will be approximately
equal.  A precise statement of this ``equality of averages'' which is
straightforward to derive is
\begin{mathletters}
\label{sameaverages}
\begin{eqnarray}
\int & d{\tilde \theta} & p^{(0)}({\tilde \theta}) \, \Sigma_{\rm
FREQ}^{ij}[ \tilde \theta; {\hat \theta}(\cdot)] \nonumber \\
\label{sameaverage1}
\mbox{}  &  = &  \int {{\cal D} {\bf
s} \over {\cal N}({\bf s})} \, \Sigma_{\rm BAYES}^{ij}[ {\bf s}, {\hat
\theta}({\bf s}) ; p^{(0)}(\cdot)] \\
\label{sameaverage2}
\mbox{}  &  = &  \int {d  {\hat \theta } \over {\bar {\cal N}}({\hat
\theta})} \,  \Sigma_{\rm BAYES}^{ij}[ {\hat \theta} ; p^{(0)}(\cdot)].
\end{eqnarray}
\end{mathletters}
Here the various matrices $\Sigma^{ij}$ are defined in
Eqs.~(\ref{sigma_frequentist}), (\ref{sigma_bayesian1}), and
(\ref{sigma_bayesian2}), respectively, and the factors ${\cal N}({\bf
s})$ and ${\bar {\cal N}}({\hat \theta})$ are the normalization
constants appearing in Eqs.~(\ref{pdf1}) and (\ref{tildegivenhat}),
respectively.  The (formal) measure ${\cal D} {\bf s}$ is defined such
that
\begin{equation}
\langle F[{\bf n}] \rangle= \int {\cal D} {\bf n} \, F[{\bf n}] \, e^{-
\left( {\bf n} | {\bf n} \right) / 2 },
\end{equation}
for any functional $F[{\bf n}]$ of the noise ${\bf n}$.

Because of this equality of averages, we conclude that either Bayesian
or frequentist errors can be used to anticipate the capabilities of
LIGO/VIRGO, essentially because one is only interested in the range of
possible errors and not their value at a fixed point in parameter
space.  Similarly, if one is using Bayesian errors, it is appropriate
to use the matrix (\ref{sigma_bayesian2}) instead of
(\ref{sigma_bayesian1}) to anticipate measurement accuracies, since
from Eqs.~(\ref{sameaverages}) the measure of error
(\ref{sigma_bayesian2}) is simply an average of
(\ref{sigma_bayesian1}) over values of ${\bf s}$ for which ${\hat
\theta}({\bf s}) = {\hat \theta}$.  This conclusion has already been
reached in a recent paper of Finn's \cite{finn1} in which he advocates
the use of what in our notation is essentially $\Sigma_{\rm
BAYES}^{ij}[ {\hat \theta}_{\rm ML} ; p^{(0)}(\cdot)]$, where ${\hat
\theta}_{\rm ML}$ is the so-called maximum-likelihood estimator (see
below).  [However, his calculation of this quantity does not
incorporate the {\it a priori} PDF quite correctly, as we show below.]
Previous analyses of parameter-extraction accuracy for gravitational
wave detectors by Echeverria \cite{Fernando1} and by Krolak and
collaborators \cite{krolak2,krolak1,krolak,krolak3} have used the
frequentist error $\Sigma_{\rm FREQ}^{ij}[ \tilde \theta; {\hat
\theta}_{\rm ML}(\cdot)]$. By contrast, in Sec.~\ref{dist_errors} of
this paper we have calculated the Bayesian error
\FL
\begin{equation}
\label{Bayes_error}
\Sigma_{\rm
BAYES}^{ij}[ {\bf s}; p^{(0)}(\cdot)] = \Sigma_{\rm BAYES}^{ij}[ {\bf
s}, {\hat \theta}_{\rm BE}({\bf s}) ; p^{(0)}(\cdot)],
\end{equation}
because, as we argue below, it is more accurate to use ${\hat
\theta}_{\rm BE}$ rather that ${\hat \theta}_{\rm ML}$.

One final important point about the two types of error is the
following well-known fact: to leading order in $1/\rho$, where $\rho$
is the signal-to-noise ratio [Eq.~(\ref{network_snr}) below], the two
approaches yield identical results.  More specifically, assuming the
Gaussian noise statistics (\ref{noise_distribution}), we have
\FL
\begin{equation}
\label{samesig}
\Sigma_{\rm FREQ}^{ij}[{\tilde \theta}; {\hat \theta}_{\rm ML}(\cdot)]
= \Sigma_{\rm BAYES}^{ij}[{\hat \theta}_{\rm ML}; p^{(0)}(\cdot)] \
\times \left[1 + O(\rho^{-1}) \right]
\end{equation}
when ${\hat \theta}_{\rm ML} = {\tilde \theta}$. Moreover the same
quantity [given by Eq.~(\ref{sigma_def0}) below] is also obtained to
leading order using the Bayes estimator (\ref{Bayesestimator}), and
also from the estimator-independent measure of error
(\ref{sigma_bayesian}).  This is essentially because to this order,
all the PDFs are Gaussian.  These assertions are straightforward to
prove using the tools developed by Finn \cite{finn1}, and moreover are
well-known in more general statistical contexts.  Hence the
distinctions that we have been drawing only matter when effects that
are non-linear in $1/\rho$ contribute significantly to the predicted
accuracies (as for example when measuring distances to coalescing
binaries), or when $\rho$ is sufficiently small that the {\it a
priori} information represented by $p^{(0)}$ becomes significant.
[However, this may be the rule rather than the exception for typical
detected gravitational wave bursts; see Sec.~\ref{complex} above.]

\subsection{Choice of data-processing algorithm ${\hat \theta}(\cdot)$}
\label{how_filter}

Given a particular measurement ${\bf s}$, the PDF (\ref{pdf1}) in
principle contains all the information contained in the measurement
about the source parameters ${\tilde \theta}$.  However in practice
one often wants to focus on a small portion of this information, by
calculating a ``best estimate'' value ${\hat \theta}({\bf s})$
together with estimates of the statistical errors.  The choice of
estimator ${\hat \theta}(\cdot)$ is determined both by practical
considerations, and by whatever criteria are adopted to judge ``good''
estimators; there is no unique choice.

One obvious criterion is to choose that estimator which minimizes the
expected error in parameter extraction.  However, one could choose to
minimize either Bayesian or frequentist errors, and also the errors
depend on the parameter values (${\tilde \theta}$ or ${\hat \theta}$).
If one minimizes the average over parameter space of the measurement
error [as given by the common value of Eqs.~(\ref{sameaverages})],
then the resulting best estimator is just the Bayes estimator
(\ref{Bayesestimator}), which we have used in Sec.~\ref{complex}.  Its
use for gravitational wave data analysis has been suggested by Davis
\cite{Davis}.  Unfortunately, as Davis indicates, calculation of the
Bayes estimator is very computationally intensive, as it typically involves a
multidimensional integral of a function whose evaluation at each point
requires the numerical calculation of an inner product of the type
(\ref{product_def}).  Our application of the Bayes estimator in
Sec.~\ref{complex} was an exception in this regard, because all the
inner products could be evaluated analytically.
It seems likely that the Bayes estimator will be used only
after preliminary estimates of the signal parameters have been made
using Wiener optimal filtering.  The use of the Bayes estimator also
goes by the name of ``non-linear filtering'' \cite{Davis}.

A simpler estimator that has been proposed by Finn \cite{finn1,finn2},
Krolak \cite{krolak3} and others in the gravitational wave
data-analysis context is the so-called maximum-likelihood estimator
${\hat
\theta}_{\rm ML}({\bf s})$.  This defined to be the value of ${\tilde
\theta}$ which maximizes the PDF (\ref{pdf1}).  It is convenient
because it is closely related to the Wiener optimal filtering method
\cite{300years} that will be used to detect the signals --- the
detection procedure outlined in Sec.~\ref{intro} will essentially
return the maximum-likelihood estimates of the source parameters (see
below).  However, once the signals have been detected, there is no
reason to only use maximum-likelihood estimation --- other estimation
methods can be used to give better results.  Hence, the quantities
$\Sigma_{\rm BAYES}^{ij}[{\hat \theta}_{\rm ML};p^{(0)}(\cdot)]$ or
$\Sigma_{\rm FREQ}^{ij}[{\tilde \theta}; {\hat
\theta}_{\rm ML}(\cdot)]$ represent the {\it potential} accuracy of
measurements only to leading order in $1/\rho$.  (If
maximum-likelihood estimation is the only estimation method used, then
they represent the actual accuracy of measurement).  We note that the
quantities ${\hat
\theta}_{\rm ML}({\bf s})$ and ${\hat \theta}_{BE}({\bf s})$ can
differ by substantial factors for detected gravitational wave signals,
as for example in Fig.~\ref{prob} above where $\langle D
\rangle = 1.44 D_0$.

Maximum-likelihood estimation also has the following disadvantages.
First, as discussed in Sec.~\ref{complex}, the maximum-likelihood
estimator for a particular variable does not necessarily maximize the
reduced PDF for that variable obtained by integrating over the other
variables.  By contrast, the value of the Bayes estimator
(\ref{Bayesestimator}) for a given variable does not depend on whether
or not other variables have been integrated out.  Second, the best-fit
point obtained by the maximum-likelihood method depends on the choice
of variables used to parametrize the waveform ${\bf h}(t;{\tilde
\theta})$.  For example, in Sec.~\ref{sec3} of this paper we could
have used as variables either the individual masses $M_1$ and $M_2$ of
the binaries components, or the chirp and reduced masses ${\cal M}$
and $\mu$.  Since probability distributions for $(M_1,M_2)$ and
$({\cal M},\mu)$ are related by a non-trivial Jacobian factor, a local
maximum of one of them will not correspond to a local maximum of the
other.  A slightly different kind of maximum-likelihood estimator,
which maximizes the likelihood ratio $\Lambda({\tilde \theta}) \propto
\exp
\big[-\big( {\bf h}({\tilde {\bf \theta}}) - {\bf s} \, \big| \, {\bf
h}({\tilde {\bf \theta}}) - {\bf s} \big)/2\big] $ instead of the PDF
(\ref{pdf1}), does not suffer from this problem.  This is the
maximum-likelihood estimator that is usually discussed in the
statistics literature.  However, it does not take into account in any
way our {\it a priori} knowledge.

We conclude that calculations of measurement accuracy using ${\hat
\theta}_{\rm ML}$ represent the true potential measurement accuracy
only to leading order in $1/\rho$.  If this leading order
approximation becomes invalid (as occurs for sufficiently small
SNR's), then one should use instead either the Bayesian error
(\ref{Bayes_error}) or $\Sigma_{\rm BAYES}^{ij}[ {\hat \theta}_{\rm
BE} ; p^{(0)}(\cdot)]$.  One could also use $\Sigma_{\rm
FREQ}^{ij}[{\tilde \theta}; {\hat \theta}_{\rm BE}(\cdot)]$, but this
is much more difficult to calculate than (\ref{Bayes_error}) when the
large $\rho$ limit does not apply.

\subsection{Relation between maximum-likelihood estimation\\
and Wiener optimal filtering}

In Sec.~\ref{sec2}, we discussed a method for finding best-fit
parameters ${\hat {\bf \theta}}$ which was based on maximizing the
overlap of the measured signal with theoretical templates
[cf.~Eq.~(\ref{sn}) above].  We now briefly indicate the relationship
of this method to the maximum-likelihood procedure.  That the two
methods are equivalent in general has been shown by Echeverria
\cite{Fernando}.

Given the measured signal ${\bf s}(t)$, define for any ${\bf \theta}$
the quantity
\begin{equation}
\label{snr_def_1}
\rho[{\bf \theta}] = { \left( {\bf h}({\bf \theta}) \, | \, {\bf s} \right)
\over \sqrt{ \left( {\bf h}({\bf \theta}) \, | \, {\bf h}({\bf \theta})
\right) } }.
\end{equation}
This is the signal-to-noise ratio (SNR) defined in Eq.~(\ref{sn}), and
can be calculated by integrating the signal ${\bf s}(t)$ against a
Wiener optimal filter whose Fourier transform is proportional to ${\bf
S}_n(f)^{-1} \cdot {\tilde {\bf h}}(f ; {\bf
\theta})$.  The quantity $\rho[{\bf \theta}]$ is a random variable with
Gaussian PDF of unit variance.  Its expected value is zero if no
signal is present, when ${\bf s}(t) = {\bf n}(t)$.  If a signal is
present, so that Eq.~(\ref{signal_present}) holds for some ${\tilde
{\bf \theta}}$, then the expected value of $\rho[\theta]$ is
\begin{equation}
\langle \, \rho[{\bf \theta}] \, \rangle = { \big( {\bf h}({\bf \theta}) \,
\big| \, {\bf h}({\tilde {\bf \theta}}) \big)
\over \sqrt{ \left( {\bf h}({\bf \theta}) \, | \, {\bf h}({\bf \theta})
\right) } }.
\end{equation}
Now if the {\it a priori} probability $p^{(0)}({\tilde \theta})$ can
be approximated to be constant, then the value ${\hat {\bf
\theta}}_{\rm ML}$ of ${\tilde {\bf \theta}}$ which maximizes the PDF
(\ref{pdf1}) for a given signal ${\bf s}$ also maximizes $\rho[{\bf
\theta}]$
\cite{Fernando}.  Hence we can find ${\hat {\bf \theta}}_{\rm ML}$ (up to the
overall amplitude of the signal \cite{caveat2}) by computing the
overlap (\ref{snr_def_1}) of the signal with various templates, and by
choosing the template which gives the maximum overlap.

When a signal is present and the signal-to-noise ratio is large, the
maximum value $\rho[{\hat {\bf \theta}}_{\rm ML}]$ of $\rho[{\bf
\theta}]$ will approximately given by
\FL
\begin{equation}
\rho[{\hat {\bf \theta}}_{\rm ML}]^2 \approx \big( {\bf h}({\tilde {\bf
\theta}}) \, \big| \, {\bf h}({\tilde {\bf \theta}}) \big) \,\,\,
\approx \big( {\bf h}({\hat {\bf \theta}}_{\rm ML}) \, \big| \, {\bf
h}({\hat {\bf \theta}}_{\rm ML}) \big).
\end{equation}
The quantity
\begin{equation}
\label{network_snr}
\rho^2 = \big( {\bf h} ({\tilde {\bf \theta}}) \, \big| \, {\bf h}
({\tilde {\bf \theta}}) \big)
\end{equation}
is what is usually referred to as the (square of) the SNR of the
signal ${\bf h}(t;{\tilde {\bf \theta}})$.  When correlated sources of
noise are unimportant so that the matrix (\ref{s_h_def}) is diagonal,
this overall SNR will be given by combining in quadrature the SNR's
for each individual detector, cf.~Eq.~(\ref{rho}) above.

\subsection{The Gaussian approximation\\
and conditions for its validity}
\label{Gaussian_section}

We now consider the high signal-to-noise limit in which many of the
subtleties that we have been discussing become unimportant.  In
particular, in this limit the Bayes and maximum-likelihood estimators
become identical.  From Eq.~(\ref{pdf1}), the maximum-likelihood
estimator ${\hat {\bf
\theta}}_{\rm ML}$ satisfies
\FL
\begin{equation}
\label{bestfit}
\big( {\bf h}_{,i}({\hat {\bf \theta}}_{\rm ML}) \, \big| \, {\bf
h}({\hat {\bf \theta}}_{\rm ML}) - {\bf s} \big) - [\ln
p^{(0)}]_{,i}({\hat {\bf \theta}}_{\rm ML}) = 0,
\end{equation}
where the subscript $,i$ means derivative with respect to $\theta^i$
for $1 \le i \le k$, and $k$ is the number of parameters.  If the {\it
a priori} information is unimportant so that the last term in
Eq.~(\ref{bestfit}) is negligible, then as outlined in Sec.~\ref{sec3}
the following simple geometric interpretation applies: Let ${\cal S}$
be the finite dimensional surface formed by the set of all signals
${\bf h}(t;{\bf \theta})$ in the space of all possible signals ${\bf
h}(t)$.  Then the measured signal ${\bf s}(t)$ will generally not lie
on the surface ${\cal S}$, and the best-fit point ${\bf h}(t;{\hat
{\bf
\theta}}_{\rm ML})$ is just that point on ${\cal S}$ that is closest to ${\bf
s}(t)$, where distance is measured using the norm $||{\bf f}||^2
\equiv \left( {\bf f} \, | \, {\bf f} \right)$
derived from the inner product (\ref{product_def}).  Correspondingly,
${\bf h}({\hat {\bf \theta}}_{\rm ML})$ can be obtained by just
dropping a perpendicular from ${\bf s}(t)$ onto the surface ${\cal
S}$, which is the content of Eq.~(\ref{bestfit}) and is illustrated in
Fig.~\ref{hypersurface}.

When the SNR $\rho$ is sufficiently large, one can find approximate
expressions for $\Sigma_{\rm FREQ}[{\tilde \theta}; {\hat \theta}_{\rm
ML}(\cdot)]$ and $\Sigma_{\rm BAYES}[{\hat \theta}_{\rm ML};
p^{(0)}(\cdot)]$.  Such a calculation has been carried out by Finn
\cite{finn1}.  We now briefly outline the calculation, and also extend
it to determine the next to leading order terms in an expansion in
$1/\rho$, in order to determine how large $\rho$ needs to be for the
leading order term to be a good approximation.  Throughout this
subsection we assume that the {\it a priori} PDF $p^{(0)}$ is
approximately constant; in subsection \ref{apriori} below we consider
the effects of non-constant $p^{(0)}$.

First we find an approximate solution to Eq.~(\ref{bestfit}).
Abbreviating ${\hat \theta}_{\rm ML}$ as ${\hat \theta}$, inserting
Eq.~(\ref{signal_present}) into Eq.~(\ref{bestfit}) and expanding in
the difference $\delta {\bf \theta} = {\hat {\bf \theta}} - {\tilde
{\bf \theta}}$, we obtain
\FL
\begin{equation}
\label{muhat_linear}
{\hat \theta}^i = {\tilde \theta}^i + \delta^{(1)} \theta^i
+ \delta^{(2)} \theta^i + \delta^{(3)} \theta^i + O({\bf n}^4).
\end{equation}
Here
\begin{equation}
\delta^{(1)} \theta^i = \left( {\bf \Gamma}({\tilde
\theta})^{-1} \right)^{ij} \, \left({\bf n} \, | \, {\bf h}_{,j}
\right),
\end{equation}
where
\begin{equation}
\label{gamma_def}
\Gamma({\tilde \theta})_{ij} \equiv \big( {\bf h}_{,i}({\tilde {\bf
\theta}}) \, \big| \, {\bf h}_{,j}({\tilde {\bf \theta}}) \big)
\end{equation}
is the so-called Fisher information matrix [cf.~Eq.~(\ref{sig})
above].  The second-order term $\delta^{(2)} \theta$ is
\FL
\begin{eqnarray}
\delta^{(2)} \theta^i & = & \left( {\bf n} \, | \, {\bf h}^i_{\,\,\,j}
\right) \, \left( {\bf n} \, | \, {\bf h}^j \right) \\
 &- & \mbox{}\left[ \left( {\bf h}^i_{\,\,\,j} \, | \, {\bf h}_k
\right) + {1 \over 2} \left( {\bf h}^i \, | \, {\bf h}_{jk} \right)
\right] \,
\left( {\bf n} \, | \, {\bf h}^j \right) \, \left( {\bf n} \, | \,
{\bf h}^k \right). \nonumber
\end{eqnarray}
In this expression and below we have for brevity omitted the commas
denoting derivatives, and all quantities are evaluated at ${\tilde
{\bf \theta}}$.  We lower and raise indices with the tensor
(\ref{gamma_def}) and its inverse, so that, for example,
\begin{equation}
{\bf h}^i_{\,\,\,j} \equiv \left( {\bf \Gamma}^{-1} \right)^{ik} {\bf
h}_{,kj}.
\end{equation}
There is a similar but more complex expression for $\delta^{(3)}
\theta^i$.

Equations (\ref{muhat_linear}) and (\ref{noise_distribution}) now
determine the PDF $p({\hat \theta} | {\tilde \theta})$.  Using
Eq.~(\ref{identity}) and its extension to fourth order moments, and
Eqs.~(\ref{muhat_linear}) and (\ref{sigma_frequentist}), we obtain
\begin{equation}
\label{sigmaf_highSNR}
\Sigma_{\rm FREQ}^{ij}[{\tilde \theta};{\hat \theta}_{\rm ML}(\cdot)]
= (\Gamma^{-1})^{ij} + {}^{(2)}\Sigma^{ij}.
\end{equation}
At leading order, $p({\hat \theta} | {\tilde \theta})$ is a
multivariate Gaussian with mean ${\tilde \theta}$ and
variance-covariance matrix proportional to $1/\rho^2$ given by the
first term in Eq.~(\ref{sigmaf_highSNR}):
\begin{equation}
\label{sigma_def0}
{\bf \Sigma} = {\bf \Gamma}^{-1},
\end{equation}
cf.~Eq.~(\ref{gauss}) above.  The correction term $\propto 1/\rho^4$ in
Eq.~(\ref{sigmaf_highSNR}) is
\FL
\begin{eqnarray}
\label{firstcorrection}
{}^{(2)}\Sigma^{ij} & = &
\langle \delta^{(2)} \theta^i \delta^{(2)} \theta^j \rangle +
\langle \delta^{(1)} \theta^i \delta^{(3)} \theta^j \rangle +
\langle \delta^{(3)} \theta^i \delta^{(1)} \theta^j \rangle \nonumber \\
& = & \left( {\bf h}^i_{\,\,\,k} \, | \, {\bf
h}^{jk} \right) - \left( {\bf h}^i_{\,\,\,k} \, | \, {\bf h}_l \right)
\, \left( {\bf h}^{jk} \, | \, {\bf h}^l \right) \nonumber \\ &+ & {1
\over 4} \left( {\bf h}^i \, | \, {\bf h}^k_{\,\,\,k}
\right) \, \left( {\bf h}^j \, | \, {\bf h}^l_{\,\,\,l} \right)
+ {1 \over 2} \left( {\bf h}^i \, | \, {\bf h}_{kl} \right) \,
\left( {\bf h}^j \, | \, {\bf h}^{kl} \right) \nonumber \\
& - & \left( {\bf h}^i \, | \, {\bf
h}^{jk}_{\,\,\,\,\,\,k} \right) - \left( {\bf h}^{ij} \, | \, {\bf
h}^k_{\,\,\,k} \right)
+ \left( {\bf h}^{ij} \, | \, {\bf h}_k \right) \, \left( {\bf h}^k \,
| \, {\bf h}^l_{\,\,\,l} \right) \nonumber \\
 & + & \left( {\bf h}^i \, | \, {\bf
h}^{jk} \right) \, \left( {\bf h}_k \, | \, {\bf
h}^l_{\,\,\,l} \right)
+ 2 \left( {\bf h}^i \, | \, {\bf h}_{kl} \right) \, \left( {\bf h}^k \,
| \, {\bf h}^{jl} \right).
\end{eqnarray}
In the case where there is only one variable so that ${\bf \theta} = (
\theta^1, \ldots , \theta^k) = (\theta^1)$, it follows from
Eqs.~(\ref{sigmaf_highSNR}) and (\ref{firstcorrection}) that
\FL
\begin{equation}
\Sigma_{\rm FREQ}^{11} = {1 \over \left( {\bf h}^\prime \, | \, {\bf h}^\prime
\right) } \left[ 1 +
 {15 \over 4} { \left( {\bf h}^{\prime \prime}
\, | \,{\bf h}^\prime \right)^2 \over \left( {\bf h}^\prime \, | \,
{\bf h}^\prime \right)^3 }
- { \left( {\bf h}^\prime \, | \,{\bf
h}^{\prime \prime \prime} \right) \over \left( {\bf h}^\prime \, | \, {\bf
h}^\prime \right)^2 } \right],
\end{equation}
where primes denote derivatives with respect to $\theta^1$.  The
correction terms in the square brackets in this expression will be
small whenever
\begin{equation}
\label{linearcondition}
|| {\bf h}^{\prime \prime} || \ll || {\bf h}^\prime ||^2,
\end{equation}
and
\begin{equation}
\label{linearconditione}
|| {\bf h}^{\prime \prime \prime} || \ll || {\bf h}^\prime ||^3.
\end{equation}
Using Eq.~(\ref{product_def}), the equation $\rho^2 =
\left( {\bf h} \, | \, {\bf h} \right)$, and assuming for simplicity
that ${\bf S}_n(f) = S_n(f) {\bf 1}$, Eq.~(\ref{linearcondition})
reduces to the condition
\begin{equation}
\label{minSNR}
\rho^2 \ \gg \  {
\langle \langle \, ({\tilde {\bf h}}^{\prime\prime\,\dagger} \cdot
{\tilde {\bf h}}^{\prime\prime}) / ({\tilde {\bf h}}^{\dagger} \cdot
{\tilde {\bf h}}) \, \rangle \rangle
\over
\langle \langle \, ({\tilde {\bf h}}^{\prime\,\dagger} \cdot
{\tilde {\bf h}}^{\prime}) / ({\tilde {\bf h}}^{\dagger} \cdot {\tilde
{\bf h}}) \, \rangle \rangle^2 },
\end{equation}
where for any function of frequency $F(f)$, we define the weighted
average $\langle \langle F(f) \rangle \rangle$ to be $\left( F {\bf h}
\, | \, {\bf h} \right) / \left( {\bf h} \, | \, {\bf h} \right)$.

Equations (\ref{linearcondition}) and (\ref{linearconditione}) give sufficient
conditions for the Gaussian approximation to be valid, when there is
only one unknown parameter $\theta^1$.  When there are several unknown
parameters, a generalization of Eq.~(\ref{linearcondition}) is
obtained by interpreting the prime to mean the operator $v^i \partial
/ \partial \theta^i$ that differentiates in some direction $v^i$ in
the space of parameters ${\bf \theta}$, and requiring the condition to
hold for all directions $v^i$.  This yields the condition
\begin{equation}
\label{linearcondition1}
|| {\bf h}_{,ij} \, v^i \,v^j || \ll || {\bf h}_{,i} \,v^i ||^2 =
\Gamma_{ij}
\, v^i \, v^j,
\end{equation}
which is required to hold for all $v^i$.  We note that, although
Eq.~(\ref{sigmaf_highSNR}) does correctly indicate the regime
(\ref{minSNR}) where the Gaussian approximation is valid, the
correction term ${}^{(2)}\Sigma^{ij}$ is {\it not} an accurate
expression for the leading order correction to the measurement
accuracy, because as we have argued above the true potential
measurement accuracy is given by using the estimator ${\hat
\theta}_{\rm BE}(\cdot)$ and not ${\hat \theta}_{\rm ML}(\cdot)$.

One frequent source of confusion about the leading order expression
(\ref{sigma_def0}) for the measurement error is the following.  A
general theorem in statistics called the Cramer-Rao inequality
\cite{krolak1,Helstrom} states that for {\it any} unbiased estimator
${\hat \theta}$ \cite{Fisher_caveat},
\begin{equation}
\label{cramer_rao}
{\bf \Sigma}_{\rm FREQ}[{\tilde \theta}; {\hat \theta}(\cdot)] \ \ge \
{\bf \Gamma}({\tilde \theta})^{-1}.
\end{equation}
Hence, one might imagine that the quantity (\ref{sigma_def0}) is a
lower bound for the accuracy obtainable by most reasonable estimators,
and also for low signal-to-noise ratios.  That this is not the case
can be seen from the following consideration, which we discuss in the
body of the paper: at degenerate points ${\tilde \theta}_0$ for which
the signal derivatives $\partial {\bf h} / \partial \theta^i$ become
linearly dependent, the matrix (\ref{gamma_def}) becomes degenerate,
and the predicted rms errors given by the matrix (\ref{sigma_def0})
become infinite.  More careful calculations of, for example, ${\bf
\Sigma}_{\rm FREQ}[{\tilde \theta}; {\hat \theta}_{\rm ML}(\cdot)]$ at
such degenerate points, going beyond linear order, give finite
results.  Hence the inverse of the Fisher matrix is {\it not} a
generic, useful lower bound.  The reason that the Cramer-Rao
inequality does not apply is that most estimators are not unbiased and
cannot easily be made so.  When one generalizes the inequality
(\ref{cramer_rao}) to incorporate the effects of bias \cite{Helstrom},
an extra factor appears on the right-hand side multiplying the Fisher
matrix, which can be small.  This can allow ${\bf \Sigma}_{\rm
FREQ}[{\tilde \theta}; {\hat \theta}(\cdot)]$ to be much smaller than
the inverse of the Fisher matrix, for some statistics ${\hat \theta}$.

\subsection{Incorporation of {\it a priori} probabilities}
\label{apriori}

We now turn to the effects of {\it a priori} information.  First, we
remark that it is not necessary for {\it a priori} information to be
very detailed or restrictive in order that it have a significant
effect on parameter-extraction accuracy.  All that is necessary is
that it be more restrictive than the information contained in the
waveform, for some of the parameters $\theta^i$.  In other words it
will be important whenever the statistical error $\langle (\Delta
\theta^i)^2 \rangle$ which we obtain from Eq.~(\ref{sigma_def0}) for some
parameter $\theta^i$ is much larger than our {\it a priori}
constraints on $\theta^i$.  For example, this would be the case if we
obtained rms errors for measurements of the dimensionless spin
parameter $a$ of a black hole to be larger than one, since we expect
$|a| \le 1$ always.  If we include such poorly determined variables in
a calculation of the variance-covariance matrix ${\bf \Sigma}$ and
neglect the {\it a priori} restrictions, then the results obtained for
the rms error in $\theta^i$ may be severely overestimated.  This is
not unexpected; what is more surprising is that due to the effects of
correlations, the rms errors obtained for the other parameters may
also be overestimated by large factors (see, e.g., Sec.~\ref{spin_sec}
above).  We now extend the approximate calculations of the previous
subsection to incorporate {\it a priori} information, and also now
calculate Bayesian as well as frequentist errors.  Our results in this
subsection correct Eq.~(3.19) of Ref.~\cite{finn1}.

Roughly speaking, {\it a priori} information will be unimportant when
the PDF $p^{(0)}$ does not vary substantially within one or two sigma
of the best-fit point ${\hat \theta}$.  This condition is logically
independent of the condition (\ref{minSNR}), although both will be
satisfied in the high $\rho$ limit.  Hence, we can treat separately
deviations from the leading order measurement accuracy
(\ref{sigma_def0}) that are due to second-order derivatives ${\bf
h}_{,ij}$ of the signal [cf.~Eq.~\ref{firstcorrection} above], and
that are due to {\it a priori} information.  In the remainder of this
subsection we therefore assume the condition (\ref{minSNR}) and
consistently neglect all second-order derivatives ${\bf h}_{,ij}$.  In
particular we treat the Fisher matrix (\ref{gamma_def}) as a constant
in this approximation.  [Note that our results will be exact in the
case where the dependence of ${\bf h}({\tilde \theta})$ on the
parameters ${\tilde \theta}$ is exactly linear, as in
Sec.~\ref{complex} above.]

We start by considering the Bayes error (\ref{sigma_bayesian}).  When
we neglect second-order derivatives of ${\bf h}$ we find that the PDF
(\ref{pdf1}) takes the form
\begin{eqnarray}
\label{pdf2}
p[{\tilde \theta} \,& | & \, {\bf s},\, \mbox{detection} ] = {\cal
N}^\prime \, p^{(0)}({\tilde \theta})
\nonumber \\
\mbox{} \times & & \exp \left[ - {1 \over 2} \Gamma_{ij} ({\tilde
\theta}^i - s^i) \, ( {\tilde \theta}^j - s^j) \right].
\end{eqnarray}
Here we have decomposed the measured signal according to
\begin{equation}
{\bf s} = s^j {\bf h}_{,j} + {\bf s}^\perp,
\end{equation}
where $\left( {\bf h}_{,i} \, | \, {\bf s}^\perp \right) =0$ for $1
\le i \le k$, and have
absorbed a factor of $\exp \left[ - || {\bf s}^\perp ||^2 /2 \right]$
into the normalization constant ${\cal N}^\prime$.  If the PDF
$p^{(0)}$ simply restricts the allowed ranges of the parameters, then
the PDF (\ref{pdf2}) is a truncated Gaussian distribution whose
variance-covariance matrix $\Sigma_{\rm BAYES}[{\bf
s};p^{(0)}(\cdot)]$ will normally be within a factor of $\sim 2$ or so
of ${\bf \Gamma}^{-1}$.  If $p^{(0)}$ is approximately Gaussian with
variance-covariance matrix ${\bf \Sigma}_0$, then we see from
Eqs.~(\ref{sigma_bayesian}) and (\ref{pdf2}) that
\begin{equation}
\label{simple_answer}
{\bf \Sigma}_{\rm BAYES}[{\bf s};p^{(0)}(\cdot)] = \left\{ {\bf
\Gamma} + {\bf \Sigma}_0^{-1} \right\}^{-1}.
\end{equation}
This is the formula which we use in Sec.~\ref{spin_sec} above to
incorporate our {\it a priori} knowledge about the spin parameter
$\beta$.

Next we calculate an approximate expression for the second type of
Bayesian error given by Eq.~(\ref{sigma_bayesian2}), which is
appropriate for the situation where we discard all information about
the measured signal ${\bf s}$ except the best estimate values ${\hat
\theta}({\bf s})$ of the parameters.  For simplicity, we assume that
$p^{(0)}({\tilde {\bf \theta}})$ is a Gaussian with mean ${\bf
\theta}_0$ and width ${\bf \Sigma}_0$.  We also use the
maximum-likelihood estimator ${\hat \theta}_{\rm ML}$; however, the
same results are obtained for the Bayes estimator ${\hat
\theta}_{\rm BE}$.  From Eqs.~(\ref{bestfit}) and
(\ref{signal_present}) and neglecting second-order derivatives of
${\bf h}$, we find
\FL
\begin{equation}
\left( {\bf \Sigma}_1^{-1} \right)_{ij} \, ({\hat \theta}_{\rm ML}^j - {\tilde
\theta}^j) = \left( {\bf h}_{,i} \, | \, {\bf n} \right) +
\left( {\bf \Sigma}_0^{-1} \right)_{ij} \, ({\theta}_0^j - {\tilde
\theta}^j),
\end{equation}
where ${\bf \Sigma}_1^{-1} \equiv {\bf \Gamma} + {\bf \Sigma}_0^{-1}$.
Together with Eq.~(\ref{identity}) this implies that
\begin{equation}
\label{phat2}
p({\hat {\bf \theta}}_{\rm ML} | {\tilde {\bf \theta}}) \, \propto \,
\exp
\left[ - {1 \over 2} {\bf v}^{\rm T} \cdot {\bf \Gamma} \cdot {\bf v}
\right],
\end{equation}
where
\begin{eqnarray}
\label{vis1}
{\bf v} & = & {\hat {\bf \theta}} - {\bf \Sigma}_1 \cdot {\bf \Gamma}
\cdot {\tilde {\bf \theta}} - {\bf \Sigma}_1 \cdot {\bf \Sigma}_0^{-1}
\cdot \theta_0  \\
\label{vis2}
\mbox{} & = & {\bf \Sigma}_1 \cdot {\bf \Gamma} \cdot \left( {\tilde
{\bf \theta}} - \mbox{ const} \right).
\end{eqnarray}
Using Eqs.~(\ref{sigma_frequentist}), (\ref{phat2}), and (\ref{vis1})
we see that the result (\ref{sigmaf_highSNR}) becomes modified to read
\begin{equation}
\label{sigmaf_highSNR1}
{\bf \Sigma}_{\rm FREQ}[{\tilde \theta};{\hat \theta}_{\rm ML}(\cdot)]
= \Gamma^{-1} + {\bf b} \otimes {\bf b},
\end{equation}
where the bias ${\bf b} = {\bf \Sigma}_1 \cdot {\bf \Sigma}_0^{-1}
\cdot ( \theta_0 - {\tilde \theta} )$.  A more interesting
quantity is the Bayesian error (\ref{sigma_bayesian2}), which from
Eqs.~(\ref{tildegivenhat}) and (\ref{vis2}) is given by
\FL
\begin{eqnarray}
\label{hard_answer}
{\bf \Sigma}_{\rm BAYES}[{\hat \theta}_{\rm ML}; p^{(0)}(\cdot)]^{-1}
& = & {\bf \Sigma}_{\rm BAYES}[{\hat \theta}_{\rm BE};
p^{(0)}(\cdot)]^{-1} \\
\nonumber
\mbox{} & = & {\bf \Sigma}_0^{-1} + {\bf \Gamma} \cdot {\bf
\Sigma}_1 \cdot {\bf \Gamma} \cdot {\bf \Sigma}_1 \cdot {\bf \Gamma}. \nonumber
\end{eqnarray}
This expression gives approximately the same results as
Eq.~(\ref{simple_answer}), the differences never being more than $\sim
25 \%$.  The variances $\Sigma^{ii}$ given by Eq.~(\ref{hard_answer})
are always larger than those given by Eq.~(\ref{simple_answer}), as a
result of our having thrown away all the information in ${\bf s}$
apart from ${\hat {\bf \theta}}({\bf s})$.

The result (\ref{hard_answer}) disagrees with a corresponding analysis
of Finn [Eq.~(3.19) of Ref.~\cite{finn1}].  The reason for the
disagreement is that Finn solves Eq.~(\ref{bestfit}) to obtain
${\tilde {\bf
\theta}}$ as a function of ${\hat {\bf \theta}}_{\rm ML}$ and ${\bf
n}$, and then invokes the PDF (\ref{noise_distribution}) of the noise
to find $p({\tilde {\bf
\theta}} | {\hat {\bf \theta}}_{\rm ML})$.  This method of calculation
[analogous to the method used for calculating $p({\hat \theta}_{\rm
ML} | {\tilde \theta})$] is invalid because it implicitly assumes that
\begin{equation}
\label{not_true}
p[{\bf n} = {\bf n}_0 \, | \, {\hat {\bf \theta}}_{\rm ML}] \, = \,
p[{\bf n} = {\bf n}_0],
\end{equation}
which is not the case.  The fact that Eq.~(\ref{not_true}) does not
hold can be seen from the joint PDF for ${\tilde {\bf \theta}}$,
${\hat {\bf \theta}}_{\rm ML}$ and ${\bf n}$, which is
\FL
\begin{eqnarray}
\label{joint_pdf}
p[{\hat {\bf \theta}}_{\rm ML},{\tilde {\bf \theta}},{\bf n}] \,&
\propto &\, p^{(0)}({\tilde {\bf \theta}}) \, e^{- \left( {\bf n} \, |
\, {\bf n} \right) /2} \nonumber \\
\mbox{} & & \times \delta( {\hat {\bf
\theta}}_{\rm ML} - {\hat {\bf \theta}}_{\rm ML}[{\bf h}({\tilde
\theta}) + {\bf n}]).
\end{eqnarray}

\subsection{Treatment of degenerate variables}
\label{degenerate}

As we have noted in Sec.~\ref{complex}, the accuracy of the linear
approximation (\ref{linearcondition1}) which yields the simple PDF
(\ref{pdf2}), depends in part on what set of variables $\theta^i$ are
used in the calculation.  A different PDF will be obtained from this
approximation if one first makes a non-linear change of co-ordinates
$\theta^i \to {\bar \theta}^i(\theta^j)$.  Hence, the PDF (\ref{pdf2})
will approximate most closely the true PDF when it is computed using
variables for which ${\bf h}_{,ij}$ is as small as possible.

Consequently, there are two qualitatively different ways in which the
linear approximation may break down.  First, for sufficiently low
signal-to-noise ratios, the extrinsic curvature of the surface ${\cal
S}$ formed by the set of waveforms ${\bf h}(t;\theta)$ may be
sufficiently large that Eq.~(\ref{linearcondition1}) is not a good
approximation for {\it any} set of coordinates $\theta^i$.  In this
case the ``Gaussian'' method breaks down completely.  Second, the
approximation may break down simply because of a bad choice of
coordinates.  This is usually the case at points of degeneracy where
the vectors $\partial {\bf h} / \partial \theta^i$ become linearly
dependent, which we discuss in Secs.~\ref{sec3} and
\ref{dist_errors} and at the end of Sec.~\ref{Gaussian_section} above.
At such points the straightforward linear approximation method breaks
down, but frequently one can find a non-linear nonlinear change of
variables of the form
\begin{equation}
\label{nonlinearchange}
{\bar \theta}^i = {\bar \theta}^i(\theta^j)
\end{equation}
such that the vectors $\partial {\bf h} / \partial {\bar \theta}^i$
are not linearly dependent.  One then obtains from from the linear
approximation a Gaussian PDF in the variables ${\bar \theta}^i$.
Substituting the relation (\ref{nonlinearchange}) into this PDF and
multiplying by the appropriate Jacobian factor gives a non-Gaussian
PDF in terms of the variables $\theta^i$ (as in Sec.~\ref{complex}
above).  From this PDF, measurement errors for the variables
$\theta^i$ can be calculated.  As has been pointed out by Markovi\'{c}
\cite{draza}, measurement errors at degenerate points in parameter
space typically scale like $1/\sqrt{\rho}$ instead of like $1/\rho$.
This is true if the lowest order derivative of ${\bf h}$ which is
non-vanishing in all directions is the second derivative, as can be
seen from, e.g., Eq.~(\ref{muhat_linear}) above.

\appendix{Approximate constancy of \\
the spin parameter $\beta$ that influences\\ the waveform's phase}
\label{beta_const}

The leading order contribution of the bodies' spins to the secular
growth of the gravitational-wave phase has been derived by Kidder,
Will and Wiseman \cite{kidder}, and is given by the term proportional
to $4 \pi - \beta$ in Eq.~(\ref{pnsdfdt}).  The quantity $\beta$ is
defined by Eq.~(\ref{defbeta}) and depends on the masses of the two
bodies $M_1$ and $M_2$, their spins ${\vec S_1}$ and ${\vec S_2}$, and
the unit vector in the direction of the orbital angular momentum
${\hat L}$.  Over the course of the inspiral $\beta$ will evolve,
because the directions of the vectors ${\hat L}$, ${\vec S}_1$, ${\vec
S}_2$ will be changing due to spin-orbit and spin-spin interactions.

Nevertheless, in our analysis in the body of the paper, we have
assumed that the factor $\chi = 4 \pi - \beta$ which appears in
Eq.~(\ref{pnsdfdt}) can be treated as constant.  This assumption is
necessary to make the analysis tractable.  In this appendix, we
present evidence which strongly suggests that $\chi$ is always
constant apart from some small amplitude oscillations, showing that
our assumption of constant $\chi$ is a reasonable one for all
coalescing binaries.  We calculate the evolution of $\chi$ by
integrating the orbit-averaged equations (\ref{spineqs}) governing the
evolution of the spins, using both analytic and numerical methods.  A
more complete discussion of the evolution of the spins and orbital
angular momentum, and of their influence on the emitted gravitational
waves, can be found in Ref.~\cite{spins_paper1}.

We start by introducing some dimensionless variables.  Let ${\hat
S}_j$ be the unit vector in the direction of ${\vec S}_j$ for $j =
1,2$, and define
\begin{eqnarray}
\label{alpha_defs}
\alpha_1 & = & {\hat S}_1 \cdot {\hat L}, \\
\alpha_2 & = & {\hat S}_2 \cdot {\hat L}, \\
\alpha_3 & = & {\hat S}_1 \times {\hat S}_2 \cdot {\hat L},
\end{eqnarray}
and
\begin{equation}
\alpha_4 = {\hat S}_1 \cdot {\hat S}_2.
\end{equation}
The $\alpha_j$'s are not all independent variables as they satisfy the
constraint
\begin{equation}
\label{alpha_constraint}
\alpha_1^2 + \alpha_2^2 + \alpha_3^2 + \alpha_4^2 = 1 + 2 \, \alpha_1
\alpha_2 \alpha_4.
\end{equation}

The reason that it is convenient to use these variables is the
following.  The spin-evolution equations (\ref{spineqs}) comprise nine
equations in nine unknowns, with three conserved quantities (the
magnitudes of the three vectors).  Thus, there are effectively six
degrees of freedom.  If we specify the three independent values of the
variables $\alpha_1, \ldots ,\alpha_4$, then the remaining three
degrees of freedom can be parametrized by an overall rotation matrix.
More precisely, given the vectors ${\hat S}_1$, ${\hat S}_2$ and
${\hat L}$, there will be a unique rotation matrix ${\bf R}$ which
takes ${\hat L}$ into ${\hat L}^\prime = {\hat e}_z$ (the unit vector
along the $z$ axis), ${\hat S}_1$ into a vector ${\hat S}_1^\prime$ in
the $x$-$z$ plane, and ${\hat S}_2$ into some ${\hat S}_2^\prime$.
The vectors ${\hat S}_1$, ${\hat S}_2$ and ${\hat L}$ are determined
by ${\bf R}$ and by the variables $\alpha_1, \ldots, \alpha_4$.
Hence, the variables $\alpha_j(t)$ for $1 \le j \le 4$ and ${\bf
R}(t)$ can be used instead of the vectors themselves to parameterize a
solution to the spin-evolution equations.  Now it turns out that the
evolution of the $\alpha_j$'s {\it decouples} from the evolution of
${\bf R}$, in the sense that each $d
\alpha_j / d t$ depends only on $\alpha_1, \ldots, \alpha_4$ and is
independent of ${\bf R}$.  This greatly simplifies our analysis.


If we use units in which $M = 1$ and define $s_j = | {\vec S}_j |$ for
$j = 1,2$, then we obtain from Eqs.~(\ref{spineqs}),
(\ref{alpha_defs}) and (\ref{drdt}) the following coupled system of
equations for $\alpha_1, \ldots, \alpha_4$:
\FL
\begin{eqnarray}
\label{alpha1_eqn}
{d \alpha_1 \over d r} & = & {- 15 \over 128 \mu} \left[ {1
\over M_2} - {s_1 \alpha_1 \over L} \right] s_2 \alpha_3 \\
\label{alpha2_eqn}
{d \alpha_2 \over d r} & = & { 15 \over 128 \mu} \left[ {1 \over M_1}
- {s_2 \alpha_2 \over L} \right] s_1 \alpha_3 \\
\label{alpha3_eqn}
{d \alpha_4 \over d r} & = & {- 15 \alpha_3 \over 128 \mu}
\left[ \left( {M_2 \over M_1} - {M_1 \over M_2}
\right) L + s_1 \alpha_1 - s_2 \alpha_2 \right].
\end{eqnarray}
Here $L \equiv \mu \sqrt{r}$ denotes the magnitude of the orbital
angular momentum, and we have changed the dependent variable from time
$t$ to orbital separation $r$.  The omitted equation for $d \alpha_3 /
d r$ can be obtained by combining Eqs.~(\ref{alpha_constraint}) --
(\ref{alpha3_eqn}).  From Eq.~(\ref{defbeta}), the spin parameter
$\beta$ is given in terms of these variables by
\begin{eqnarray}
\label{defbeta1}
\beta & = & {113 \over 12} ( s_1 \alpha_1 + s_2 \alpha_2) \nonumber \\
\mbox{} & & + {25 \over 4 M_1 M_2} \left( M_2^2 s_1 \alpha_1 + M_1^2
s_2 \alpha_2 \right).
\end{eqnarray}

We have numerically integrated the equations (\ref{alpha1_eqn}) --
(\ref{alpha3_eqn}) for various initial spin and angular momentum
directions, for the cases of NS-NS, NS-BH and BH-BH binaries.  We
assumed all neutron stars have masses of $1.4 \, M_\odot$, and black
holes have masses $10 \, M_\odot$.  We integrated inwards, starting at
that value of $r$ at which the emitted waves enter the LIGO/VIRGO
waveband at $10 \, {\rm Hz}$, and ending at $r = 6 M$ near the last
stable circular orbit \cite{lastorbit}.  In the special case that one
of the spins vanishes, it can be seen from Eqs.~(\ref{defbeta}) and
(\ref{spineqs}) that $\beta$ will be conserved \cite{spins_paper1}.
Hence we took both spins to be non-vanishing.  We also assumed that
their magnitudes are maximal, so that $s_j = M_j^2$ for $j = 1,2$
[cf.~Sec.~\ref{spin_sec}], as these are the values which can be
expected to give the largest changes in $\beta$.

Typical results are shown in Figs.~\ref{nsns} - \ref{bhbh}.  The
factor $\chi = 4 \pi - \beta$ undergoes small oscillations with an
amplitude of order $0.1$ which is small compared to the mean value of
$\chi$.  This mean value depends on the mass ratio and on the initial
spin directions, but always lies between $4 \pi - \beta_{\rm max}
\approx 3$ and $4 \pi + \beta_{\rm max} \approx 22$, where $\beta_{\rm
max}$ is as given in Sec.~\ref{spin_sec}.  The angles between the
vectors given by $\alpha_1, \ldots, \alpha_4$ also oscillate, all with
the same frequency.  [This frequency is {\it not} the frequency with
which the total spin ${\vec S} = {\vec S}_1 + {\vec S}_2$ precesses
around ${\vec L}$ \cite{spins_paper1}, as that precession does not
change the angles between the vectors, and thus is not described by
Eqs.~(\ref{alpha1_eqn}) -- (\ref{alpha3_eqn})].

Some insight into the behavior of the general solutions to Eqs.
(\ref{alpha1_eqn}) -- (\ref{alpha3_eqn}) can be gained by considering
the special case when the magnitude of one of the spins (say ${\vec
S}_1$) is small, so that $s_1 \equiv |{\vec S}_1 | / M^2 \ll 1$.  This
condition will sometimes be satisfied by NS-NS and BH-BH binaries, but
will always be satisfied by NS-BH binaries since all compact bodies
satisfy $|{\vec S}_j| \alt M_j^2$.  Below we find analytic solutions
to first order in $s_1$.  As we now describe, in the approximation
$s_1 \ll 1$ the amplitude of the oscillations of $\beta$ (and hence
also of $\chi$) is always smaller than $\sim 1/4$, for all initial
spin directions and for all mass-ratios.  Although rapidly spinning
NS-NS and BH-BH binaries will not satisfy $s_1 \ll 1$, nevertheless we
find that amplitudes of the oscillations of $\beta$ in the numerical
solutions agree roughly with those predicted by the small spin
approximation.  [For some special initial spin directions, such as
$\alpha_1 = \alpha_3 = 0$, the analytic solutions are poor
approximations to the numerical solutions, but in all such cases that
we have checked, the amplitudes of the $\beta$ oscillations are still
$\alt 0.2$]

The solutions to first order in $s_1$ can be written as
\begin{equation}
\alpha_j(r) = \alpha_j^{(0)}(r) + \alpha_j^{(1)}(r) \, s_1 + O(s_1^2),
\end{equation}
for $1 \le j \le 4$.  Now as we have already mentioned, it can be seen
from Eqs.~(\ref{defbeta}) and (\ref{spineqs}) that when $s_1 = 0$, the
angle between ${\vec S}_2$ and ${\vec L}$ is conserved, so that
$\beta$ is constant.  However, in this case the angles between the
small spin ${\vec S}_1$ and the other two vectors will not be
conserved.  Thus, the zeroth order solutions $\alpha_j^{(0)}$ will be
non-constant.  We start by deriving these solutions.

Substituting $s_1 = 0$ into Eqs.~(\ref{alpha1_eqn}) --
(\ref{alpha3_eqn}) we find that $\alpha_2^{(0)}$ is constant, i.e.,
$\alpha_2^{(0)} = \alpha_{2,i} \equiv \alpha_2(r_i)$, where $r_i$ is
the initial orbital separation, and that
\begin{eqnarray}
{d \alpha_1^{(0)} \over d r} & = & - h_1 \alpha_3^{(0)} \\ {d
\alpha_4^{(0)} \over d r} & = & - h_4 \alpha_3^{(0)}.
\end{eqnarray}
Here
\begin{eqnarray}
\label{h1_def}
h_1 & = & - {15 s_2 \over 128 \mu M_2}, \\
\label{h4_def}
h_4 & = & {15 \over 128 \mu} \left( s_2 \alpha_{2,i} + L \, \delta
\right),
\end{eqnarray}
and $\delta \equiv (M_1^2 - M_2^2) / (M_1 M_2)$.  The coefficient
$h_4$ is non-constant as $L = \mu \sqrt{r}$ depends on $r$.  However,
since it will typically vary slowly compared to the oscillations in
the angles, we can approximate it to be constant.  [The evolution of
$h_4$ gives rise to a slow evolution in the amplitude and frequency of
the oscillations of the $\alpha_j^{(0)}$'s].  Defining
\begin{equation}
\label{pmdef}
\alpha_\pm \equiv h_4 \alpha_1^{(0)} \pm h_1 \alpha_4^{(0)}
\end{equation}
we find that $\alpha_-(r)$ is constant, $\alpha_-(r) = \alpha_{-,i}
\equiv \alpha_-(r_i)$, and
\begin{equation}
\label{lastalpha}
{d \alpha_+ \over d r} = - 2 h_1 h_4 \alpha_3^{(0)}.
\end{equation}
This equation can be solved by combining it with the constraint
(\ref{alpha_constraint}).  To zeroth order in $s_1$, the constraint
can be expressed using Eq.~(\ref{pmdef}) in the form
\begin{equation}
\label{alpha_constraint1}
\alpha_3^{(0)}(r)^2 + \nu^2 \left[\alpha_+(r) - {\hat \alpha}_+
\right]^2 = \kappa^2,
\end{equation}
where
\begin{eqnarray}
\label{miscdefs}
\nu & = & {\nu_0 \over 2 h_1 h_4}, \\
\nu_0^2 & = & h_1^2 + h_4^2 - 2 h_1 h_4 \alpha_{2,i}, \\
\kappa^2 & = & 1 - \alpha_{2,i}^2 - (1 +
\alpha_{2,i}^2) {\alpha_{-,i}^2 \over \nu_0^2},\\
{\hat \alpha}_+ & = & -(h_1^2 - h_4^2) {\alpha_{-,i} \over \nu_0^2}.
\end{eqnarray}
Combining Eqs.~(\ref{lastalpha}) and (\ref{alpha_constraint1}) yields
the solutions
\FL
\begin{eqnarray}
\label{alpha3_soln}
\alpha_3^{(0)}(r) & = & \alpha_{3,i} \cos \Phi + \nu (\alpha_{+,i} - {\hat
\alpha}_+) \sin
\Phi \\
\label{alphaplus_soln}
\alpha_+(r) & = &{\hat \alpha}_+ - {\alpha_{3,i} \over \nu} \sin \Phi +
(\alpha_{+,i} - {\hat \alpha}_+ ) \cos \Phi,
\end{eqnarray}
where $\alpha_{+,i} = \alpha_+(r_i)$, $\alpha_{3,i} = \alpha_3(r_i)$,
and
\begin{equation}
\Phi = \nu_0 (r - r_i).
\end{equation}
Note that $\nu_0$ is the frequency of oscillation of the
$\alpha_j^{(0)}$'s --- frequency with respect to changing orbital
radius $r$, not changing time $t$.

Analytic expressions for the functions $\alpha_1^{(0)}, \ldots ,
\alpha_4^{(0)}$ can now be obtained by combining Eqs.~(\ref{h1_def})
-- (\ref{pmdef}) and (\ref{miscdefs}) -- (\ref{alphaplus_soln}).
These expressions depend in a complex way on all of the initial spin
direction parameters $\alpha_{1,i}$, $\alpha_{2,i}$ and
$\alpha_{4,i}$, and also on $s_2$, on the mass ratio $M_1 /M_2$, and
on the initial orbital separation $r_i / M$.  For the equal mass case
$M_1 = M_2$, the ``frequency'' $\nu_0$ is given by
\begin{equation}
\label{nu0equalmass}
\nu_0^2 = {s_2^2 \over M_2^2} \left[ {255 \over 4096} + {1125 \over
16384} \alpha_{2,i}^2 \right].
\end{equation}
Values of $\nu_o$ for $M_1 \ne M_2$ are typically much larger than
this.

The first-order corrections $\alpha_j^{(1)}(r)$ can be obtained using
the zeroth order solutions and Eqs.~(\ref{alpha1_eqn}) --
(\ref{alpha3_eqn}).  However, we are only interested in determining
the leading order behavior of $\beta$, and for this purpose we need
only evaluate $\alpha_2^{(1)}$.  From Eqs.~(\ref{alpha2_eqn}) and
(\ref{alpha3_soln}), this is given by
\FL
\begin{eqnarray}
\label{alpha2_soln}
\alpha_2^{(1)}(r)  &=& {15 \over 128 \mu \nu_0} \left[ {1 \over M_1}
- {s_2 \alpha_{2,i} \over L} \right] \nonumber \\
\mbox{} & & \times \left[ \alpha_{3,i} \sin \Phi - \nu \left(
\alpha_{+,i} - {\hat \alpha}_+ \right) \left( \cos \Phi - 1 \right)
\right].
\end{eqnarray}
Substituting Eqs.~(\ref{alpha2_soln}) and (\ref{miscdefs}) --
(\ref{alphaplus_soln}) into (\ref{defbeta1}) gives a result of the
form
\begin{equation}
\beta(r) = A + B \cos \Phi + C \sin \Phi,
\end{equation}
where the constants $B$ and $C$ are first order in $s_1$.  The
resulting expression for the total amplitude of oscillation ${\cal A}
= \sqrt{B^2 + C^2}$ in terms of the variables $\alpha_{1,i}, \ldots,
\alpha_{4,i}$, $M_1/M_2$ and $r$ is complicated and not very
illuminating, so we do not reproduce it here.  Instead we show in
Figs.~\ref{amp1} and \ref{amp2} the quantity ${\cal A}_{\rm max} =
{\cal A}_{\rm max}[\alpha_{2,i},\alpha_{4,i}]$ obtained in the
following way: (i) Use Eq.~(\ref{alpha_constraint}) to eliminate
$\alpha_{3,i}$ in terms of $\alpha_{1,i}$, $\alpha_{2,i}$ and
$\alpha_{4,i}$.  (ii) Numerically maximize over values of
$\alpha_{1,i}$ that lie in the range between the values $\alpha_{2,i}
\, \alpha_{4,i} \pm \sqrt{ ( 1 - \alpha_{2,i}^2) \,(1 -
\alpha_{4,i}^2)}$ allowed by Eq.~(\ref{alpha_constraint}).  (iii)
Choose the maximal spin magnitudes $s_1 = M_1^2$, $s_2 = M_2^2$.  (iv)
Choose the final orbital separation $r = 6 M$, the value for which the
amplitude ${\cal A}$ will most likely be largest.  It can be seen from
these plots that for all choices of initial angles, ${\cal A} \le
0.25$.

In the special case that $M_1 = M_2$, the formulae simplify and we
find that ${\cal A} \propto 1 / \sqrt{r}$ (this is not true in
general).  Specifically we find in this case that
\begin{eqnarray}
B & = & {376 \over 384 \sqrt{r}} \bigg[ \alpha_{1,i} \, \alpha_{2,i} -
2
\alpha_{4,i} \nonumber \\
\mbox{} & & + {15 \over 128} \left( 4 - \alpha_{2,i}^2 \right) \left(
\alpha_{1,i} \, \alpha_{2,i} + 2 \alpha_{4,i} \right) \nu_0^{-2} \bigg],
\end{eqnarray}
and
\begin{equation}
C = - { 235 \over 512}\, { \alpha_{2,i} \, \alpha_{3,i} \over \sqrt{r}
\nu_0},
\end{equation}
where $\nu_0$ is given by Eq.~(\ref{nu0equalmass}).

To summarize, we have determined the evolution of the quantity $\chi =
4 \pi - \beta$ both numerically, for a wide range of initial
conditions, and analytically, in the regime where $|{\vec S}_1| \ll
M^2$.  In all cases we find that the amplitude of the oscillations of
$\chi$ is $\le 0.25$.

\appendix{The decoupling of phase and \\
amplitude parameters in the Fisher \\ information matrix}
\label{decouple}

The phase $\Psi(f)$ of the Fourier transform of the waveform can be
written in the form
\begin{equation}
\Psi(f) = \sum_{n= 1,2,\cdots} c_n (f/f_0)^{\alpha_n},
\end{equation}
where $(\alpha_1, \alpha_2, \alpha_3, \ldots) = (0,1,-5/3, -1,-2/3,
\ldots)$, and the parameters $c_1$, $c_2$, $c_3$, $c_4$ etc. are
simply related to the parameters $\phi_c$, $t_c$, ${\cal M}$, $\mu$,
$\beta$ etc., via Eq.~(\ref{pnspsi}).  The number of variables $c_n$
will depend on the post-Newtonian order to which $\Psi(f)$ is
calculated; the following analysis holds for any number of these
variables.  We can make a linear transformation to new variables
\begin{equation}
d_m = U_m^{\,\,\, n} c_n
\end{equation}
in such a way that
\begin{equation}
{\partial {\bf h} \over \partial d_1} = {\partial {\bf h} \over
\partial c_1} = i {\bf h},
\end{equation}
and that for $m \ge 2$ \cite{caveat5},
\FL
\begin{eqnarray}
\label{vanish}
\bigg( i {\bf h} \, \bigg| \, {\partial {\bf h} \over \partial d_m} \bigg) &
\propto &  \sum_n ({\bf U}^{-1})^n_{\,\,\,m}
\,\,\,\int_0^\infty df \, {|{\tilde h}_0|^2 \over S_n(f) } (f /
f_0)^{\alpha_n} \\ & = & 0 \nonumber.
\end{eqnarray}
The key point now is that the inner product $$
\Gamma_{a m} = \bigg( {\partial {\bf h} \over \partial \mu^a} \,
\bigg| \, {\partial {\bf h} \over \partial d_m} \bigg),
$$ where $\mu^a$ is any of the ``amplitude'' parameters $D$, $\psi$,
$v$ and $d_1$, will also be proportional to the right-hand side of
Eq.~(\ref{vanish}) for $m \ge 2$, and so will vanish.  This can be
seen from the structure of Eq.~(\ref{fullexpr}).  Consequently, in the
new variables $D$, $v$, $\psi$, and $d_m$, $m = 1,2,
\ldots$, the Fisher matrix (\ref{gamma_def}) will be block diagonal,
which establishes the result stated in Sec.~\ref{draza_approx}.

\newpage

\figure{Gravitational waveforms from coalescing compact binaries are
completely specified by a finite number of parameters $\theta =
(\theta^1, \ldots, \theta^k)$, and so form a surface ${\cal S}$
in the vector space $V$ of all possible measured detector outputs $s =
s(t)$.  The statistical properties of the detector noise endow $V$
with the structure of a infinite-dimensional Euclidean space.  This
figure illustrates the relationships between the true gravitational
wave signal $h({\tilde \theta})$, the measured signal $s$, and the
``best-fit'' signal $h({\hat \theta})$.  Given a measured detector
output $s = h({\tilde \theta}) + n$, where $n = n(t)$ is the detector
noise, the most likely values ${\hat {\bf \theta}}$ of the
binaries parameters are just those that correspond to the point
$h({\hat \theta})$ on the surface ${\cal S}$ which is closest [in the
Euclidean distance $(s - h \, | \,s - h)\,$] to $y$.
\label{hypersurface}}

\figure{This plot shows how the total signal-to-noise squared $S^2/N^2$
for a detected coalescing-binary waveform is distributed in frequency
$f$, assuming the detector noise curve (\ref{snf}).  Most of the
signal-to-noise ratio comes not near $70 \, {\rm
Hz}$ where the detector sensitivity $S_n(f)^{-1}$ is highest, but
rather at a somewhat lower frequency of $\sim 50 \, {\rm Hz}$, because
more cycles per unit frequency are received at lower frequencies.
\label{snrf}}

\figure{This plot shows the curve of constant probability on the
$\mu\beta$ plane for a NS-BH binary, where $\mu$ is the binary's reduced mass
and
$\beta$ is a dimensionsless spin-related parameter, such that the true
values of these parameters lie inside the ellipse with $95 \%$
confidence.  The strong correlation between possible values of $\mu$
and $\beta$ is evident.  To a good approximation, the chirp mass
${\cal M}$ is measured to arbitrarily high accuracy.  Hence in the
three-dimensional space of the parameters $({\cal M},\mu,\beta)$, the
true values of these parameters are confined with high confidence to a
thin strip of the above ellipsoidal shape in a plane of constant
${\cal M}$.
\label{mubeta}}

\figure{A diagram showing the information obtained from the gravitational
wave signal, constraining the individual masses $M_1$ and $M_2$ of the binary
components, in various cases.  Because of the highly accurate
measurement of the chirp mass ${\cal M}$ in each case, the individual
masses are essentially constrained to lie on a curve of constant
${\cal M} \equiv (M_1 M_2)^{3/5}(M_1
+M_2)^{-1/5}$ in the $M_1 \, M_2$ plane.  The measured
value of ${\cal M}$ provides a sharp lower bound for the larger mass
$M_1$, and a sharp upper bound for the smaller mass $M_2$, since the
constant-${\cal M}$ curves terminate sharply at the forbidden, hatched
region.  The measurement of the reduced mass $\mu$ gives some (but not
much) information about where along the constant ${\cal M}$ curve the
binary is most likely to be located.  In each case, the solid circles
show the true values of $M_1$ and $M_2$, the solid curve denotes the
$68 \%$ (1 sigma) confidence interval, and the dashed extension
denotes the $95 \%$ (2 sigma) confidence interval.  The detector noise spectrum
(\ref{snf}) was assumed.
\label{m1m2}}

\figure{The amplitude sensitivity function $\sigma_D({\bf n})$, as a
function of position on the sky parametrized by the Earth-fixed
coordinates $\theta$ and $\varphi$, for the detector network
consisting of the two LIGO detectors in Hanford, Washington and
Livingston, Louisiana, and the VIRGO detector in Pisa, Italy.  The
axis $\theta=0$ is the Earth's axis of rotation, and $\varphi=0$ is
$0^\circ$ longitude.  Only sky positions over the northern hemisphere
are shown, because $\sigma_D$ takes the same values at antipodal
points.  The function $\sigma_D({\bf n})$ has the following meaning:
for a source of waves in the direction ${\bf n}$, the combined
signal-to-noise ratio of the whole network, averaged over all
polarization angles $\psi$ of the source (equivalently, averaged over
rotations of the source in the plane perpendicular to the line of
sight), will be proportional to $\sigma_D({\bf n})$.  The thick black
lines indicate the positions of the three detectors.  This plot can be
generated by combining Eqs.~(\ref{sigmaDdef}) and (\ref{ptensordef})
of the text with the network parameters given after
Eq.~(\ref{ptensordef}).
\label{ligovirgo1}}

\figure{The polarization sensitivity function $1 - \varepsilon_D({\bf
n})$, for the LIGO/VIRGO detector network; see caption of
Fig.~\ref{ligovirgo1}.  This plot can be generated by combining
Eqs.~(\ref{bigS}), (\ref{tensorprod}), (\ref{varepsilonDdef}), and
(\ref{ptensordef}) of the text.  The quantity $\varepsilon_D({\bf n})$
essentially measures the ``skewness'' or assymetry in the
sensitivities of the network to the two independent polarization
components of waves propagating in the direction ${\bf n}$.  When
$\varepsilon_D \approx 0$, the network has roughly equal sensitivity
to both polarization components.  When $\varepsilon_D \approx 1$, on
the other hand, one polarization component can be measured far more
accurately than its orthogonal counterpart.  In this case the
signal-to-noise ratio for incident, strongly linearly polarized bursts
of waves (e.g., those from edge-on coalescing binaries) will depend
sensitively on the polarization axis, i.e., it would vary by large
factors if the source were rotated in the plane perpendicular to the
line of sight.  Note that the polarization sensitivity is poor ($\alt
0.2$) for directions directly overhead the two LIGO detectors (because
the two detectors are nearly parallel), and is typically $\alt 0.3$
over most of the sky.  Good sensitivity is achieved in isolated
regions.
\label{ligovirgo2}}

\figure{The dependence of the distance measurement accuracy $\Delta D
/ D$ on the sky location ${\bf n}$, the polarization angle $\psi$, and
the cosine $v$ of the angle of inclination of the orbit to the line of
sight is approximately given by $\Delta D / D \propto \Upsilon({\bf
n},v,\psi)$, where the dimensionless function $\Upsilon$ is defined in
Eq.~(\ref{upsilon}).  Here we plot for the LIGO/VIRGO detector network
the quantity $\Upsilon_{\rm max}$
obtained by maximizing $\Upsilon$ over all polarization angles $\psi$,
at $v^2 = 1/2$, as a function of $\theta$ and $\varphi$.  Higher
values of $\Upsilon$ indicated by regions of lighter shading
correspond to poorer measurement accuracy.
\label{upmax}}

\figure{The quantity $\Upsilon_{\rm min}$ which is obtained by
minimizing $\Upsilon({\bf n},v,\psi)$ over $\psi$, at $v^2 = 1/2$; see
caption of Fig.~\ref{upmax}
\label{upmin}}

\figure{The quantity $\Omega(\varepsilon_D)$, which is the solid angle
on the sky for which the polarization sensitivity is less than $1-
\varepsilon_D$, for two different detector networks.  The solid line
indicates the LIGO/VIRGO detector network, and the dashed line a
4-detector network consisting of the LIGO and VIRGO detectors together
with a hypothetical detector in Perth, Australia.  These plots were
generated using 1000 randomly chosen directions ${\bf n}$.  The great
improvement in polarization sensitivity due to the additional detector
is apparent: e.g., the polarization sensitivity is $\le 0.2$ over $\sim
60\%$ of the sky for the 3-detector network, but only over $\sim 20\%$
of the sky for the 4-detector network.
\label{aus}}

\figure{An example illustrating the necessity of going beyond the Gaussian
approximation.  Consider a neutron-star neutron-star binary merger in
the direction given by $(\theta,\varphi) = (50^\circ, 276^\circ)$.
The LIGO/VIRGO network parameters for this direction are $\sigma_D =
1.03$ and $\varepsilon_D = 0.74$.  Suppose that an experimenter
determines from the measured signal the following ``best-fit''
(maximum-likelihood) parameters: distance $D_0 = 432 \, {\rm Mpc}$
[corresponding to a signal-to-noise ratio of $\rho = 12.8$, assuming
the advanced detector sensitivity level (\ref{snf})], masses $M_1 =
M_2 = 1.4 M_\odot$, cosine of inclination angle $v_0 = 0.31$, and
polarization angle ${\bar \psi}_0 = 56.5^\circ$.  Then the distribution
that she would infer by a Bayesian analysis for the distance to the
source is shown by the solid curve; it is given by
Eqs.~(\ref{PDFvcalD}), (\ref{PDFcalD}), and (\ref{r0}) of the text.
The Gaussian approximation [Eq.~(\ref{final_ans}) of the text] to this
distribution is shown by the dashed curve.  The distance measurement
accuracy is atypically poor in this example; see Fig.~\ref{histogram}
below.
\label{prob}}

\figure{The solid line shows the distance measurement accuracy $\Delta
D / D$ for the binary merger discussed in the caption of
Fig.~\ref{prob} (for which $\Delta_1=0.10$, $\Delta_2 = 0.057$), as a
function of the cosine of the angle of inclination, $v_0$.  The
improvement in accuracy at high values of $v_0$ is due in part to an
increased signal-to-noise ratio there.  The dashed curve shows the
prediction (\ref{final_ans}) of the linear error-estimation theory,
which diverges as $v_0 \to 1$.
\label{vdep}}

\figure{The distance measurement accuracies that result from displacing
along the line of sight to the Earth, to various distances $D_0$, the
binary merger of Fig.~\ref{prob}.  As in Fig.~\ref{vdep}, the dashed curve
shows the approximate linear estimate (\ref{final_ans}), and the solid
curve shows the more accurate estimate (\ref{final_ans1}).  The curves
terminate at that distance ($\sim 700 \, {\rm Mpc}$) beyond which the
merger is no longer visible, assuming the detector sensitivity level
(\ref{snf}) and a combined signal-to-noise threshold of $8.5$.
\label{ddep}}

\figure{The distances $D_0$ for 1000 NS-NS binaries whose locations and
orientations were randomly chosen, and the corresponding predicted
signal-to-noise ratios $\rho$ for the LIGO/VIRGO network.
The lower dashed line is the signal-to-noise threshold of 8.5, below
which sample points were discarded; the upper dashed line shows the
maximum possible value (\ref{rhomax}) of $\rho$ at a given distance.
Six points with $\rho$ between $50$ and $90$ are not shown.  The number
of sources with $\rho$ larger than a given value $\rho_*$ is
proportional to $\rho_*^{-3}$.  The detection
of this many binary inspirals with the advanced LIGO/VIRGO detectors
would take several years, if merger rates are as currently estimated
\cite{Narayan,sterl}, and assuming the detector sensitivity level
(\ref{snf}).
\label{scatter1}}

\figure{The distance measurement accuracy $\Delta D /D$ computed from
Eq.~(\ref{final_ans1}) for the same 1000 NS-NS binaries, versus the
distance $D_0$; see caption of Fig.~\ref{scatter1}.  The spread in the
values of $\Delta D/D$ is due to different source directions and
orientations.  Note that the accuracy for sources within $200 \,{\rm
Mpc}$ (of which there are estimated to be $\sim 3$ per year
\cite{Narayan,sterl}) can vary between $\sim 2 \%$ and $\sim 25 \%$.
For the most distant detectable sources (at $\sim 1200 \, {\rm Mpc}$),
the accuracy can sometimes be as good as $\sim 20 \%$.  The dashed
line shows the theoretical lower bound (\ref{deltadmin}) derived using
the linear error-estimation formalism; points below this line mostly
have values of $v_0$ close to one for which value the linear
error-estimation theory fails.
\label{scatter2}}

\figure{Distance measurement accuracy versus signal-to-noise for the
same 1000 NS-NS binaries; see caption of Fig.~\ref{scatter1}.  The
dashed line shows the theoretical lower bound (\ref{deltadmin1}).  As
in Fig.~\ref{scatter1}, six points with $\rho$ between $50$ and $90$
are not shown.
\label{scatter3}}

\figure{The frequency of occurrence of different ranges of $\Delta D /
D$, out of a total of 1000 signals, for the LIGO/VIRGO detector
network.  It can be seen that $\sim 8\%$ of detected signals will
have distance measurement accuracies of better than $15\%$, while
$\sim 60\%$ of them will have accuracies of better than $30\%$.  These
conclusions are insensitive to the overall scale of the detectors'
intrinsic noise, which essentially sets the event-detection rate.  By
contrast, they {\it are} sensitive to the number of detectors in the
detector network, and to their orientations; see text and also
Fig.~\ref{aus}.
\label{histogram}}

\figure{Distance measurement accuracy versus the detector network
polarization sensitivity $1 - \varepsilon_D({\bf n})$, for 1000 NS-NS
binaries; see caption of Fig.~\ref{scatter1}.  The strong correlation
between very poor distance-measurement accuracy and low polarization
sensitivity is evident --- essentially all points with $\Delta D / D >
0.5$ have $1 - \varepsilon_D \alt 0.2$.
\label{scatter4}}

\figure{During the last few minutes of inspiral, the angles between the
bodies' spins ${\vec S}_1$, ${\vec S}_2$ and the orbital angular momentum
${\vec L}$ all oscillate, in
addition to and separately from the precession of the total spin
${\vec S} = {\vec S}_1 + {\vec S}_2$ around ${\vec L}$.  This
oscillation gives rise to an oscillation of the parameter $\chi = 4
\pi - \beta$ which governs the contribution of the spins to the
accumulated phase of the emitted gravitational waves
[cf.~Eq.~(\ref{pnsdfdt})].  Here we show a typical plot of $\chi$ as a
function of the orbital separation $r$, for a NS-NS
binary.  The spin and orbital angular momentum
directions were taken to be ${\vec S}_1 \propto {\vec i} + {\vec k}$,
${\vec S}_2 \propto - {\vec j}$ and ${\vec L} \propto {\vec i} + {\vec
j}$ at the initial gravitational wave frequency of $10 \, {\rm Hz}$.  The
spin of each neutron star was assumed have the maximal magnitude of
$(1.4 M_\odot)^2$, corresponding to a rotation period of a few
miliseconds.  It can be seen that in each case the amplitude of oscillation of
$\chi$ is very small compared to its mean value, so that to a good
approximation we can take $\chi = \mbox{constant}$.
\label{nsns}}

\figure{The evolution of $\chi$ for a NS-BH binary; see caption of
Fig.~\ref{nsns}.  The black hole was assumed to be maximally rotating.
\label{nsbh}}

\figure{The evolution of $\chi$ for a BH-BH binary; see caption of
Fig.~\ref{nsns}.  Both black holes were assumed to be maximally rotating.
\label{bhbh}}

\figure{When one of the spins is small, the evolution of the parameter
$\chi = 4 \pi - \beta$ is approximately given by
$
\chi = \chi_0 + {\cal A} \, \cos \left[ \nu_0 r + \mbox{const} \right],
$
where the ``frequency'' $\nu_0$ and amplitude ${\cal A}$ are slowly
varying functions of $r$.  Here we show the amplitude ${\cal A}$ for
an equal-mass binary, at an orbital separation of $r = 6 M$, as a
function of $\alpha_{2,i} \equiv {\hat S}_2 \cdot {\hat L}$ and
$\alpha_{4,i} \equiv {\hat S}_1 \cdot {\hat S}_2$, where the maximum is
taken over the remaining angles.  ${\hat S}_1$, ${\hat S}_2$ and
${\hat L}$ are unit vectors in the directions of the initial spins and
the initial orbital angular momentum.
\label{amp1}}

\figure{As in Fig.~\ref{amp1}, but for a NS-BH binary with $M_1 / M_2
= 1.4/10$.
\label{amp2}}

\newpage

\begin{table}
\caption{The rms errors for signal parameters and the correlation
coefficient $c_{{\cal M} \mu}$, calculated assuming spins are
negligible. The results are for a single ``advanced'' detector, the
shape of whose noise curve is given by Eq.~(\ref{snf}).  $M_1$ and
$M_2$ are in units of solar masses, while $\Delta t_c$ is in units of
msec. The rms errors are normalized to a signal-to-noise ratio of $S/N
= 10$; the errors scale as $(S/N)^{-1}$, while $c_{{\cal M} \mu}$ is
independent of $S/N$.
\label{table1}}
\begin{tabular}{lllllll}
$M_1$&$M_2$&$\Delta \phi_c$&$\Delta
t_c $&${{\Delta {\cal M}}/{\cal M}}$&${{\Delta \mu}/
{\mu}}$&$ c_{{\cal M} \mu}$ \\
\tableline
2.0& 1.0 & 1.31 & 0.721 & 0.0038 \% & 0.39 \% & 0.899\\
1.4& 1.4 & 1.28 & 0.713 & 0.0040 \%& 0.41 \% & 0.906\\
10& 1.4 & 1.63 & 1.01 & 0.020 \% & 0.54 \% & 0.927\\
15& 5.0 & 2.02 & 1.44 & 0.113 \% & 1.5 \% & 0.954\\
10& 10 & 1.98 & 1.43 & 0.16 \% & 1.9 \% & 0.958\\
\end{tabular}
\end{table}

\mediumtext
\begin{table}
\caption{The rms errors for signal parameters and the correlation
coefficients $c_{{\cal M} \mu}$, $c_{{\cal M} \beta}$, and $c_{\mu
\beta}$, calculated using spin-dependent waveforms.  The results are
for a single ``advanced'' detector, the shape of whose noise curve is
given by Eq.~(\ref{snf}).  For the rows marked with a $\dag$ (and {\it
only} for those rows), the variance-covariance matrix has been
``corrected'' to approximately account for the fact that the spin
parameter $\beta$ must
satisfy $|\beta| < \beta_{max} \approx 8.5$.  The rms errors are
normalized to a signal-to-soise ratio of $S/N = 10$.  Except for rows
marked with a $\dag$, errors scale as $(S/N)^{-1}$, while the
correlation coefficients are independent of $S/N$.  Except for rows
marked with a $\dag$, if $\beta$ had been chosen non-zero with $M_1$
and $M_2$ unchanged, then $\Delta {\cal M}/{\cal M}$, $\Delta
\mu/{\cal M}$, and $c_{{\cal M} \mu}$ would have been unchanged (but
$\Delta \beta$, $c_{{\cal M} \beta}$, and $c_{\mu \beta}$ would have
been altered).  As in Table \ref{table1}, $M_1$ and $M_2$ are in units
of $M_\odot$, while $\Delta t_c$ is in msec.  The results for the
LIGO/VIRGO network of detectors, for a signal with combined
signal-to-noise ratio from all the detectors of $10$, will be
approximately the same as those shown here; see text.
\label{table2}}
\begin{tabular}{lllllllllll}
$M_1$ & $M_2$& $\beta$ &$\Delta \phi_c$& $\Delta
t_c $ &${{\Delta {\cal M}}/{\cal M}} $ & ${{\Delta \mu}/
{\mu}}$ &$ \Delta \beta $&$ c_{{\cal M} \mu}$
&$c_{{\cal M} \beta} $ &$c_{\mu \beta} $\\
\tableline
2.0& 1.0& 0 & 4.13 & 1.14 & 0.034 \% & 8.44 \% & 1.04 & -0.988 & 0.993
& -0.9989  \\
1.4& 1.4& 0 & 4.09 & 1.13 & 0.034 \% & 9.65 \% & 1.24 & -0.988 & 0.993
& -0.9991  \\
10& 1.4& 0 & 6.24 & 2.03 & 0.19 \% & 15.2 \% & 1.99 & -0.990 & 0.994
& -0.9994 \\
5& 1.4& 0 & 4.89 & 1.44 & 0.10 \% & 13.4 \% & 1.73 & -0.989 & 0.994
& -0.9992 \\
15& 5& 0 & 9.26 & 3.53 & 1.06 \% & 76.4 \% & 11.4 & -0.992 & 0.994
& -0.99980 \\
15& 5& 0 \dag & 5.77 & 2.40 & 0.64 \% & 45.8 \% & 6.81 & -0.978 & 0.984
& -0.9995 \\
10& 10& 0 & 9.26 & 3.53 & 1.42 \% & 125 \% & 19.5 & -0.992 & 0.994
& -0.99988 \\
10& 10&  0 \dag& 4.13 & 1.92 & 0.59 \% & 49.9 \% & 7.79 & -0.953 & 0.964
& -0.9992 \\
\end{tabular}
\end{table}

\begin{table}
\caption{Measurement accuracies, including spins, as in Table
\ref{table2} except that we the take shape of the noise curve to be given
by Eq.~(\ref{snf2}).
\label{table3}}
\begin{tabular}{lllllllllll}
$M_1$ & $M_2$& $\beta$ &$\Delta \phi_c$& $\Delta
t_c $ &${{\Delta {\cal M}}/{\cal M}} $ & ${{\Delta \mu}/
{\mu}}$ &$ \Delta \beta $&$ c_{{\cal M} \mu}$
&$c_{{\cal M} \beta} $ &$c_{\mu \beta} $\\
\tableline
2.0& 1.0& 0 & 2.21 & 0.47 & 0.021 \% & 5.12 \% & 0.641 & -0.986 & 0.992
& -0.9988  \\
1.4& 1.4& 0 & 2.19 & 0.46 & 0.021 \% & 5.84 \% & 0.757 & -0.986 & 0.992
& -0.9990  \\
10& 1.4& 0 & 3.90 & 1.06 & 0.13 \% & 10.1 \% & 1.33 & -0.990 & 0.994
& -0.9993 \\
5& 1.4& 0 & 2.81 & 0.66 & 0.065 \% & 8.47 \% & 1.11 & -0.988 & 0.993
& -0.9992 \\
15& 5& 0 & 6.72 & 2.40 & 0.75 \% & 55.1 \% & 8.18 & -0.992 & 0.994
& -0.99980 \\
15& 5& 0 \dag & 4.95 & 1.85 & 0.55 \% & 39.7 \% & 5.90 & -0.985 & 0.989
& -0.9996 \\
10& 10& 0 & 6.72 & 2.40 & 1.00 \% & 90.0 \% & 14.0 & -0.992 & 0.994
& -0.99988 \\
10& 10&  0 \dag& 3.71 & 1.49 & 0.53 \% & 46.7 \% & 7.27 & -0.972 & 0.978
& -0.9995 \\
\end{tabular}
\end{table}

\newpage

\begin{references}
\bibitem{Narayan} R. Narayan, T. Piran, and A. Shemi, Astrophys. J.
            {\bf 379}, L17 (1991).
\bibitem{sterl} E.~S. Phinney, Astrophys. J. {\bf 380}, L17 (1991).
\bibitem{ligoscience} A. Abrmovici, W.~E. Althouse, R.~W.~P. Drever,
            Y. G{\" u}rsel, S. Kawamura, F.J. Raab, D. Shoemaker, L.
Sievers, R.~E. Spero, K.~S. Thorne, R.~E. Vogt, R. Weiss, S.~E.
Whitcomb, and M.~E. Zucker, {\it LIGO: The Laser Interferometer
Gravitational-wave Observatory}, Science {\bf 256}, 325 (1992).
\bibitem{virgo} C. Bradaschia {\it et al.}, Nucl. Instrum. \& Methods
{\bf A289}, 518 (1990); also in {\it Gravitation: a Banff Summer
Institute}, ed. R. Mann and P. Wesson (World Scientific, Singapore,
1991).
\bibitem{tutukov} A.~V. Tutukov and L.~R. Yungelson, Mon. Not. R.
Astron. Soc. {\bf 260}, 675 (1993).
\bibitem{schutz} B.F. Schutz, Nature {\bf 323}, 310 (1986); B.F.
Schutz, Class. Quantum Gravity {\bf 6}, 1761 (1989).
\bibitem{tinto} Y. Gursel and M. Tinto, Phys. Rev. D {\bf 40}, 3884
(1990).
\bibitem{cutleretal} C. Cutler, T.~A. Apostolatos, L. Bildsten, L.~S.
Finn, \'E.~E. Flanagan, D. Kennefick, D.~M. Markovi\'{c}, A. Ori, E.
Poisson, G.~J. Sussman, and K.~S. Thorne, Phys. Rev. Lett. {\bf 70},
2984 (1993).
\bibitem{draza} D. Markovi\'{c}, Phys. Rev. D. {\bf 48}, 4738 (1993).
\bibitem{schutz0} B.~F. Schutz, {\it Hubble Constant from Gravitational
Wave Observations}, in H. Sato and T. Nakamura, Eds., {\it
Gravitational Collapse and Relativity}, (World Scientific, Singapore),
pp. 350-368.
\bibitem{finn_cosmo} D.F. Chernoff and L.S. Finn, Astrophys. J.
{\bf 411}, 5 (1993).
\bibitem{300years} K.S. Thorne, in {\it 300 Years of Gravitation},
            ed. S.W. Hawking and W. Israel (Cambridge University
Press, Cambridge, 1987), pp. 330-458.
\bibitem{cutler_finn} C. Cutler, L.~S. Finn,
E. Poisson and G.~J. Sussman, Phys. Rev. D {\bf 47}, 1151 (1993).
\bibitem{finn2} L.~S. Finn and D.~F. Chernoff, Phys. Rev. D {\bf 47},
2198 (1993).
\bibitem{krolak2} P. Jaranowski and Andrzej Krolak, {\it Optimal
Solution to the Inverse Problem for the Gravitational Wave Signal of a
Coalescing Compact Binary}, in preparation.
\bibitem{peters} P.~C. Peters, Phys. Rev. {\bf 136}, B1224 (1964).
\bibitem{quinlan} However eccentricities may not be negligible for
binaries formed in dense star-clusters in galactic nuclei; see
G.D.~Quinlan and S.L. Shapiro, Astrophys. J {\bf 321}, 199 (1987).
\bibitem{Bildsten_Cutler}L. Bildsten and C. Cutler, Astrophys. J.
{\bf 400}, 175 (1992).
\bibitem{Kochanek}C. Kochanek, Astrophys. J., {\bf 398}, 234 (1992).
\bibitem{yes_but} A rapidly rotating neutron star will be somewhat
oblate, and therefore, compared to a point-mass, its potential energy
will be modified by a term proportional to its quadrupole moment times
the second derivative of the gravitational potential at its
center-of-mass
\cite{Bildsten_Cutler}.  However this correction to the orbital energy
is of second post-Newtonian order, and thus is of higher order than
the other post-Newtonian effects considered in this paper.
\bibitem{WZ} L.~A. Wainstein and V.~D. Zubakov,{\it Extraction of
Signals from Noise} (Dover Publications, Inc., New York, 1962).
\bibitem{flanagan_thorne} E. Flanagan and K.~S. Thorne, in preparation.
\bibitem{Kip} K.S. Thorne, private communication.
\bibitem{newnoise}  The noise spectrum to which Eq.~(\ref{snf2}) is an
approximate analytic fit is given by the sum of the following terms
from Ref.~\cite{finn2}: Eq.~(4.1), with parameters $\eta I_0 = 60 \,
{\rm W}$, $L = 4 \, {\rm km}$, $f_c = 130 \, {\rm Hz}$, $\lambda = 5.1
\times 10^{-7} \, {\rm m}$, and $A^2 = 5 \times 10^{-5}$; Eq.~(4.3)
with the modifications to Eq.~(4.2) of $f_0 \to f^2/f_0$ in the
numerator and $f f_0 / Q_0 \to f_0^2 / Q_0$ in the denominator, and
with parameters $Q_0 = 10^9$, $T = 300 \, {}^\circ {\rm K}$, $f_0 = 1
\, {\rm Hz}$; and Eq.~(4.4), similarly modified, with parameters
$f_{\rm int} = 14 \, {\rm kHz}$ and $Q_{\rm int} = 10^6$.  Eqs.~(4.2)
and (4.4) are modified in order to describe structural damping
\cite{Saulson} which is now thought to be the likely dominant damping
mechanism in the thermal modes \cite{Kip}.
\bibitem{Saulson}  P.R. Saulson, Phys. Rev. D {\bf 42}, 2437 (1990).
\bibitem{finn1} L.~S. Finn, Phys. Rev. D {\bf 46}, 5236 (1992).
\bibitem{krolak1} S.V. Dhurandhar, A. Krolak, B.F. Schutz and W.J. Watkins,
 {\it Gravitational Wave Astronomy with Broadband Detectors. I.
Extraction of Coalescing Binary Signals}, in preparation.
\bibitem{caveat} Note that this definition differs by a factor of two
from that found in Ref.~\cite{finn1}.  Our definition is chosen to
correspond to the quadratic form ``$x_i ({\bf
\Sigma}^{-1})_{ij} x_j$'' which appears in finite-dimensional
Gaussian PDFs, cf.~Eq.~(\ref{noise_distribution}) above.
\bibitem{Kip_Amos} L.S.~Finn, A.~Ori, and K.S.~Thorne, unpublished.
\bibitem{lastorbit} The last stable circular orbit is not exactly at $r =
6 M$ because of the fact that $r$ is the orbital separation in de
Donder gauge and not the Schwarschild radius, and also because of the
finite mass-ratio.  However, this orbit will be close to $r = 6 M$;
see L.~E. Kidder, C.~M. Will and A.~G. Wiseman, Class. Quant.  Grav.
{\bf 9}, L125 (1992).
\bibitem{innerprod} The calculation is simplified if one first defines
the moments $$ {\overline {f^k}} \equiv { \left( f^k h \,|\, h \right)
\over \left( h
\,|\, h \right)},
$$ where the inner product $\left( \,\, | \,\, \right)$ is given by
Eq.~(\ref{inner}). As pointed out by Finn and Chernoff \cite{finn2},
all the elements of $\Gamma_{ij}$ can expressed in terms of the
orbital parameters and a few of the moments ${\overline {f^k}}$. This
continues to hold true when post-Newtonian corrections are added to
the signal.  A useful identity is $\left( f^j h \, | \, i f^k h\right)
=0$ for all real $j$ and $k$, which follows from Eq.~(\ref{inner}).
\bibitem{wagoner1}R.~V. Wagoner and C.~M. Will, Astrophys. J
{\bf 210}, 764 (1976); {\bf 215}, 984 (1977).
\bibitem{wiseman} A.~G. Wiseman, Phys. Rev. D {\bf 46}, 1517 (1992).
\bibitem{eric} E. Poisson, Phys. Rev. D {\bf 47}, 1497 (1993).
\bibitem{krolak} A. Krolak in {\it Gravitational Wave Data Analysis},
ed. B.~F. Schutz (Kluwer Academic Publishers, 1989), p. 59.
\bibitem{wagoner2}R. Epstein and R.~V. Wagoner, Astrophys. J
{\bf 197}, 717 (1975), and {\bf 215}, 984 (1977).
\bibitem{k_var} In fact, due to the spin-induced precession of the
orbital plane described in Sec.~\ref{spin_sec}, in general $k$ will be
a slowly varying function of time instead of a constant.  We neglect
this small effect.
\bibitem{Damour} L. Blanchet and T. Damour, Phys. Rev. D {\bf 46}, 4304 (1992).
\bibitem{cutoff} Somewhat inconsistently, we neglect the
delta-function contribution to the derivatives $\partial \tilde h(f)/
\partial {\cal M}$ and  $\partial \tilde h(f)/ \partial \mu$
that comes from varying the cut-off frequency.  While a sharp cut-off
gives a acceptable approximation to $h(f)$ (and is easy to work with
analytically), it gives a terrible approximation for the derivatives
of $h(f)$.  Our somewhat careless attitude towards the high-frequency
end of the waveform is justified by the fact that the detector noise
$S_n(f)$ rises steeply at high frequency, so very little
signal-to-noise is accumulated there.
\bibitem{correlation_effects} The pattern of how the predicted rms
errors change, when one includes extra variables in the calculation of
the Fisher matrix (\ref{sig}) which are strongly correlated with the
original variables, can be simply understood by considering the
approximation in which all but two of the variables are fixed.  The
predicted measurement accuracy for a variable $x$, when its
correlations with other variables are neglected, is $\delta x =
(h_{,x} | h_{,x})^{-1/2}$.  When we include the effects of
correlations with another variable $y$, described by the correlation
coefficient $$ c_{xy} = {\Sigma^{xy} \over \sqrt{\Sigma^{xx}
\Sigma^{yy}}} = - { (h_{,x} | h_{,y}) \over \sqrt{ (h_{,x} | h_{,x})
\, (h_{,y} | h_{,y} ) }}, $$ then from Eqs.~(\ref{sig}) and
(\ref{bardx}) we find that: (i) The rms error in $x$ is now $\Delta x
= \delta x / \sqrt{1 - c_{xy}^2}$, and thus is increased by a large
factor if $|c_{xy}|$ is close to one; and (ii) if $1 - c_{xy}^2 \ll
1$, the eigenvalues of the variance-covariance matrix
(\ref{sigma_def0}) are approximately $(\delta x^{-2} + \delta
y^{-2})^{-1}$ and $(\delta x^2 + \delta y^2) / (1 - c_{xy}^2)$, where
$\delta y = (h_{,y} | h_{,y})^{-1/2}$.  Thus, if $\delta x$ and
$\delta y$ are comparable, one linear combination of $x$ and $y$ will
be measurable with an accuracy comparable to $\delta x$, i.e., that
accuracy predicted when correlations are neglected; and the orthogonal
linear combination will have an rms error that is larger than this by
a factor $\sim 1 / \sqrt{1 - c_{xy}^2}$.
\bibitem{kidder} L.~E. Kidder, C.~M. Will and A.~G. Wiseman, Phys. Rev. D
{\bf 47}, R4183 (1993).
\bibitem{spins_paper} B.~M. Barker and R.~F. O'Connell, Gen. Rel. and
Grav. {\bf 11}, 149 (1979).
\bibitem{spins_paper1} T.~A. Apostolatos, C. Cutler, G.~J.
Sussman, and K.~S. Thorne, in preparation.
\bibitem{Meers_all}  A. Krolak, J.~A. Lobo, B.~J. Meers, Phys. Rev.
 D {\bf 43}, 2470 (1991).
\bibitem{Forward} R.~L. Forward, Phys. Rev. D {\bf 17}, 379 (1978).
\bibitem{caveat4} In this section we interpret the parameter $t_c$ to
be the time at which the coalescence would be observed by a
hypothetical detector at the origin of spatial coordinates ${\bf x}$.
\bibitem{confusion} ~
\bibitem{timing} Other detector network parameters, such as the
distances between the detectors, affect strongly the angular
resolution $\Delta {\bf n}$ of measurements of sky location, but
affect only weakly the distance measurement accuracies.  This is due
to the decoupling discussed in Sec.~\ref{draza_approx} and Appendix
\ref{decouple}.
\bibitem{caveat5} We have here ignored the fact that a factor of ${\cal
M}^{5/6}$ appears in the signal amplitudes; this is unimportant
because ${\cal M}$ will be measured to much higher relative accuracy
($\sim 10^{-3}$) than the amplitudes ($\sim 10^{-1}$).  In other words
we in fact calculate $\Delta D_1 / D_1$, where $D_1 \equiv D {\cal
M}^{-5/6}$; for all practical purposes this is the same as $\Delta D /
D$.
\bibitem{drazacorr} In fact we have compared the values of $\Delta D$
given by the approximation used in Ref.~\cite{draza} to those given by
Eq.~(\ref{final_ans}), and found that they never differ by more than
$10 \%$ for any values of $\sigma_D, \, \varepsilon_D, \, v$, and
$\psi$.
\bibitem{alternative} Another way to think about this is to associate a
signal-to-noise ratio with each polarization component (with respect
to fiducial axes determined by the detector network,
cf.~Eq.~(\ref{deltapsi}) above) of the incident waves; the distance
measurement accuracy will essentially be determined by the smaller of
the two signal-to-noise ratios.
\bibitem{prefactor} It is simplest to calculate this prefactor using
the variables $\alpha$ and $\beta$ defined by Eq.~(\ref{transform})
instead of $D$ and $v$.  In fact it diverges at $v=1$,  because of the fact
that $\partial {\bf h} / \partial \psi \, \propto \, \partial {\bf h}
/ \partial \phi_c$ in this limit.  This divergence would seem to
contradict our claim that the prefactor does not depend strongly on
$v$ and $D$.  In the special case where $\varepsilon_D=0$, a more careful
calculation of the integral over $\psi$ and $\phi_c$ (without first
expanding to quadratic order in $\psi - \psi_0$ and $\phi_c -
\phi_{c0}$) shows that the effective prefactor remains finite at $v
=1$; we expect similar behavior for $\varepsilon_D \ne 0$.  In the
case $\varepsilon_D =0$ we obtain
$$
p(v,D) \,  \propto  \, p^{(0)}(v,D) F(\omega {\hat \omega}) \,
F(\zeta {\hat \zeta}) \, e^{-(\Delta \omega^2 + \Delta
\zeta^2)/2},
$$
where $\omega = (\alpha + \beta) / \sqrt{2 \sigma_D} r_0$, $\zeta =
-(\alpha - \beta) / \sqrt{2 \sigma_D} r_0$, and ${\hat \omega}$,
$\Delta \omega$ etc. are similarly defined in terms of ${\hat
\alpha}, {\hat \beta}$ [cf.~Eqs.~(\ref{hatalpha}) and (\ref{hatbeta})
above] and $\Delta \alpha \equiv \alpha - {\hat \alpha}$, $\Delta
\beta \equiv \beta - {\hat \beta}$.  The prefactor function $F$ is
\nonumber
\begin{eqnarray}
\nonumber
F(x) & \equiv & {1 \over 2 \pi} \int_0^{2 \pi} e^{-x (1 - \cos \theta)}
d\theta \nonumber \\
\mbox{} & \approx & \left\{ \begin{array}{ll}
			1 \over \sqrt{2 \pi x}   & \mbox{\ \ \ $x \gg
1$} \nonumber \\
			1 - x                    & \mbox{\ \ \ $x \ll
1$}. \nonumber
		\end{array}
	\right.
\end{eqnarray}
Thus the prefactor is regular and slowly varying despite the apparent
divergence $\propto \,1 / \sqrt{\zeta} \propto 1 / (1 - v)$ that would
be obtained by doing a Gaussian integral over the angles $\psi$ and
$\phi_c$.
\bibitem{caveat7} Note however that these non-linear correlation
effects are typically $\alt 10 - 20 \, \%$ in the special case
$\varepsilon_D=0$, as can be shwon from Ref.~\cite{prefactor}; we
expect similar behavior for $\varepsilon_D \ne 0$.
\bibitem{loredo} T. Loredo, in {\it Statistical Challenges in Modern
Astronomy}, Eds. E.D. Feigelson and G.J. Babu, Springer-Verlag, New
York (1992).
\bibitem{Davis} M.~H.~A. Davis, in {\it Gravitational Wave Data
Analysis}, Ed. B.F. Schutz, Kluwer Acad. Pub. (1989).
\bibitem{DS} B.~S. Sathyaprakash and S.~V. Dhurandhar, Phys. Rev. D
            {\bf 44}, 3819 (1991).
\bibitem{Helstrom} C.~W. Helstrom, {\it Statistical Theory of Signal
Detection} (Pergamon Press, 2nd ed., 1968).
\bibitem{Schutz2} B.~F. Schutz, in {\it The Detection of Gravitational
Radiation} (Cambridge University Press, Cambridge, England, 1989).
\bibitem{Fernando1} F. Echeverria, Phys. Rev. D {\bf 40}, 3914 (1989).
\bibitem{Fernando} F. Echeverria, unpublished Ph.D. thesis, California
Institute of Technology, addendum to Ch. 2 (1993).
\bibitem{caveat2} In fact this method can determine all the parameters
${\bf \theta} = (\theta^1, \ldots , \theta^k)$, {\it except} for the overall
amplitude of the signal which drops out of Eq.~(\ref{snr_def_1}).  To
obtain the overall amplitude one must use Eq.~(\ref{bestfit});
see Ref.~\cite{Fernando} for more details.
\bibitem{Fisher_caveat} Throughout this paper we use the term {\it
Fisher matrix} to refer to the matrix (\ref{gamma_def}); strictly
speaking, this term as defined in, e.g., Ref.~\cite{Helstrom} refers to
a quantity which coincides with the matrix (\ref{gamma_def}) only when the
noise is Gaussian.  The Cramer-Rao inequality is usually stated in
terms of this more general Fisher matrix.
\bibitem{krolak3} For an example of such a Monte-Carlo simulation, see
K.~Kokkotas, A.~ Krolak, and G.~Tsegas, in preparation.
\end{references}
\end{document}